\documentclass[fleqn,usenatbib]{mnras}

\usepackage[T1]{fontenc}

\DeclareRobustCommand{\VAN}[3]{#2}
\let\VANthebibliography\thebibliography
\def\thebibliography{\DeclareRobustCommand{\VAN}[3]{##3}\VANthebibliography}

\usepackage{graphicx}	
\usepackage{amsmath}	
\usepackage{amssymb}	
\usepackage{dsfont}

\usepackage{newtxtext,newtxmath}

\usepackage[normalem]{ulem}

\newcommand{\klammer}[1]{\ensuremath{\left(#1\right)}}
\newcommand{\abs}[1]{\ensuremath{\left\vert#1\right\vert}}
\newcommand{\dif}{\ensuremath{\text{d}}}
\newcommand{\mathcases}[1]{\ensuremath{\begin{cases}#1\end{cases}}}
\newcommand{\outerprod}{\ensuremath{\otimes}}
\newcommand{\e}[1]{\ensuremath{\cdot 10^{#1}}}

\newcommand{\myvector}[1]{\ensuremath{\mathbfit{#1}}}
\newcommand{\mymatrix}[1]{\ensuremath{\mathbfss{#1}}}
\newcommand{\kernel}{\ensuremath{\mathcal{W}}}

\title[MFM in \textsc{OpenGadget3}]{The Cosmological Simulation Code \textsc{OpenGadget3} -- Implementation of Meshless Finite Mass}

\author[F. Groth et al.]{
Frederick Groth,$^{1}$\thanks{E-mail: fgroth@usm.lmu.de}
Ulrich P. Steinwandel,$^{2}$
Milena Valentini,$^{1,3,4}$
and Klaus Dolag$^{1,5}$
\\
$^{1}$Universit{\"a}ts-Sternwarte, Fakult{\"a}t für Physik, Ludwig-Maximilians-Universit{\"a}t M{\"u}nchen, Scheinerstr. 1, 81679 M{\"u}nchen, Germany \\
$^{2}$Center for Computational Astrophysics, Flatiron Institute, 162 Fifth Avenue, New York, NY 10010, USA\\
$^{3}$Astronomy Unit, Department of Physics, University of Trieste, via Tiepolo 11, I-34131 Trieste, Italy\\
$^{4}$ INAF - Osservatorio Astronomico di Trieste, via Tiepolo 11, I-34131 Trieste, Italy\\
$^{5}$Max-Planck-Institut f{\"u}r Astrophysik, Karl-Schwarzschild-Stra{\ss}e 1, 85741 Garching, Germany
}

\date{Accepted XXX. Received YYY; in original form ZZZ}

\pubyear{2023}

\begin{document}
\label{firstpage}
\pagerange{\pageref{firstpage}--\pageref{lastpage}}
\maketitle

\begin{abstract}
Subsonic turbulence plays a major role in determining properties of the intra cluster medium (ICM). We introduce a new Meshless Finite Mass (MFM) implementation in \textsc{OpenGadget3} and apply it to this specific problem.
To this end, we present a set of test cases to validate our implementation of the MFM framework in our code. These include but are not limited to: 
the soundwave and Kepler disk as smooth situations to probe the stability, a Rayleigh-Taylor and Kelvin-Helmholtz instability as popular mixing instabilities, a blob test as more complex example including both mixing and shocks, shock tubes with various Mach numbers, a Sedov blast wave, different tests including self-gravity such as gravitational freefall, a hydrostatic sphere, the Zeldovich-pancake, and a $10^{15}M_{\odot}$ galaxy cluster as cosmological application.
Advantages over SPH include increased mixing and a better convergence behavior. We demonstrate that the MFM-solver is robust, also in a cosmological context.
We show evidence that the solver performs extraordinarily well when applied to decaying subsonic turbulence, a problem very difficult to handle for many methods. MFM captures the expected velocity power spectrum with high accuracy and shows a good convergence behavior. Using MFM or SPH within \textsc{OpenGadget3} leads to a comparable decay in turbulent energy due to numerical dissipation. When studying the energy decay for different initial turbulent energy fractions, we find that MFM performs well down to Mach numbers $\mathcal{M}\approx 0.01$.
Finally, we show how important the slope limiter and the energy-entropy switch are to control the behavior and the evolution of the fluids.
\end{abstract}




\section{Introduction}

Turbulence plays a key role in a variety of astrophysical systems at all scales, ranging from stellar structure, star-formation in the interstellar medium (ISM) all the way up to the ICM. It leads to enhanced small-scale mixing, and contributes to the global pressure of a system.
While being mostly supersonic in the ISM, turbulence is mainly subsonic in the ICM \citep[compare, e.g.][for observations on the Coma cluster]{Schuecker+2004}.
A theoretical framework for subsonic turbulence has been provided by \citet{Kolmogorov1941}, assuming isotropy.
Simulations are an essential tool to better understand physical properties of astrophysical turbulence as well as its influence on local observables such as star formation in the ISM or its contribution to heating in the ICM.

Historically, there exist different methods to solve the hydrodynamical equations in co-moving/cosmological context. Hereby, one has the option to discretize the hydrodynamic equations by mass or volume. The former leads to the concept of ``Lagrangian'' (particle based) codes and the concept of Smoothed Particle Hydrodynamics (SPH), and the more recent Meshless Finite Mass (MFM) and Meshless Fintie Volume (MFV). The latter gives rise to the concept of ``Eulerian'' (grid based) codes and the Godunov finite volume approach.

Popular SPH codes include \textsc{Gadget} in the different versions including \textsc{Gadget-1} \citep{Springel+2001}, \textsc{Gadget-2} \citep{Springel2005}, and \textsc{Gadget-4} \citep{Springel+2021}, \textsc{Phantom} \citep[][]{Lodato&Price2010,Price+2018} and \textsc{gasoline} \citep[][]{Wadsley+2004,Wadsley+2017}. An improved SPH scheme with non-standard enhancements has been implemented in \textsc{magma2} \citep{Rosswog2020}.
MFM has been implemented in e.g. \textsc{gizmo} \citep{Hopkins2015}, \textsc{GANDALF} \citep{Hubber+2018}, \textsc{Gadget-3} \citep{Steinwandel+2020}, and \textsc{pkdgrav-3} \citep{AlonsoAsensio+2023}.

Mesh codes exist in two flavors: either as a stationary mesh, possibly with adaptive mesh refinement, as implemented e.g. in \textsc{Zeus} \citep[][]{Stone&Norman1992}, \textsc{TVD} \citep{Ryu+1993,Ryu+1998}, \textsc{Enzo} \citep{Bryan+1995, Bryan+2014}, \textsc{FLASH} \citep{Fryxell+2000}, \textsc{RAMSES} \citep{Teyssier2002}, \textsc{athena} \citep{Stone+2008}, and \textsc{athena++} \citep{Stone+2020} or as a moving mesh as in \textsc{Arepo} \citep{Springel2010,Weinberger+2020} and \textsc{shadowfax} \citep[][]{Vandenbroucke&DeRijcke2016}. The latter have the advantage of being Pseudo-Lagrangian.
While mesh codes as well as MFM employ a Godunov-method and calculate fluxes between neighbors \citep{Godunov1959}, SPH directly retrieves the hydrodynamical fluid vectors from the kernel density estimation that is obtained by adopting a weighted sum over a certain (typically non-constant) number of neighbors.

All of them can be used for computations of turbulence, with earlier calculations primarily carried out in the supersonic regime, relevant in the ISM for regulating star formation. Many results have been obtained assuming driven turbulence in which an energy input at large scales is provided during the whole simulation.
In contrast to driven turbulence, we expect decaying turbulence to be present in galaxy clusters. Turbulence is injected at large scales for example due to collapse of large scale structure and subsequent merger activity \citep{Roettiger&Burns1999,Subramanian+2006}, after which it energy is transported down to the smaller scales (``turbulent cascade'') on which it is dissipated (generally below the resolution scale of any given code).

In the series of papers by \citet{Federrath+2008,Federrath+2009,Federrath+2010}, they have used a stationary grid code to calculate turbulent boxes with driven turbulence. They found that the choice of the driving scheme plays an important role in determining properties of the resulting turbulence, leading to significant differences in the density statistics. Their results suggest a different mixture of driving-mechanisms for different star forming regions. Overall, they found good agreement with observations as well as other results, independent of the driving-mechanism employed. More recently, \citet{Federrath+2021} increased the resolution to even resolve the sonic scale, starting from supersonic turbulence with a resolution of $\sim 10000^3$ cells.

\citet{Kitsionas+2009} and \citet{Price&Federrath2010} also compared the performance of different implementations of SPH and hydro schemes with a stationary mesh, and find good agreement between these two methods at high Mach numbers. Mesh codes are more efficient to obtain volumetric statistics such as the power spectrum, while SPH recovers the high-density tail better due to automatically adapting the resolution.

While all these methods work well in the supersonic turbulent regime, they have problems dealing with subsonic turbulence.
Going to smaller Mach numbers ($\mathcal{M}$) \citet{Padoan+2007} showed that SPH performs sub-optimum when compared to finite volume methods.
Based on this work, \citet{Bauer&Springel2012} studied the capabilities of SPH for subsonic turbulence at $\mathcal{M}=0.3$. They found that classic (vanilla) SPH fails in reproducing the expected velocity power spectrum as well as the dissipation range. Reasons are mainly the artificial viscosity scheme used and velocity noise introduced by the kernel.
These results raised the general question of whether SPH can deal with subsonic turbulence to begin with.

An answer has been provided by \citet{Price2012} who showed that these limitations are not intrinsic to SPH, but rather a consequence of some SPH setups adopted to study subsonic turbulence.
In contrast to what previous studies reported, SPH can capture the expected power spectrum by using more modern formulations of SPH that are able to reduce artificial viscosity in subsonic regimes.

The role of subsonic turbulence in galaxy clusters has been analyzed both from observational and theoretical perspectives. Simulations of turbulence in the ICM have been carried out mostly using grid codes \citep{Vazza+2009,Vazza+2018,Mohapatra+2021,Mohapatra+2022,Iapichino&Niemeyer2008a,Iapichino+2017}. 
\citet{Miniati2014,Miniati2015} found a lack of turbulent energy at small scales depending on the refinement technique. In addition, they discussed the importance of microphysics for the evolution of turbulence. A possible improvement for modeling turbulence  has been presented by \citet{Maier+2009a} combining AMR with large eddy simulations.
Simulations by \citet{Dolag+2005b} have shown that also SPH can model turbulence in galaxy clusters when properly reducing artificial viscosity.

In addition to the impact on gas dynamics, turbulence is responsible for amplifying magnetic fields through a turbulent dynamo. 
Simulations by \citet{Schekochihin+2001,Schekochihin+2004} and \citet{Steinwandel+2021} have focused on this turbulent dynamo, analyzing its growth. Another work of \citet{Kritsuk+2020} has focused again on turbulent boxes with stochastic forcing, comparing different hydrodynamical methods and finding reasonably good convergence but significant differences in computational costs.

More recently, \citet{Sayers+2021} have compared simulated clusters to observed ones. Especially, there should be a difference depending on the dynamical state, with more relaxed clusters showing less turbulence. Simulations, however, do not always find such a difference. Thus, it is important to accurately capture the turbulent cascade and the decay in turbulent energy. While the latter would require including additional microphysics such as viscosity, the former also depends on the hydro-scheme.

We use Meshless Finite Mass (MFM) as an alternative, newer method to the aforementioned ones to study subsonic turbulence. MFM combines ideas of SPH with those of a moving mesh and thus aims at solving several of their individual issues.
The development of MFM goes back to first ideas presented by \citet{Vila1999} and Godunov SPH  \citep{Inutsuka2002,Cha&Whitworth2003}, which was still unstable, and to a Meshless Finite Element Method suggested by \citet{Idelsohn+2003}, until the nowadays used version first formulated by \citet{Lanson&Vila2008,Lanson&Vila2008a}.
We present a new implementation in the \textsc{Gadget} derivative \textsc{OpenGadget3}, originally based on that in the code \textsc{GANDALF}, where the skeleton of the MFM implementation has been originally taken from their code-base and then adjusted. Several extensions allow its use in cosmological simulations compared to the implementation in \textsc{GANDALF} that is focused on star and planet formation. This allows for a stable baseline framework for applications on  scales of star and planet formation that we extend into the cosmological integration framework of \textsc{OpenGadget3}, which is a re-base of \textsc{Gadget-2} with the ability to be compiled with C++ compilers, and making vast use of templating.
It comes with modules containing state-of-the-art physics and sub-resolution models, as for instance: self-interacting dark matter \citep{Fischer+2022}, magneto-hydrodyanamics (MHD) \citep{Dolag&Stasyszyn2009,Stasyszyn+2013}, thermal conduction \citep{Arth+2014}, cosmic rays \citep{Boss+2023}, star formation and stellar/blackhole feedback according to the \textit{Magneticum}-model \citep[][]{Springel&Hernquist2003a,Tornatore+2003,Tornatore+2004,Tornatore+2007,Hirschmann+2014,Steinborn+2015,Dolag2015} or with the MUPPI (MUlti Phase Particle Integrator) extension for non-equilibrium star formation \citep{Murante+2010,Murante+2015,Valentini+2017,Valentini+2020}. These extensions are so far coupled only to the SPH hydro-solver. Further work will be required to couple them also to MFM.

To make use of modern computer architectures, \textsc{OpenGadget3} includes a hybrid MPI-OpenMP parallelization.
In addition, calculations of gravity, density, SPH hydro-force and thermal conduction, can be carried out on GPUs. These modules requiring most of the runtime \citep{Ragagnin+2020} GPU offloading can be useful for some applications, leading to a speed up by a factor of a few (2-4, depending on the exact application). The long-term goal is to have a fully publicly available updated \textit{Gadget} version for OpenMP and OpenACC.

Before the introduction of this paper the code was solving the hydrodynamical equations using modern SPH as formulated by \citet{Springel&Hernquist2002}, including modern, time-dependent artificial viscosity \citep{Beck+2016} and conduction \citep{Price2008}. With the new implementation of MFM as a modern meshless method, we can combine advantages both of this method and efforts previously made to optimize the pre-existing code base. This also involves a treatment in order to evolve strong shocks for which we need the timestep limiter to be non-local which is ensured by a wakeup scheme \citep{Saitoh&Makino2009,Pakmor2010,Pakmor+2012}. \textsc{OpenGadget3} closely follows the implementation described by \citet{Beck+2016}.

A main goal of this paper is to use Meshless Finite Mass to study decaying, subsonic turbulence, as present in galaxy clusters.
To this end, we present a new implementation in the cosmological simulation code \textsc{OpenGadget3} as an alternative hydro-solver to the currently implemented SPH.

This paper is structuered as follows. We first describe the codebase of \textsc{OpenGadget3} including its SPH implementation in Sec.~\ref{sec:numerics}.
We continue with a brief overview on MFM and a description of our MFM implementation in Sec.~\ref{sec:MFM}.
In Sec.~\ref{sec:tests}, we use a suite of test cases, each probing specific aspects and properties of the code, to validate the performance of our MFM implementation. All settings are kept exactly the same between test cases, independent of the individual test case, without further tuning. We continue with an analysis of decaying subsonic turbulence with our new implementation presented in Sec.~\ref{sec:turbulence}.
In all cases, comparisons between different codes and methods are provided, including MFM and SPH in \textsc{OpenGadget3}, MFM in \textsc{gizmo} and a moving and stationary mesh in the publicly available \textsc{Arepo} version.
We analyze the effect of specific numerical parameters in Sec.~\ref{sec:parameters}.
Our main findings are discussed in Sec.~\ref{sec:discussion}.

Additional material such as the formulation of the slope-limiters and a comparison of the Riemann solvers implemented are presented in App.~\ref{app:limiters} and~\ref{app:Riemann}, respectively. 

\section{\textsc{OpenGadget3} -- Backbone} \label{sec:numerics}

Solving the system of differential equations describing the evolution of the gas requires discretizing them. In the temporal dimension a sufficiently small timestep $\Delta t$ is introduced.
The spatial discretisation can be obtained using various different approaches. In \textsc{OpenGadget3}, hydrodynamics is discretized either using Smoothed Particle Hydrodynamics (SPH) or with the newly implemented Meshless Finite Mass (MFM). Gravity is solved by a TreePM method.

\subsection{Integrator and Timestepping} \label{sec:integrator}
For the time integration, we employ a Leapfrog scheme in kick-drift-kick (KDK) form to achieve second order accuracy \citep[compare, e.g.,][]{Hernquist&Katz1989} in the implementation following \citet{Verlet1967,Springel2005}.

Starting from values at timestep number $n$, velocities $\myvector{v}$ are updated in a first half-step kick. It is followed by drifting the positions $\myvector{r}$, and another, second half-step kick:
\begin{align}
    \myvector{v}^{n+1/2} =~& \myvector{v}^{n} + \frac{1}{2} \myvector{a}^{\tilde n} \Delta t\\
    \myvector{r}^{n+1} =~& \myvector{r}^{n} + \myvector{v}^{n+1/2} \Delta t\\
    \myvector{v}^{n+1} =~& \myvector{v}^{n+1/2} + \frac{1}{2}\myvector{a}^{\widetilde{n+1}} \Delta t.
\end{align}
The acceleration $\myvector{a}^{\tilde n}=\myvector{a}_{\text{hydro}}^{\tilde n}+\myvector{a}_{\text{grav}}^{n}$ consists of hydrodynamical accelerations $\myvector{a}_{\text{hydro}}$ and gravitational accelerations $\myvector{a}_{\text{grav}}$. Following the operator splitting approach, they are calculated separately.
Gravity and hydrodynamical accelerations are evaluated between the drift and the second half-kick. Gravitational interactions depend only on the position, and can thus be calculated at timestep $n$. While for traditional SPH entropy does not change, such that also a calculation of the hydrodynamical forces at timestep $n$ is possible, there is an evolution of entropy and also a velocity dependence for the viscous terms of modern SPH. Thus, we use predicted values based on the changes calculated during the previous timestep, updated at the drift. The dependence of the predicted variables is indicated by $\tilde n$.

For SPH, the entropic function 
\begin{align}
    A =~& (\gamma-1)U/\rho^{\gamma-1}
\end{align}
is integrated in two half-steps at the kicks
\begin{align}
    A^{n+1/2} = A^{n} + \frac{1}{2}\klammer{\frac{\dif A}{\dif t}}_{\text{hydro}}^{\tilde n}\Delta t,\\
    A^{n+1} = A^{n+1/2} + \frac{1}{2}\klammer{\frac{\dif A}{\dif t}}_{\text{hydro}}^{\widetilde{n+1}}\Delta t.
\end{align}
While traditional SPH is conserving entropy, it is allowed to change in the modern SPH implementation due to e.g. artificial viscosity and conductivity.

\textsc{OpenGadget3} uses hierarchical timestepping to ensure synchronization, while allowing adaptive timesteps, depending on different timestep limiters such as a Courant-like timestep criterion
\begin{align}
    \Delta t_i^{\text{Courant}} =~& \frac{C_{\text{Courant}}ah_i}{c_{\max,i}}
\end{align}
with maximum signal velocity over the neighbors $c_{\max,i}$, scale factor $a$, smoothing length $h_i$, and free parameter $C_{\text{Courant}}$,
as described by \cite{Springel2005}.
Timesteps are chosen as the largest timestep that fulfills $\Delta t_i = 2^{-n}\Delta t_{\text{global}}\leq\Delta t_i^{\text{Courant}}$ with timebin $n\in \mathbb{N}_0$. 

Accelerations are calculated only for active particles, which are in synchronization with the current time-step, while they are not modified for particles located on a smaller time bin, corresponding to larger time-steps.

For strong shocks large differences can occur between the timesteps of close-by particles. This is avoided by a wake-up scheme, described in more detail by \citet{Pakmor2010,Pakmor+2012,Beck+2016}.
\textsc{OpenGadget3} uses a criterion based on the signal velocity. If for any neighbor $j$, the signal velocity varies strongly $c_{\max,i}>f_w c_j$ with tolerance factor $f_w=3$, wake-up is triggered. In this case, the particle is considered active and moved to a shorter timestep, such that synchronization is still ensured.

While this scheme will break conservation, it works reasonable well and avoids numerical errors in strong shocks.

\subsection{Gravity Solver -- TreePM} \label{sec:treepm}
The accurate treatment of gravity is of great importance for cosmological simulations \citep{Springel2010}. In principle, it can be solved accurately by a direct summation, which is, however, computationally expensive ($\mathcal{O}(N^2)$). Instead, we follow the much more efficient combined Oct-Tree-Particle Mesh (PM) approach \citep[][]{Xu1995,Bode+2000, Springel2005, Springel2010, Springel+2021}. \textsc{OpenGadget3} mainly follows the implementation in \textsc{Gadget-2}, which has been extensively described by \citet{Springel2005}.
In the following, we briefly review the main concept. The potential is split into short-range and long-range contributions. Short-range forces are calculated following the oct-tree algorithm, while long-range forces are calculated using a particle mesh. The idea of a tree algorithm has been proposed by \citet{Appel1985} and \citet{Barnes&Hut1986}. Nodes of an oct-tree are constructed by splitting the domain into a sequence of cubes. Force-contributions of nodes satisfying an opening angle criterion are calculated. For numerical reasons to keep the equation linear with respect to adding and removing particles from nodes, only the monopole contributions are taken into account. The implementation in \textsc{Gadget} has been described by \citet{Springel+2001}.
The total gravitational acceleration of particle $i$ from other nodes/particles $j$ with mass $m_j$ at location $\myvector{r}_{ij}$ relative to particle $i$ and with (gravitational) softening length $\epsilon_j$ is given by
\begin{align}
    \myvector{a}_{\text{grav},i} =~& G\sum_{j}^{N_\text{tot}}\myvector{r}_{ij}\mathcases{\frac{m_j}{r_{ij}^3}~&\text{if }r_{ij}>\epsilon_j\\
    \frac{m_j}{\epsilon_j^3}\text{Corr}(r_{ij}/\epsilon_j)~&\text{if }r_{ij}\le \epsilon_j,}
\end{align}
with total number of particles and nodes $N_{\text{tot}}$.
$\text{Corr}$ is a correction term, taking into account the softening. $G$ is the gravitational constant. For the particle mesh \citep[][]{Eastwood&Hockney1974}, all particles are assigned to grid-cells, such that a discrete Fourier-transformation can be calculated, with the gravitational potential $\Phi_k$ in Fourier space at wavenumber $k$ being calculated as
\begin{align}
    -k^2\Phi_k =~& 4\pi G \rho_k.
\end{align}
Corrections for small-range truncation as well as periodic boundaries are applied by multiplications in Fourier space. The gravitational potential in real space is calculated as inverse Fourier-transform, and is interpolated to the original particle positions to finally obtain gravitational accelerations.
\textsc{OpenGadget3} uses the more modern FFTW3 (``Fastest Fourier Transform in the West'') library \citep[][]{FFTW3} instead of FFTW2 for the implementation of the Fourier transform.

\subsection{Hydrodynamical Solver -- SPH} \label{sec:SPH}

For Smoothed Particle Hydrodynamics (SPH), the domain is decomposed into a finite number of ``particles''. The physical quantities at each point are represented by contributions of close-by (neighboring) particles weighted by a kernel $\kernel_i(r_i,h_i)$, depending on the distance $r_i$ from particle $i$, and its smoothing length $h_i$.  The kernel has to be continuous, radially symmetric, have compact support and fulfill the limit $\lim_{h\to 0}\kernel = \delta $, but otherwise can be chosen arbitrarily. \textsc{OpenGadget3} offers the choice between different commonly used kernels, including a cubic spline \citep{Monaghan&Lattanzio1985}, quintic spline \citep{Morris1996}, or a Wendland C2/C4/C6 kernel \citep{Wendland1995,Dehnen&Aly2012}. The effective volume of each particle is well approximated by $V_i^{-1}=\kernel(r_{i})$, such that the density follows as
\begin{align}
    \rho(\myvector{r}_i) =~& \sum_{j\in\text{Ngb}} m_j \kernel\klammer{\abs{\myvector{r}_i-\myvector{r}_j},h_i}, \label{eq:density}
\end{align}
summing over the neighboring particles (Ngb).
We allow for adaptive smoothing, automatically increasing resolution in high-density regions compared to low-density ones.
Smoothing length and effective neighbor number $N_{\text{Ngb}}$ are related to the density via:
\begin{align}
    \frac{4\pi}{3}\rho_ih_i^3 =~& \bar m N_\text{Ngb} \label{eq:sph_smoothing}
\end{align}
with mean neighbor mass $\bar m$.
As Eqns.~(\ref{eq:density}) and~(\ref{eq:sph_smoothing}) are coupled for fixed neighbor number, one solves for smoothing length and density iteratively via finding roots. Quantities other than the density, labeled with $X$, are approximated via
\begin{align}
    X(\myvector r_0) \approx~& \sum_{i\in\text{Ngb}}\frac{X_i}{\rho_i}\kernel(\abs{\myvector r_0-\myvector r_i},h_{i})m_i.
\end{align}

Different formulations of the hydrodynamical acceleration can be derived. In \textsc{OpenGadget3} the fully conservative formulation for the hydrodynamical acceleration \citep{Springel&Hernquist2002}
\begin{align}
    \myvector a_{\text{hydro},i} =~& -\sum_{j\in\text{Ngb}} m_j\left( f_i\frac{P_i}{\rho_i^2}\nabla_i\kernel_{ij}(h_i) + f_j\frac{P_j}{\rho_j^2}\nabla_i \kernel_{ij}(h_j)\right),\\
    f_i =~& \left(1 +  \frac{h_i}{3\rho_i}\frac{\partial\rho_i}{\partial h_i}\right)^{-1}
\end{align}
is utilized. Instead of calculating gradients of physical quantities, all spatial derivatives are expressed by gradients of the kernel function. Traditional SPH has problems dealing with shocks, as well as reproducing mixing instabilities \citep[][]{Morris1996,Agertz+2007}. These issues can be resolved by including artificial viscosity and conductivity.
In \textsc{OpenGadget3}, time and spatial dependent artificial viscosity \citep{Beck+2016} and artificial conductivity \citep{Price2008} are utilized, minimizing their impact in regions where they are not desired.

\section{Meshless Finite Mass} \label{sec:MFM}

As a second, newly implemented option, the hydrodynamical equations can be discretized and solved following the Meshless Finite Mass (MFM) approach. This method conceptually combines SPH with a moving mesh, calculating fluxes between neighboring particles in a scheme otherwise similar to SPH, including weighting by a kernel. Thus, it is combining advantages of both methods. In contrast to SPH, the domain associated to a particle is not spherical, but rather corresponds to a smoothed Voronoi tesselation \citep{Hopkins2015}.

\subsection{Basic Hydrodynamical Equations} \label{sec:hydro}
The evolution of any ideal fluid is described by three main equations. Mass conservation leads to the continuity equation.
The second equation is an equation of motion (Euler's equation), corresponding to Newton's second law. Energy conservation is ensured by the first law of thermodynamics. Within an inertial frame of reference, all these equations can be combined into
\begin{align}
    \frac{\dif\myvector{U}}{\dif t} + \nabla\cdot\klammer{\mymatrix{F}-\myvector{v}_\text{frame}\outerprod \myvector{U}} =~& \myvector{S}
    \label{eq:hydro}
\end{align}
with outer product $\outerprod$ and, for pure hydrodynamics, field vector $\myvector{U}=(\rho,\rho \myvector{v},\rho e)$, flux $\mymatrix{F}=(\rho \myvector{v}, \rho \myvector{v}\outerprod\myvector{v} + P\mathds{1}, \klammer{\rho e + P}\myvector{v})$ and vanishing source terms $\myvector{S}=\myvector{0}$. 

In total, Eqn.~(\ref{eq:hydro}) provides 5 constraints for 6 variables: fluid density $\rho$, energy density $e$, pressure $P$, and the three components of the velocity $\myvector{v}$. The missing constraint is provided by an equation of state, connecting the pressure to the internal energy density $u$. For an ideal gas it takes the form
\begin{align}
    P =~& \klammer{\gamma-1}\rho u
\end{align}
where the adiabatic index $\gamma$ amounts to $5/3$ if the gas is monoatomic.

\subsubsection{Equations in an Expanding Universe}
In a cosmological context, the expansion of the universe has to be taken into account. One possibility is to re-write Eqn.~(\ref{eq:hydro}) for a universe with scale factor $a$, accounting for these effects, as realized e.g. in \textsc{Gadget-1}:
\begin{align}
    \frac{\partial\myvector{v}}{\partial t} + \frac{1}{a}\klammer{\myvector{v}\cdot\nabla}\myvector{v} + \frac{\dot a}{a}\myvector{v} =~& -\frac{1}{a\rho}\nabla P - \frac{1}{a}\nabla \Phi, \\
    \frac{\partial \rho}{\partial t} + \frac{3\dot a}{a}\rho + \frac{1}{a}\nabla\cdot\klammer{\rho\myvector{v}} =~& 0, \\
    \frac{\partial}{\partial t}\klammer{\rho u} + \frac{1}{a}\vec v\cdot\nabla\klammer{\rho u} =~& -\klammer{\rho u+P}\klammer{\frac{1}{a}\nabla\cdot\myvector{v} + 3\frac{\dot a}{a}}.
\end{align}

In \textsc{OpenGadget3} we follow a different approach, and do calculations using the so called super-co-moving coordinates, as first introduced by \citet{Martel&Shapiro1998}. Code units (denoted by subscript $c$) are related to physical units ($p$) via
\begin{align}
    x_c =~& a^{-1}x_p \label{eq:code-phys_x} \\
    \rho_c =~& a^3 \rho_p \\
    v_c =~& a v_p \\
    P_c =~& a^{3\gamma} P_p \\
    u_c =~& a^{3(\gamma-1)}u_p, \label{eq:code-phys_u}
\end{align}
such that Eqn.~(\ref{eq:hydro}) keeps the same form when written in code units except for an additional contribution in the energy and momentum evolution due to the Hubble expansion. As we use peculiar velocities, the explicit dependence on the Hubble flow for momentum is absorbed into the choice of units.

\subsection{MFM Discretization}
Mathematically, Eqn.~(\ref{eq:hydro}) is discretized by multiplying by a partition function 
\begin{align}
    \psi_i =~& \frac{1}{\sum_{j\in\text{Ngb}} \kernel_j}\kernel_i
\end{align}
where $\kernel_k=\kernel(\abs{r-r_k},h_k)$ and integrating over the volume. A more detailed derivation  has been provided by \citet{Lanson&Vila2008,Gaburov&Nitadori2011}. In this paper, we only focus on the key results relevant for the implementation. For every particle $i$ changes in the quantities $\myvector{U}_i=(\rho_i, \rho_i v_i, \rho_ie_i)$ are given by source terms $\myvector{S}_i$, which vanish for pure hydrodynamics, and pairwise fluxes $\mymatrix{F}_{ij}$ with the neighbors $j$
\begin{align}
    \frac{\dif}{\dif t}\klammer{V_i\myvector{U}_i}^{\tilde n} + \sum_{j\in\text{Ngb}}\klammer{\mymatrix{F}_{ij}^{\tilde n}\cdot \myvector{A}_{ij}^{\text{eff},n}} =~& \myvector{S}_i^{\tilde n}V_i^{n}. \label{eq:mfm_main}
\end{align}
Calculating pairwise fluxes automatically ensures mass, momentum and energy conservation.
The effective interface area $\myvector{A}_{ij}^{\text{eff}}$ depends on the partition function and effective volume $V_i$,
\begin{align}
    \myvector{A}_{ij}^\text{eff} =~& V_i\tilde{\boldsymbol{\psi}}_j - V_j\tilde{\boldsymbol{\psi}}_i.\label{eq:mfm_face_area}
\end{align}
where
\begin{align}
    \tilde\psi_j^\alpha(\myvector{x}_i) =~& B_i^{\alpha\beta}(\myvector{x}_j-\myvector{x}_i)^\beta\psi_j(\myvector{x}_i) \label{eq:tildepsi}
\end{align}
with Einstein summation convention over $\beta$ in Eqn.~(\ref{eq:tildepsi}). The matrix $\mymatrix{B}$ is chosen in order to be second order accurate \citep{Lanson&Vila2008}
\begin{align}
    \mymatrix{B}_i =~& \mymatrix{E}_i^{-1}\\
    E_i^{\alpha\beta} =~& \sum_{j\in\text{Ngb}}(\myvector{x}_j-\myvector{x}_i)^\alpha (\myvector{x}_j-\myvector{x}_i)^\beta \psi_j(\myvector{x}_i).
\end{align}

Also the effective volume depends on the integrated partition function and can be expressed in terms of the number density $n_i$:
\begin{align}
    V_i =~& \int \psi_i \approx n_i^{-1}
\end{align}

For highly unisotropic particle arrangements, the matrix $\mymatrix{E}$ can become ill-conditioned, preventing an accurate numerical matrix inversion. As described by \citet{Hopkins2015} we use the condition number $N_{\text{cond},i}=N_{\text{dimensions}}^{-1}\sqrt{\abs{\abs{\mymatrix{E}_i^{-1}}}\abs{\abs{\mymatrix{E}_i}}}$ as measure of how well-conditioned the matrix is. For $N_{\text{cond},i}>100$ gradients are calculated only first order in an SPH-like way.

Most importantly, no tessellation has to be calculated explicitly as it would be necessary for a moving mesh, but an SPH-like neighbor search is used, drastically reducing the computational costs compared to the mesh reconstruction.

In contrast to SPH, for which the mass density is estimated according to Eqn.~(\ref{eq:sph_smoothing}), for MFM the number density $n_i$ is estimated together with the smoothing length in an iterative process, solving 
\begin{align}
    n(\myvector{r}_i) =~& \sum_{j\in\text{Ngb}} \kernel\klammer{\abs{\myvector{r}_i-\myvector{r}_j},h_i}, \\
    \frac{4\pi}{3}n_ih_i^3 =~& N_\text{Ngb}.
\end{align}
Also for MFM, $N_{\text{Ngb}}$ corresponds to the effective neighbor number.

The flux in Eqn.~(\ref{eq:mfm_main}) is calculated numerically using a Riemann solver, where we use an exact Riemann solver, following the implementation by \citet{Toro2009} with a tolerance of $10^{-4}$, and a maximum of $8$ iterations. Alternatively, we implemented the Riemann-solver that provides an exact solution to the linearized system of equations \citep[Roe-solver,][]{Roe1981}, as well as the two most common flavors of a Harten-Lax-van-Leer solver (HLL) 
and HLLC \citep[][]{Toro2009}. For all these, the exact Riemann solver is used as fallback in case the faster, approximate solver fails. The effect of the choice of the solver is discussed in more detail in App.~\ref{app:Riemann}.

\begin{figure}
    \centering
    \includegraphics[width=0.8\linewidth]{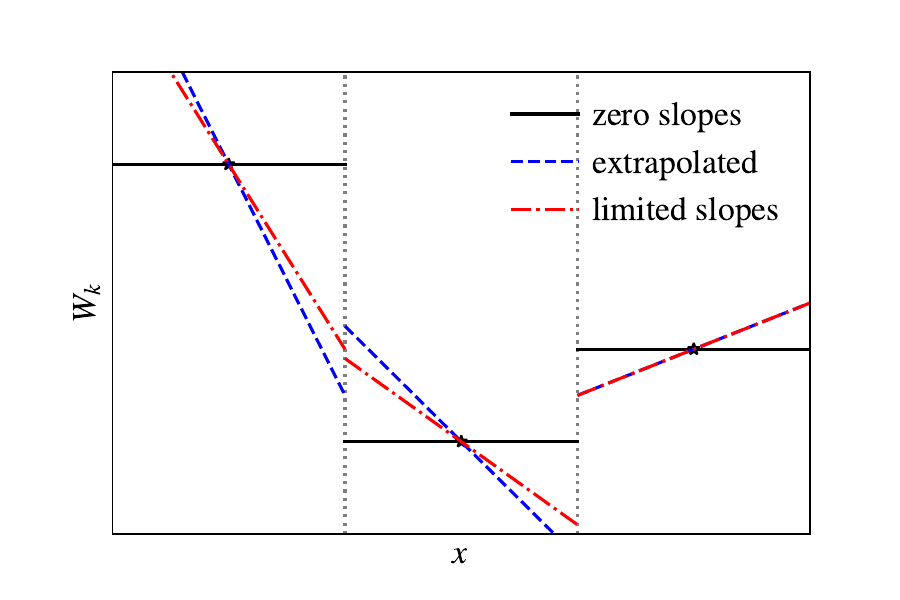}
    \caption{Sketch of extrapolation from central particle/cell values to face values. Using the central values corresponds to a zeroth order interpolation, leading to a first order scheme (black solid lines). It can be extended to be second order by extrapolating using a slope defined by neighboring particles/cells (blue dashed line), which however can lead to over-/undershooting at the faces (see left face) or even negative densities/pressures (see right face). This issue can be solved by limiting the slopes using different procedures (red dash-dot line). See text for further details.}
    \label{fig:slope_limiter}
\end{figure}

The Riemann solver requires knowledge about velocity, density and pressure values at the interfaces, summarized in the primitive fluid vector
\begin{align}
    \myvector{W} =~& \begin{pmatrix} \rho \\ \myvector{v} \\ P \end{pmatrix}.
\end{align}
In principle, values at the particle center can be used directly, following a zeroth order interpolation. While such a scheme would be stable, it is only first order accurate and very diffusive \citep{Godunov1959,Barth&Jespersen1989}. To this end, we follow a two-step approach, as illustrated in Fig.~\ref{fig:slope_limiter}, similar to what is usually done for grid-based methods and in other MFM implementations.
In a first step, gradients of the primitive fluid vector are calculated using a second-order accurate matrix gradient estimator
\begin{align}
    (\nabla\outerprod\myvector{W})_i^\alpha =~& \sum_{j\in \text{Ngb}} \klammer{\myvector{W}_j - \myvector{W}_i}\tilde\psi_j^\alpha(\myvector{x}_i).
\end{align}
The position and velocity of the face is estimated via
\begin{align}
    \dif \myvector{r}_{ij}^\text{frame} =~& \dif \myvector{r}_{ij} s_i, \\
    \myvector{v}_{ij}^\text{frame} =~& s_j \myvector{v}_j + s_i \myvector{v}_i,
\end{align}
where we set
\begin{align}
    s_{i} =~& \frac{h_i}{h_i+h_j}
\end{align}
to be second order accurate instead of $s_i=1/2$ for a first-order accurate interpolation.

By choosing the reference frame corresponding to the rest-frame of the interface, the scheme becomes Lagrangian. In MFM, also the boundaries are assumed to deform in a Lagrangian way, eliminating mass fluxes between neighbors. As the actual face velocity and deformation does not exactly correspond to the one assumed during a timestep, second order errors are introduced \citep{Hopkins2015}.
An alternative is allowing for mass fluxes using the Meshless Finite Volume (MFV) method, which, however, also is only second order accurate. In addition, it has been shown that MFV can run into problems by draining the mass for particles accelerated into low density environments in cosmological simulations \citep{AlonsoAsensio+2023}. For this reason, we do not use this scheme here but focus on the MFM method.
An additional advantage of MFM and finite volume schemes in general over SPH is that no additional dissipation terms are necessary.

The face-values are extrapolated according to
\begin{align}
    \myvector{W}_i^\text{frame} =~& \myvector{W}_i + \dif \myvector{r}_i^\text{frame}\cdot\nabla \outerprod\myvector{W}_i. \label{eq:mfm_face_value}
\end{align}
To avoid over- or undershooting or even unphysical, negative densities or pressures when strong gradients are present in the fluid, these gradients are reduced by a factor $\nabla W_{i,k}\to \alpha_{i,k}\nabla W_{i,k}$, $0\le \alpha_{i,k}\le 1$ in a second step in the face interpolation where $\alpha_{i,k}$ can be different for each particle $i$ and component $k$.
We implement different options for such a slope-limiter, including a total variation diminishing (TVD) one \citep{Duffell&MacFadyen2011}, the one from \textsc{Arepo} \citep{Springel2010} originally presented by \citet{Barth&Jespersen1989}, the scalar limiter from the \textsc{GANDALF} code \citep{Hubber+2018}, and the one used in the \textsc{gizmo} code \citep{Hopkins2015}, described further in App.~\ref{app:limiters}. In addition, the pairwise limiter according to the \textsc{gizmo} code can be used.

In a third, final step the Riemann solver is used to calculate fluxes, which can then be converted to hydrodynamical acceleration and energy changes.\footnote{As the Riemann solver requires physical units instead of (co-moving) code units, variables have to be converted accordingly \citep[compare also][App. H5]{Hopkins2015}. As flux calculations are done at the interface, no Hubble expansion has to be taken into account for the momentum changes.} All these steps are only applied to particles which currently reside in an active time-bin. While this workflow is computationally convenient, it makes the scheme less exact, as old gradients are used for the flux calculation. Nevertheless, the scheme still performs accurate enough in practical applications, as also argued by \citet{Hopkins2015}.

In addition, in our implementation fluxes are updated only for the active particle, which breaks conservation. This could be improved by updating fluxes for both particles and only considering neighbors on lower timebins. As we found no significant disadvantage for practical applications, we kept the computationally more convenient version.

\subsection{Energy-Entropy Switch} \label{sec:switch}
While the Riemann solver outputs total energy changes, the rest of the code requires internal energies. Total energy itself is never used in the code. The total energy change can straightforwardly be converted into internal energy change starting from Eqn.~(26) of \citet{Gaburov&Nitadori2011}, rewriting it as a difference equation, as we have small, but finite timesteps
\begin{align}
    \klammer{\frac{\dif U}{\dif t}}^{n} =~& \klammer{\frac{\dif E_{\text{tot}}}{\dif t}}^{n} - \klammer{\frac{\dif }{\dif t}\klammer{\frac{1}{2}m\myvector{v}^2}}^{n} \\
    \approx~& \klammer{\frac{\dif E_{\text{tot}}}{\dif t}}^{n} - \frac{1}{2}m^{n}\klammer{\frac{\klammer{\myvector{v}+\dif\myvector{v}}^2-\myvector{v}^2}{\dif t}}^{n}\\
    \approx~& \klammer{\frac{\dif E_\text{tot}}{\dif t}}^{n} - m^n\klammer{\myvector{v}^{n}+\frac{1}{2}\klammer{\frac{\dif \myvector{v}}{\dif t}}^{n}\Delta t^n}\cdot \klammer{\frac{\dif \myvector{v}}{\dif t}}^{n}. \label{eq:mfm_energy_conversion}
\end{align}
The velocity change can be calculated directly from the momentum change returned by the Riemann solver, as for MFM the mass is kept constant. Thus, both time derivatives of total energy and velocity can be obtained from the Riemann solver output.
We introduce the additional term $\frac{1}{2}\klammer{\frac{\dif \myvector{v}}{\dif t}}^{n}\Delta t^n$ in the bracket, which is a second order correction and improves the accuracy in the discretized equation, which is a result of discrete timesteps. While this transformation from total to internal energy does not conserve total energy to machine-precision, it increases the precision in the evolution of the internal energy itself. For very cold flows, the internal energy evolution is still dominated by numerical errors. This is avoided by assuming purely adiabatic changes in these rare cases.  We follow the idea of the implementation in the \textsc{gizmo} code, where the switch is only active for specific test problems such as the Zeldovich pancake. If active, internal energy 
\begin{align}
    U_{\text{est},i} =~& U_i + \dif U_i
\end{align}
is compared to potential and/or kinetic energy
\begin{align}
    E_{\text{pot},i} =~& m_ia_\text{grav}\cdot 0.5h_i, \\
    E_{\text{kin},i} =~& 0.5 m_i \max_{j\in \text{Ngb}}\klammer{\myvector{v}_j-\myvector{v}_i}^2.
\end{align}
If the internal energy is small enough compared to other energy contributions
\begin{align}
    U_{\text{est},i} <~& \alpha_1 E_{\text{pot},i} + \alpha_2 E_{\text{kin},i} \label{eq:switch}
\end{align}
in physical units, the new internal energy for particle $i$ is instead calculated assuming adiabatic expansion or contraction.
The parameters $\alpha_{1/2}$ have to be tuned to only affect the evolution of particles where necessary. We provide a comparison between different values in Sec.~\ref{sec:switch_effect}.

The internal energy is updated similarly to the entropy for SPH in two half-steps at the kicks, following a second order time integration similar to the entropy in SPH
\begin{align}
    U^{n+1/2} =~& U^{n} + \frac{1}{2}\klammer{\frac{\dif U}{\dif t}}^{\tilde n}\Delta t,\\
    U^{n+1} =~& U^{n+1/2} + \frac{1}{2}\klammer{\frac{\dif U}{\dif t}}^{\widetilde{n+1}}\Delta t.
\end{align}
For cosmological simulations, additional adiabatic contributions due to the Hubble flow are added.

\subsection{Switching between SPH and MFM in \textsc{OpenGadget3}}

To substitute SPH with MFM, the general code-structure does not have to be altered. Mainly, the SPH specific force calculation has to be replaced by the three steps of the MFM calculation, consisting of gradient calculations, slope-limiting and the actual flux calculation. As the Riemann solver both requires and outputs physical quantities, while the rest of the code deals with code units, these units have to be converted according to Eqn.~(\ref{eq:code-phys_x}) to~(\ref{eq:code-phys_u}) just before the flux calculation. At all places, where results of that calculation, including the hydrodynamical acceleration, are used, they first have to be converted back to physical units. 

Also, MFM calculates internal energy changes following the output of the Riemann solver, while in SPH the entropy is evolved.

\subsection{Differences to previous implementations of MFM}

While the general concept of MFM with respect to the implementations introduced in \textsc{gizmo} and \textsc{GANDALF} stays the same, there are several differences compared to these previously made implementations. 
Our implementation is based on the one in \textsc{GANDALF}, which is originally intended to be well suited for star and planet formation. We expand this implementation by including co-moving integration and other extensions such as an energy-entropy switch to be used for cosmological applications.
In addition, we change the time integration scheme from a second-order accurate MUSCL-Hancock to a second-order accurate Leapfrog KDK, consistent with SPH in \textsc{OpenGadget3}.

The main difference of \textsc{OpenGadget3} compared to \textsc{gizmo} is that fluxes are by default calculated using an iterative, exact Riemann solver compared to an approximate HLLC Riemann solver used in \textsc{gizmo}, with an exact Riemann solver only used as fallback.

In comparison to \textsc{pkdgrav-3} \citep{AlonsoAsensio+2023}, the energy-entropy switch and the implementation of how to deal with anisotropic particle distributions is different and more similar to \textsc{gizmo}.

In addition, there are a few minor differences such as the second-order correction in Eqn.~(\ref{eq:mfm_energy_conversion}). We also made the pairwise limiter Lagrangian, as described in App.~\ref{app:limiters}, which was independently done also in \textsc{pkdgrav-3}.
The convergence of the density calculation is slightly different between the codes. We follow the same implementation as for SPH in \textsc{OpenGadget3}, just replacing the mass density by the number density. Finally, our implementation employs a hybrid MPI-OpenMP parallelization as done for other modules of \textsc{OpenGadget3}.

\section{Test cases} \label{sec:tests}

We use several test cases to probe the ability of the different hydro-methods to accurately follow gas evolution. Only few tests have an analytical solution, including soundwaves and different shocks. Also a set of MHD waves with analytical solution has been presented by \citet{Berlok2022a}, which could be used for tests of later MHD extensions. Many other tests can be used for a more qualitative analysis. All of them explore specific numerical aspects important for cosmological simulations.

We use these tests to compare our new MFM implementation in \textsc{OpenGadget3} to SPH in \textsc{OpenGadget3}, MFM in the public \textsc{gizmo}\footnote{Obtained from \url{https://bitbucket.org/phopkins/gizmo-public/src/master/} February 2021} version and the publicly available version of the moving mesh code \textsc{Arepo}\footnote{Obtained from \url{https://gitlab.mpcdf.mpg.de/vrs/arepo} June 2021}.

\subsection{Settings}
We aim for a fair comparison of the different codes throughout the paper but adopt a general setting for slope limiters, Riemann-solvers (MFM) as well as the artificial diffusion terms (SPH) that one would adopt in cosmological simulations. While this leads to overall good performance of all solvers on almost all test cases, there are a few test problems (e.g. the square test in Sec~\ref{sec:square}) for which this is not working ideal and we will discuss this in detail in the remainder of the paper.
If not otherwise mentioned, we assume an ideal gas with $\gamma=5/3$ and all code operate on adaptive time steps for all tests (i.e. we never force a small constant time step to improve the accuracy of the results).

MFM is used with a cubic spline kernel and 32 (24) neighbors in 3d (2d). The slope limiter from \textsc{gizmo} in combination with their pairwise limiter, both as presented in App.~\ref{app:limiters}, is used. Consistent settings are chosen between \textsc{OpenGadget3} and \textsc{gizmo}. 
For SPH, a Wendland C6 kernel, including bias correction \citep[][]{Dehnen&Aly2012}, with 295 (64) neighbors in 3d (2d) is used. The modern, time-dependent artificial viscosity scheme of \citet{Beck+2016} and artificial conductivity \citep{Price2008} are included. 
For \textsc{Arepo} we use additional mesh regularization based on the center of mass, and the ``roundness'' of the cells. 
An overview of all settings is made publicly available\footnote{\url{https://github.com/fgroth/hydro_tests}}.
If not otherwise stated, the initial conditions (ICs) are created with equal particle masses. In most cases, particles are arranged in a (perturbed) regular grid in order to reduce noise introduced by the initial particle distribution. 

\subsection{Stability}
\subsubsection{Soundwave} \label{sec:soundwave}
\begin{figure}
    \centering
    \includegraphics[width=\columnwidth]{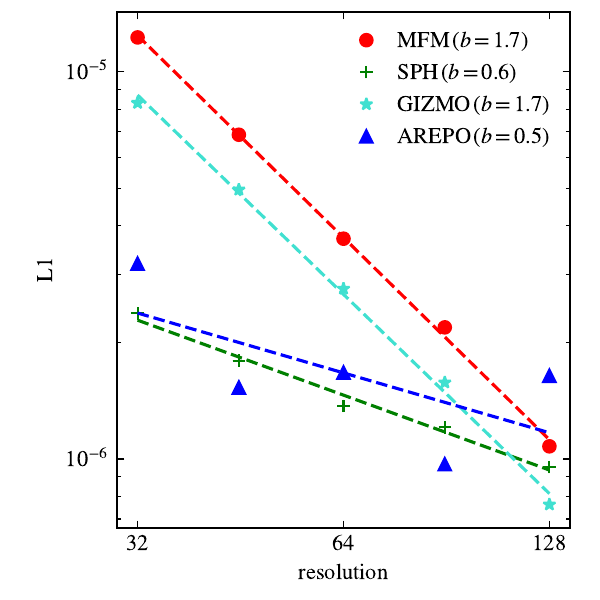}
    \caption{L1 norm for the soundwave at different resolutions. The order of convergence $b$ is obtained from a fit. While MFM in both implementations shows between first and second order convergence, SPH and the moving mesh have a convergence even below first order.}
    \label{fig:soundwave_l1}
\end{figure}

As a first test we adopt a sinusodial soundwave with density $\rho=1$ and small perturbation amplitude $\Delta\rho=10^{-4}$ in a box of length $1$ in $x$-direction and $0.75$ in $y/z$-direction. The particles are arranged in a perturbed hexagonal close packed (hpc) grid with varying resolution. The number of particles is ranging from $64^3\cdot0.75^2$ up to $128^3\cdot0.75^2$. In the following, we will define the resolution by the number of particles per unit-length in $x$ direction. We adopt a wavenumber $k=2\pi$ and a speed of sound of $c_s=2/3$. For this test there is an analytic solution $\rho(x,t) = \rho_0 + \Delta\rho\sin(k(x+c_st))$, which makes this test well suited to perform a convergence analysis. For this purpose, we measure the L1 error norm $\frac{1}{N_{\text{tot}}}\sum_i^{N_\text{tot}}\left\vert\rho_i-\rho(x,t)\right\vert$, shown in Fig.~\ref{fig:soundwave_l1}. 
All methods are able to evolve the soundwave, while the accuracy as well as the precise convergence behavior differ among the codes. We observe a similar convergence between the MFM implementation in \textsc{gizmo} and \textsc{OpenGadget3} being between first and second order. While theoretically second order convergence would be expected, the slope-limiter reduces the order of convergence, as discussed by \citet{AlonsoAsensio+2023}. The convergence for SPH in \textsc{OpenGadget3} and the \textsc{Arepo} code are similar, but even below first order. For \textsc{Arepo}, the main reasons for this low-order convergence are the mesh regularization, which introduces small numerical noise, and the fact that faces which contribute by less than $10^{-5}$ to the total face area are neglected (R. Pakmor, 2023, priv. comm.). The convergence can be improved fixing these two points, as shown in App.~\ref{app:soundwave_arepo}, leading to a similar convergence as for MFM much closer to second order. Nevertheless, these changes would make the code more unstable in cosmological simulations.

\begin{figure}
    \centering
    \includegraphics[width=\columnwidth]{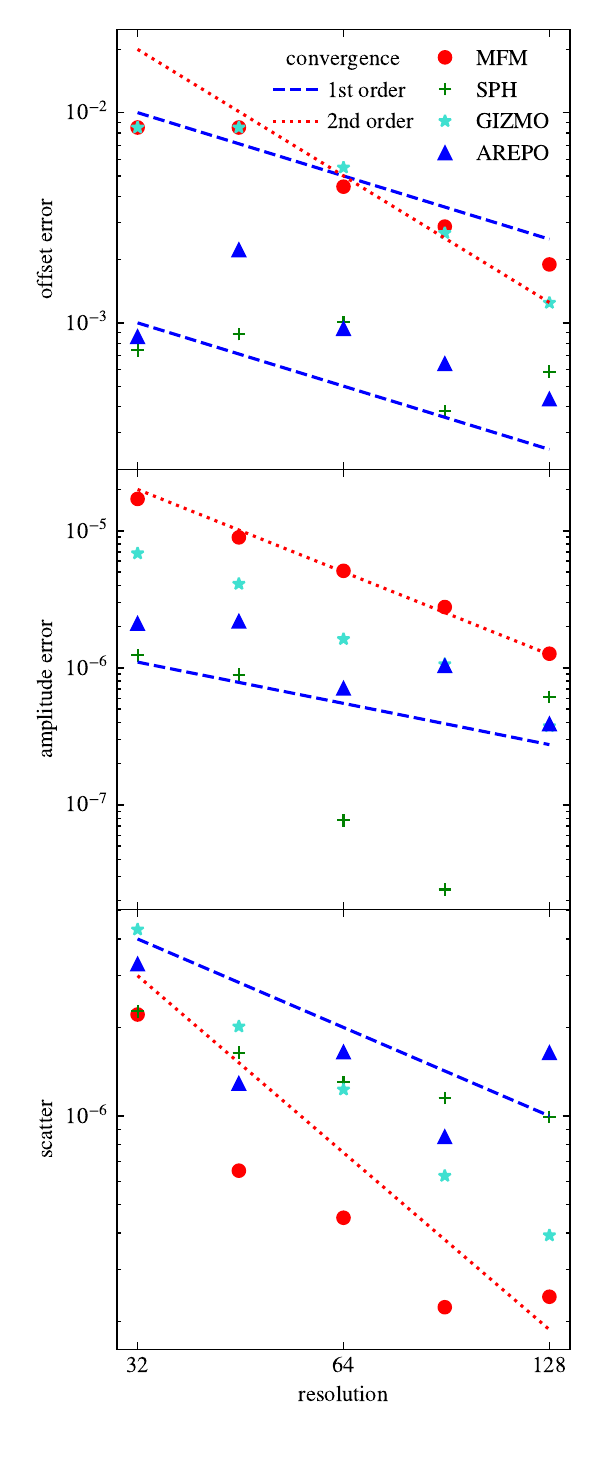}
    \caption{Offset-, amplitude- and scatter-errors of the density of a soundwave at $t=\frac{2}{c_s}$ calculated with MFM and SPH in \textsc{OpenGadget3}, MFM in \textsc{gizmo} and a moving mesh in \textsc{Arepo} at different resolutions. The scatter converges second order for all methods, while other errors show different convergence behavior. MFM shows between first and second order convergence for all error-components.}
    \label{fig:soundwave}
\end{figure}

In order to get a more detailed analysis, we split the error between errors in the position, in the amplitude, and scatter as shown in Fig.~\ref{fig:soundwave}.
A sinusodial soundwave is fit to the density distribution, such that the offset and amplitude differences to the expected wave are obtained. The remaining deviations, mostly being scatter, are then quantified by an L1-norm.

Deviations from the expected sound speed are related to dispersion errors, and will lead to an offset compared to the analytical solution. This offset error is shown in the upper panel. We observe for MFM in both implementation the convergence to be between first and second order, consistent between both codes. For SPH and \textsc{Arepo}, the overall error is roughly one order of magnitude smaller at the lowest resolution, but having a convergence even worse than first order. For SPH, this trend can be explained by low-order errors, which are prominent for traditional SPH, and still partly left for modern SPH.

The error in the amplitude, shown in the middle panel, is related to numerical diffusion. 
As we see also in other tests, the Riemannn solver and the slope-limiter introduce numerical diffusivity for MFM, which thus has the largest error. Differences between the different MFM implementations can be explained by different Riemann solvers used. SPH and \textsc{Arepo} show much lower errors. The convergence behavior, however, is again better for MFM compared to the other methods. In both implementations, it is roughly second order, while for the other methods it appears to be approximately first order. 

Finally, it is worth to note that the resulting soundwave does not have perfect sinusodial shape but shows scatter in the amplitude. This is mainly a result of the smoothing length/density iteration and the threshold chosen for the value to be taken as converged. We quantify this error by the L1 error norm, shown in the bottom panel of Fig.~\ref{fig:soundwave}. All methods show roughly second order convergence, while the amplitude of the error is different. Differences between MFM and SPH in \textsc{OpenGadget3} can be explained by the different kernel used, while other codes have differences in the iteration and treat parameters for convergence slightly differently. The large error for \textsc{Arepo}, even at higher resolution makes the values for the other errors more uncertain. In addition to the errors already mentioned, the soundwave deforms and steepens up due to non-linear terms in the evolution. This non-linearity will lead to an additional, small but constant term in the scatter error in the bottom panel of Fig.~\ref{fig:soundwave}. A reduction could be achieved by reducing the amplitude, which would also make scatter errors be more significant or the convergence more expensive. 
As non-linear contribution are expected to become important when $L1\approx (\Delta\rho/\rho)^2=10^{-8}$ in our setup, this term will not be relevant for the resolutions considered.

\subsubsection{Kepler Disk}
The Kepler disk is an important test case for cosmological simulations, allowing to study the ability of the code to conserve angular momentum and maintain stable orbits over time. Especially, the effect of viscosity can be analyzed.
To this end we initialize a two-dimensional box sufficiently large to contain all particles. The ICs are taken from \citet{Hopkins2015} and are initialized with 48240 gas particles with equal masses, arranged in a grid-like structure and setup with vanishing pressure of $P=10^{-6}$. The gas surface density distribution is given via:
\begin{align}
    \Sigma =~& 0.01+\mathcases{ (r/0.5)^3~&\text{if }r<0.5\\
    1~&\text{if }0.5\le r\le 2\\
    (1+(r-2)/0.1)^{-3}~&\text{if }2<r.
    }
\end{align}
For the \textsc{Arepo} run, we adopt a low density mesh with vanishing pressure at a resolution of $16$ particles per unit length distributed around the disk as well as inside the central hole of the disk. 

\begin{figure}
    \centering
    \includegraphics[width=\columnwidth]{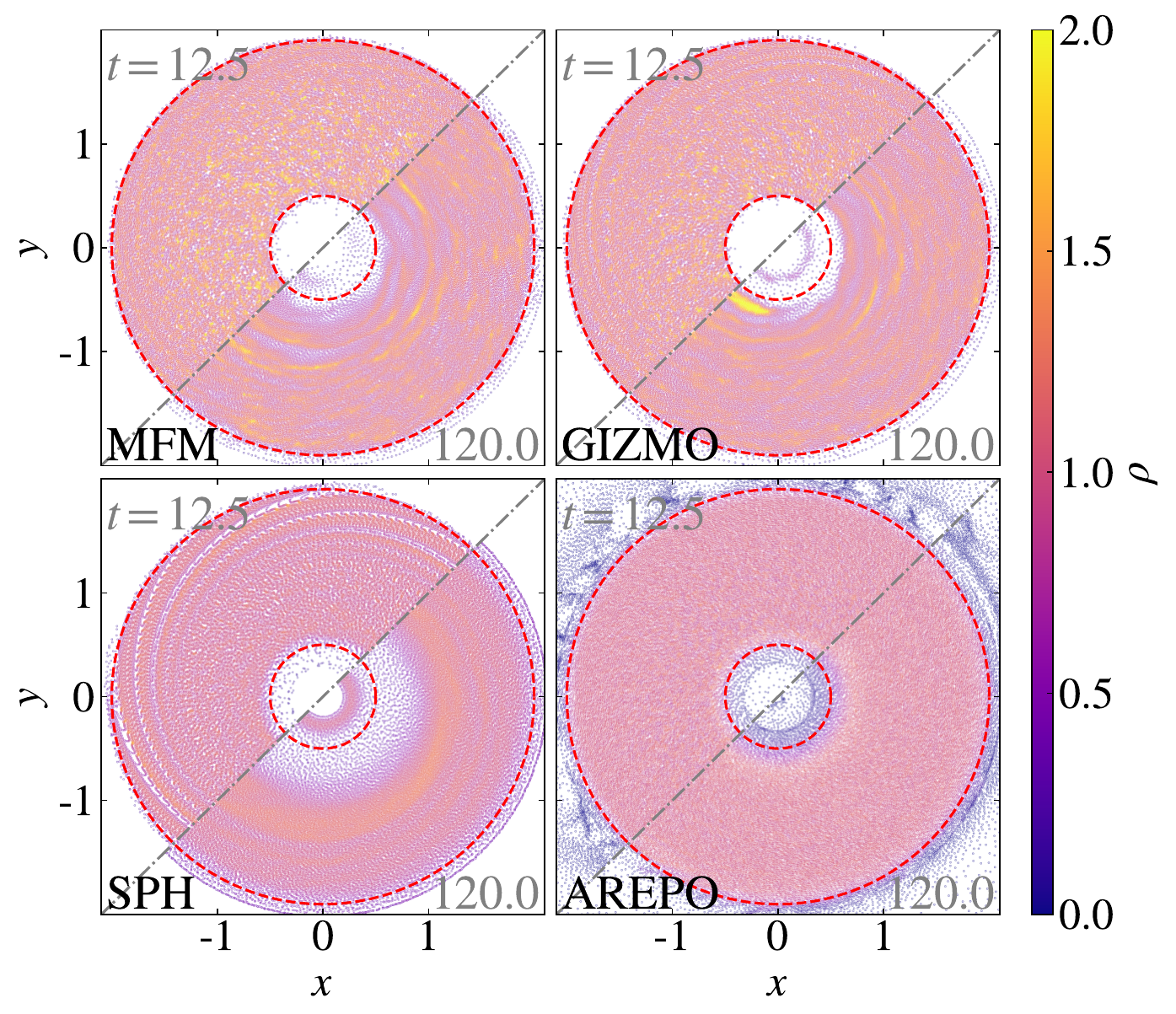}
    \label{fig:kepler_disk}
    \caption{Evolution of the Kepler disk using different hydro-methods. Surface density at two times per method: $t=12.5$ (upper left) and $t=120$ (lower right). In general, all methods are able to evolve a stable disk. Initial perturbation introduced by the ICs, however, evolve differently for the different methods.}
\end{figure}

We adopt an external potential $\Phi = -(r^2+\epsilon^2)^{-1/2}$ with resulting gravitational acceleration of the form
\begin{align}
    \myvector g =~& -\myvector r\mathcases{\klammer{\frac{\klammer{r/0.35}^2}{\klammer{\myvector r^2}^{1.5}}-\frac{\klammer{0.35 - r}/0.35}{\klammer{\myvector r^2}^{1.5}}}~ &\text{if }r\le 0.35\\
    \klammer{\frac{1}{\klammer{\myvector r^2}^{1.5}}} &\text{if }0.35<r<2.1\\
    \klammer{\frac{1+\klammer{r-2.1}/0.1}{\klammer{\myvector r^2}^{1.5}}} &\text{if }2.1\le r.}
\end{align}
We follow the evolution of the disk until $t=120$, corresponding to $\approx 20$ orbits at $r=1$. The resulting density at $t=120$ and $t=12.5$ is shown in Fig.~\ref{fig:kepler_disk}. 
Initially, all methods produce spirals as a result of perturbations in the ICs. While for more traditional SPH with Balsara viscosity switch \citep{Balsara1998} these lead to a destruction of the disk after only a few orbits, consistent with the results of \citet{Beck+2016}, the modern SPH implementation in \textsc{OpenGadget3} with the improved viscosity scheme of \citet{Beck+2016} drastically increases the stability of the disk. While the inner and outer region still show some decay, the main part of the disk is stable for the whole evolution considered.
For MFM the disk remains stable for more than $20$ orbits. We observe that the inner and outer parts of the disk degrade much less compared to SPH. The initial perturbations are diffused throughout the disk, which shows slightly larger perturbations in the main part compared to the SPH calculation. Both, our implementation and the one in \textsc{gizmo}, show qualitatively similar results. 
The \textsc{Arepo} run turns out to produce the most stable disk. Only a slight degeneration at the boundaries can be observed. Further studies would be needed to analyze whether this is a numerical effect or due to interaction with the ambient medium not present in the other calculations.

\subsection{Tests for Fluid Mixing Instabilities}
Mixing occurs in a variety of cosmological situations, most prominently during ram-pressure-stripping. To this end, we analyze the ability of the different codes and methods to evolve such mixing instabilities.

\subsubsection{Rayleigh-Taylor Instability}\label{sec:rt}
One popular fluid-mixing test is the Rayleigh-Taylor instability. It can be used to explore how well the code can describe unstable, growing modes.
The setup we use is taken from \citet{Hopkins2015}. The calculations are preformed in a two-dimensional periodic box with side-lengths $1$, populated with $65536$ particles where the particles at $y<0.1$ and $y>0.9$ are fixed as boundary conditions. In contrast to the other codes, for \textsc{Arepo} the boundary particles are not fixed but instead a relflective boundary condition is used.

A fluid of high density ($\rho=2$) is placed on top of a low-density medium ($\rho=1$) in hydrostatic equilibrium. For this test-case, we take $\gamma=1.4$, as for a diatomic gas, such as molecular hydrogen and apply the constant gravitational acceleration:
\begin{align}
    \myvector{a}_{\text{grav}} =~& -0.5\hat y.
\end{align}
To allow the instability to grow, a small velocity perturbation at the phase boundary is introduced \citep[for more details see][]{Hopkins2015}. 

\begin{figure*}
    \centering
    \includegraphics[width=\textwidth]{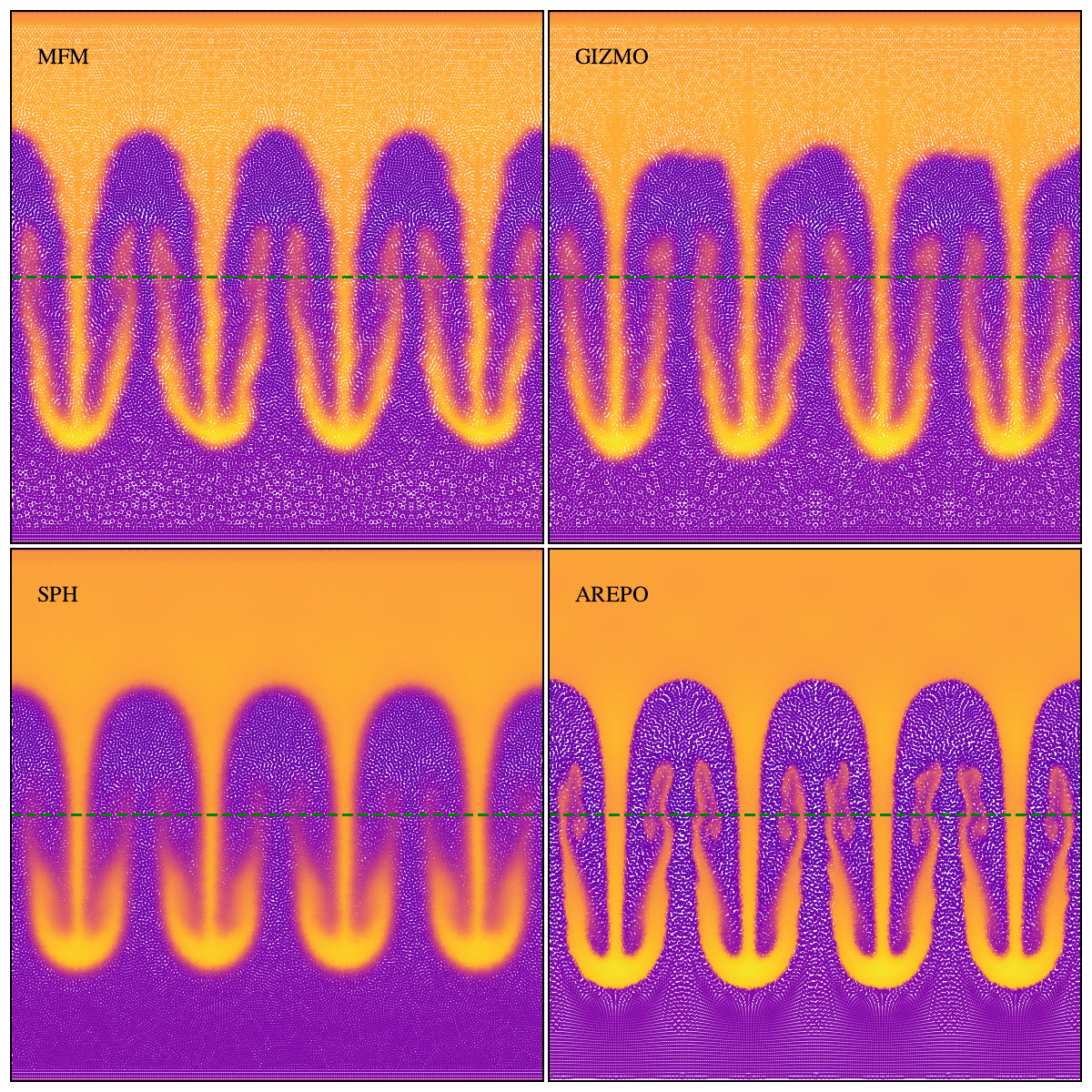}
    \caption{Rayleigh-Taylor instability at time $t=3.6$. Comparison between the different hydro-methods. Vertical line marks the initial position of the phase boundary. Differences are mainly the presence or absence of secondary instabilities.}
    \label{fig:rt}
\end{figure*}

In Fig.~\ref{fig:rt} we show that all methods are perfectly able to evolve the instability.
A major difference between the different methods is the presence of asymmetries and secondary instabilities. While these can be seen clearly for MFM, both in \textsc{OpenGadget3} and \textsc{gizmo}, and are also present in the \textsc{Arepo} calculation where they appear more symmetric, we find that they are absent from the SPH calculation, due to the smoothing over the larger kernel and the effectively lower spatial resolution \citep[e.g.][for a more detailed discussion of the occurance of secondary insatbilties and their physical meaning]{Marin-Gilabert+2022}. 
The results of \textsc{Arepo} indicate the sharpest boundary and highest density in the tip, followed by MFM. The particles close to the boundary for \textsc{Arepo} show still a clear imprint of the initial grid-like particle distribution. We note that the numerical diffusivity within modern SPH causes the boundary of the instability to have a shallower gradient and smears out initial asymmetries. In addition, the effective spatial resolution is lower by a factor of $\approx 2$ compared to MFM due to the larger neighbor number and thus SPH reaches a much lower density in the tip of the instability.

\subsubsection{Kelvin-Helmholtz Instability}
\begin{figure*}
    \centering
    \includegraphics[width=\textwidth]{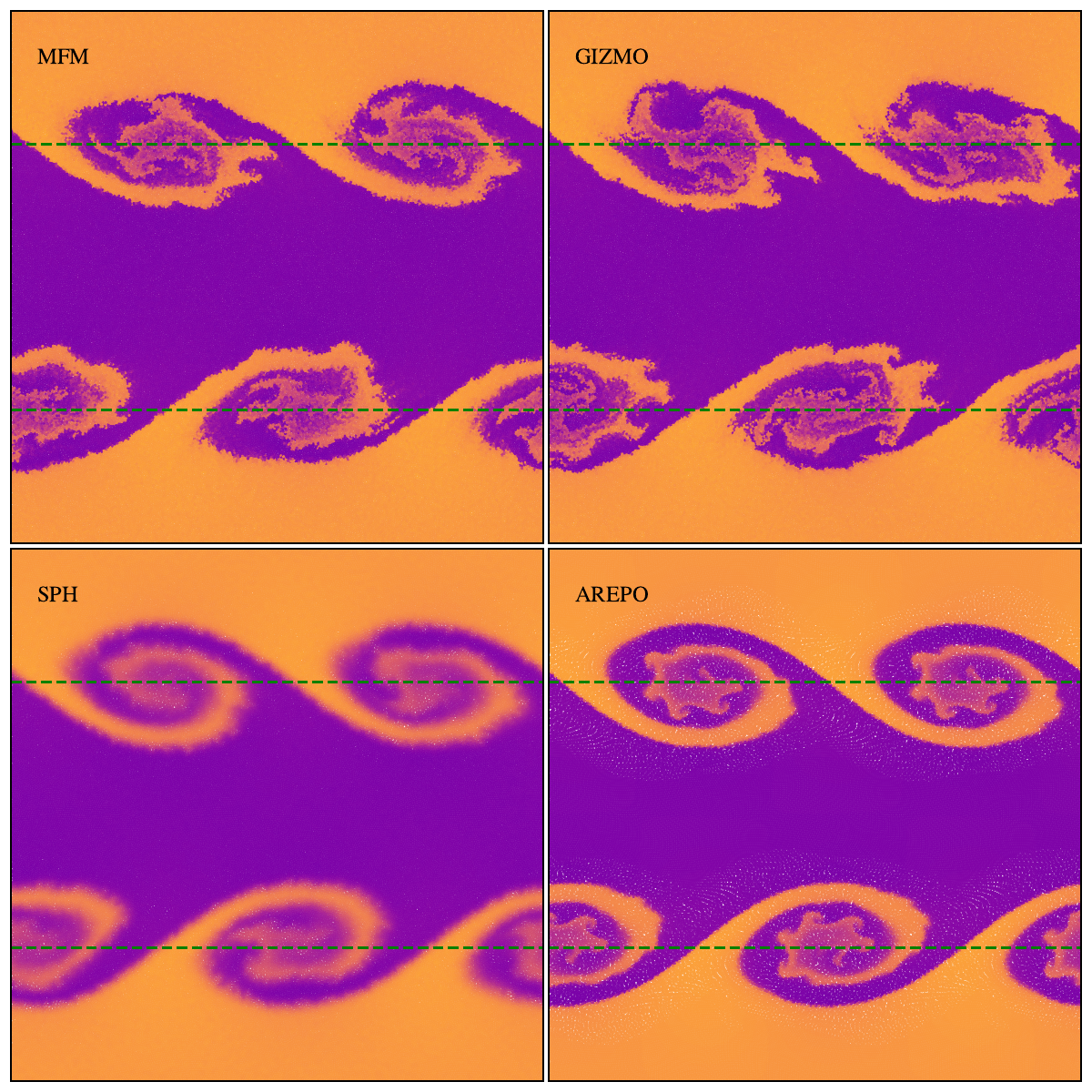}
    \caption{Build-up of a 2d Kelvin-Helmholtz instability at $t=2.5$ comparing different methods. Horizontal dashed lines mark the initial position of the phase boundary. All methods produce the roll, but with differences in their inner structure.}
    \label{fig:kh}
\end{figure*}

Similar to the Rayleigh-Taylor instability, also the Kelvin-Helmholtz instability is a famous example for fluid mixing. Again, we use the setup provided by \citet{Hopkins2015}. Two fluids of densities $\rho_1=1$ and $\rho_2=2$ in hydrostatic equilibrium are initialized in a 2d periodic box, with initial velocities $\myvector{v}_{1}=0.5\hat x$, $\myvector{v}_{2}=-0.5\hat x$ and a small perturbation following \citet{McNally+2012a}. The setup includes in total $774144$ particles. At time $t=2.5$ corresponding to $\approx 1.2\tau_{\text{KH}}$ in units of the Kelvin-Helmholtz timescale $\tau_\mathrm{KH}=\frac{\lambda}{\Delta v_x}\frac{\rho_1+\rho_2}{\sqrt{\rho_1\rho_2}}$ \citep[compare, e.g.,][]{Junk+2010}, the instability has produced a roll for all methods, as shown in Fig.~\ref{fig:kh}.

Differences are present in the inner structure of the roll. Overall the qualitative results are very similar to those for the Rayleigh-Taylor instability.
SPH is smoothing the roll, showing no secondary instabilities and evolving more smoothly towards later times. Compared to that, MFM in both implementations shows a clear separation between the higher-density roll and the less dense medium, with the presence of secondary instabilities. A more detailed analysis of the Kelvin-Helmholtz instability, also using our new MFM implementation, has been done by \citet{Marin-Gilabert+2022}. They also show that the secondary instabilities can be avoided by using a higher neighbor number in combination with a higher-order kernel. This will increase the intrinsic viscosity and prevent mixing in form of secondary instabilities.
Also \textsc{Arepo} shows secondary instabilities, present especially inside the roll.
When present, these perturbations will finally dominate the evolution over the build-up of the roll for $t\gtrsim 3$.

\subsubsection{Hydrostatic Square}\label{sec:square}
As both aforementioned fluid-mixing tests contain sharp boundaries that deform due to instabilities, the understanding of the evolution of such boundaries evolving without perturbations imprinted in the ICs is important.
The Hydrostatic Square tests this behavior, as it is well suited to study the stability of edges related to numerical surface tension.
Similar tests have been performed e.g. by \citet{Hess&Springel2010} and \citet{Hopkins2013,Hopkins2015}. 

\begin{figure}
    \centering
    \includegraphics[width=\columnwidth]{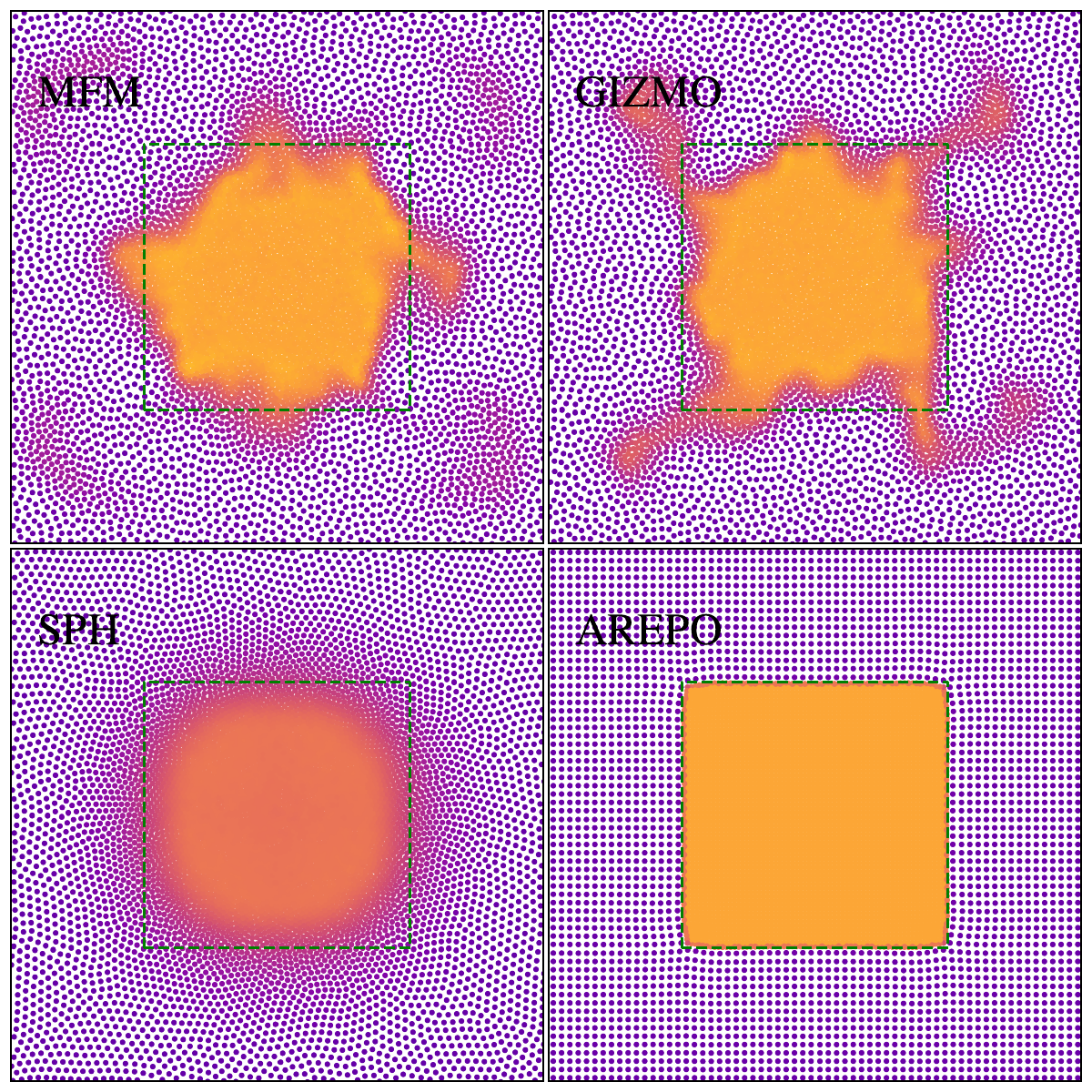}
    \caption{Density of the hydrostatic square evolved until $t=10$ using different methods. The initial location of the high density ``square'' region is overplotted as contour. Only \textsc{Arepo} is able to keep the initial square shape, while other methods lead to deformation of the square.}
    \label{fig:square}
\end{figure}

\begin{figure}
    \centering
    \includegraphics[width=\columnwidth]{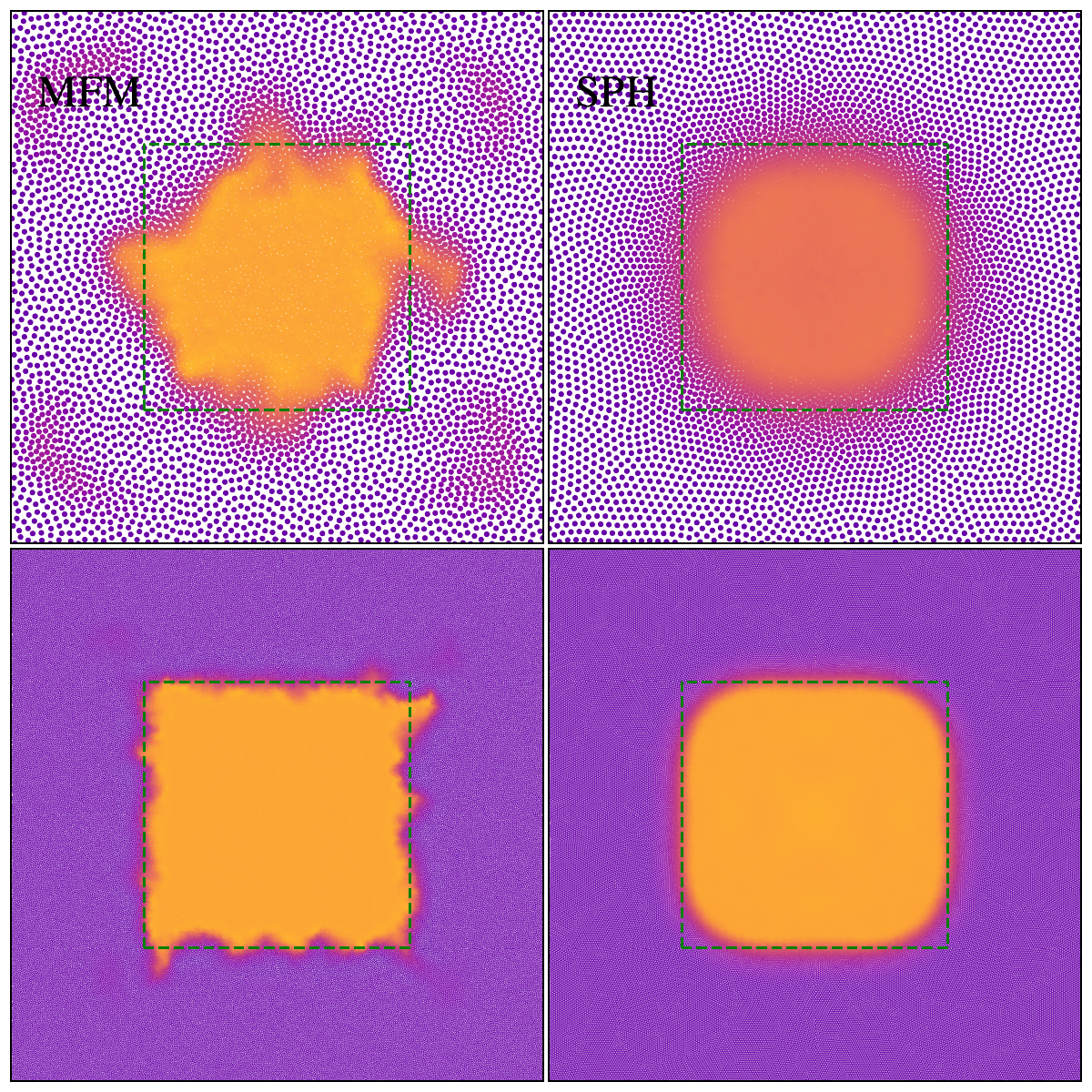}
    \caption{Hydrostatic Square at $t=10$. Comparison of MFM and SPH at two different resolutions, Top: 7168 particles, Bottom: 114688 particles (increase in resolution by factor $4$. Both, MFM and SPH, show convergence of the shape of the square.}
    \label{fig:square_res}
\end{figure}

We set up a two-dimensional box of size $L=1$ with periodic boundary conditions. It is filled with 7168 gas particles with equal masses, arranged in two regular grids, one grid for the ambient medium ($\rho_a$=1, $P_a=2.5$) and one for the square with side-length $L/2$ with increased density $\rho_s=4$ in hydrostatic equilibrium ($P_s=P_a$). In Fig.~\ref{fig:square}, we compare the resulting density distribution at time $t=10$, evolved with the different methods.
As the ICs are set in hydrostatic equilibrium, we would expect no changes to occur. This ideal state is only achieved using the moving mesh code \textsc{Arepo}.
Theoretically, we would expect the same to be true for MFM, as shown by \citet{Hopkins2015}. They use, however, a strongly idealized setup compared to ours. Especially, they use a regular grid for all particles, and increased particle masses within the square. For our setup, the gradient estimate at the boundary does not conserve linear gradients. Instead, it is biased by the in-homogeneous particle distribution due to two separate grids, especially in combination with the slope-limiter. A more detailed analysis of the effect of the slope limiter is provided in Sec.~\ref{sec:slope_limiter}, where we have shown that the amount of surface tension and resulting deformation of the square strongly depends on the slope-limiter.
We observe, using both our MFM implementation and \textsc{gizmo}, that for MFM the edges of the square start to deform, followed by some numerical instability, which leads to a more asymmetric deformation.
Increasing the resolution by a factor of 4, as shown in Fig.~\ref{fig:square_res}, this instability occurs slower and the square preserves its shape much better.
Also using SPH, the square deforms. As expected, it becomes more circular, caused by numerical errors, which behave as surface tension \citep[compare, e.g.,][]{Price2008}. For traditional SPH, these errors should be low-order. We observe, however, that this effect can be drastically reduced by increasing the resolution, as shown in Fig.~\ref{fig:square_res} indicating that modern SPH implementations, as used in \textsc{OpenGadget3}, reduce low-order errors and improve convergence.
Overall, for this specific test surface tension for SPH, but also for MFM can be observed.
A moving mesh performs best, preserving the situation perfectly.
MFM at later times shows some numerical errors leading to a more asymmetric deformation, which converge away with increasing resolution. 

\subsubsection{The ``Blob'' test}
A more complex problem is the blob test. It is designed to mimic ram-pressure stripping by an interplay of the evolution of shocks and fluid-mixing instabilities.
We use the setup described by \citet{Hopkins2015} \citep[compare also][]{Agertz+2007}. A three-dimensional box with side-length $2000$ in $x$ and $y$-direction and $6000$ in $z$-direction is populated with $9641651$ particles.
A cloud of higher density $\rho_\text{cloud}=10\rho_\text{wind}$ is placed into a wind tunnel with supersonic flow at $\mathcal{M}=2.7$ and density $\rho_{\text{wind}}=2.6\e{-8}$. Both phases are setup in pressure equilibrium.

\begin{figure*}
    \centering
    \includegraphics[width=\textwidth]{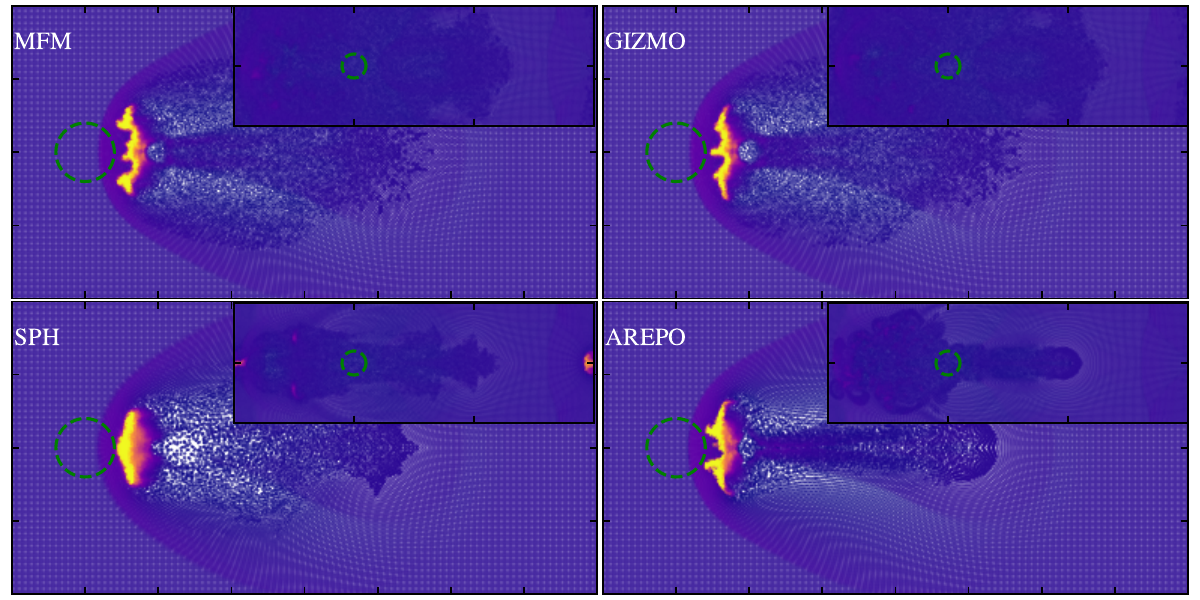}
    \caption{Blob at $t=\tau_\text{KH}$ and $t=4\tau_{\text{KH}}$ as small insertion comparing different hydro-methods. At the earlier time, SPH leads to much less deformation due to less instabilities building up, while MFM in both implementations as well as \textsc{Arepo} agree qualitatively. At late time, MFM and \textsc{Arepo} are fully mixed, while SPH still has some structure remaining.}
    \label{fig:blob}
\end{figure*}

\begin{figure}
    \centering
    \includegraphics[width=\columnwidth]{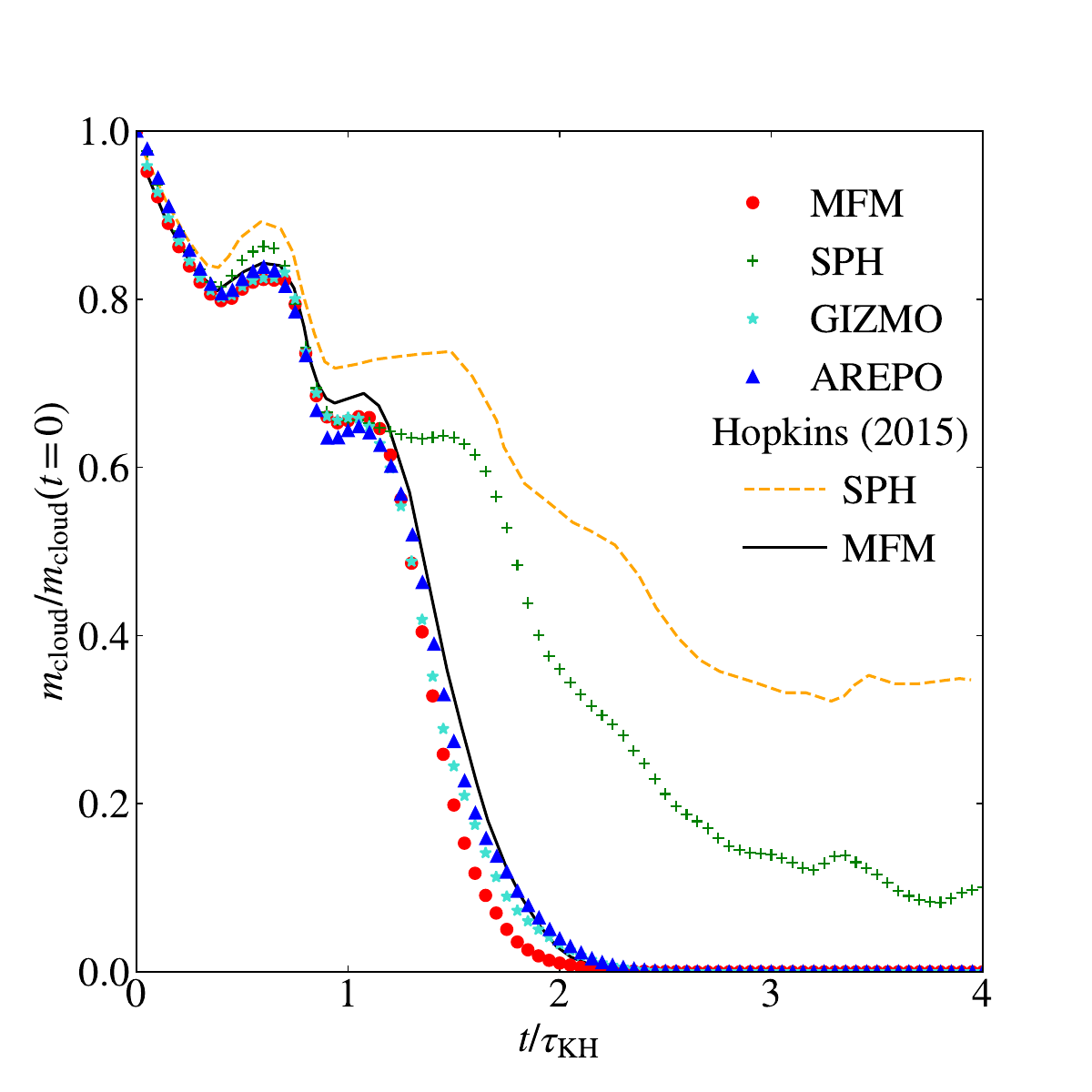}
    \caption{Decay of the cloud fraction surviving for the different methods. In the background, comparison lines of the results by \citet{Hopkins2015} for MFM (black, solid) and (traditional) SPH (orange dashed) are shown. MFM and \textsc{Arepo} agree very well, while SPH shows less mixing.}
    \label{fig:blob_decay}
\end{figure}

The resulting density in a slice through the cloud at $t=\tau_{KH}$ and $t=4\tau_{\text{KH}}$ is shown in Fig.~\ref{fig:blob}. 
In front of the cloud, a bow shock forms. At the Kelvin-Helmholtz timescale $\tau_\text{KH}=2$, the cloud has developed instabilities.
These are much more pronounced for MFM and \textsc{Arepo}, while for SPH the cloud deforms, without showing instabilities. The precise form of the cloud differs between our MFM implementation, that in \textsc{gizmo} and the moving mesh code \textsc{Arepo}. Nevertheless, the cloud mass, defined by the particles obeying $\rho>0.64\rho_{\text{cloud},i}$ and $u<0.9u_{\text{amb},i}$, is very similar for all methods until $\tau_{KH}$, shown in Fig.~\ref{fig:blob_decay}. As expected, the MFM calculations line up with the calculations done by \citet{Hopkins2015}.
The periodic bumps are a result of the self-interaction of the shock due to the choice of boundary conditions.

At later times the evolution strongly deviates. While for MFM as well a moving mesh secondary instabilities build up and lead to a disruption of the cloud, it is more stable in SPH. Compared to the more traditional SPH results of \citet{Hopkins2015}, however, we find the blob to decay stronger, as modern SPH with time-dependent artificial viscosity and conductivity is able to evolve instabilities much better, thus allowing for more mixing.

\subsection{Tests for Shock-capturing}
\subsubsection{Sod Shock-tubes}
\begin{figure*}
    \centering
    \includegraphics[width=\textwidth]{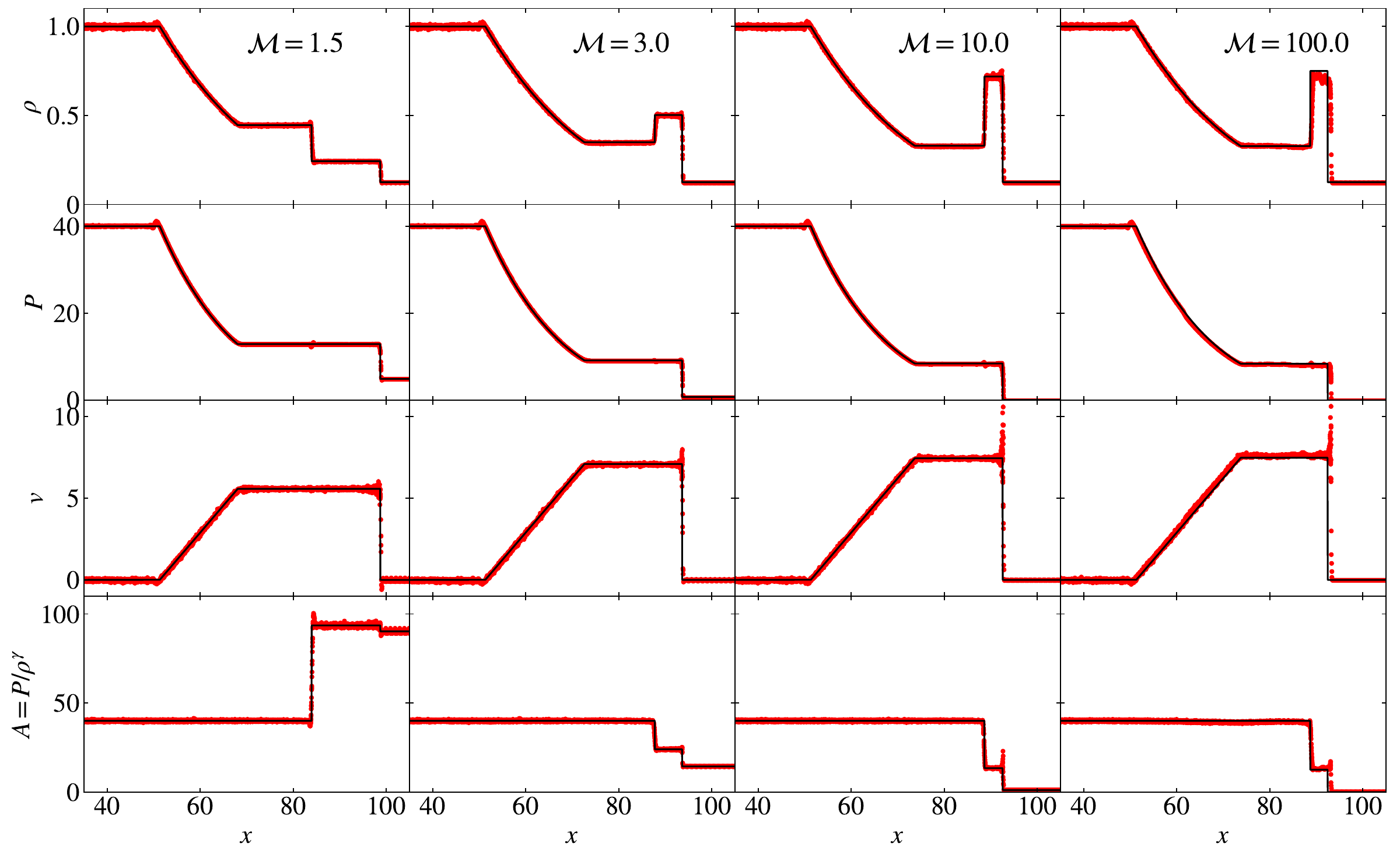}
    \caption{Density, pressure, velocity and entropy profile of the shock tube at $t=2.5$ calculated with our MFM implementation, comparison between different Mach numbers. MFM is able to reproduce the general structure of the shocks. Artifacts of surface tension introduced by the slope-limiter are visible at higher Mach numbers. The scatter is a result of the choice of ICs.}
    \label{fig:shock_tube_mfm}
\end{figure*}

Another important capability of the code is to capture strong shocks of (arbitrarily) large Mach number. We begin testing this on a simple Sod shock-tube based on the setup of \citet{Sod1978}. The test is preformed in a three dimensional periodic box of size $L_x$=140, $L_y=L_z=1$ with two fluids of different density and pressure ($\rho_1=1,P_1=1$; $\rho_2=1/8, P_2=0.1$ for $\gamma=1.4$) that are initialized in a glass-like configuration of in total $216090$ particles.
When the two phases start interacting, a shock begins to move to the right. In Fig.~\ref{fig:shock_tube_mfm}, we show the resulting structure at $t=2.5$ for the MFM calculations at different Mach number and compare them to the analytic solution.
The expected profiles are matched very well, for all the Mach numbers adopted in this work, ranging from a very low $\mathcal{M}=1.5$ shock to a strong $\mathcal{M}=100$ shock. This ability is directly connected to the accuracy of the Riemann solver. For higher Mach numbers, increasing peaks in velocity and entropy at the shock front are present as a result of the non-TVD slope-limiting procedure, which has also been reported by \citet{Hopkins2015}. We note that this peak and nearby oscillations would be even larger of no limiter was used, and can be avoided even better by using a TVD-limiter, which has more disadvantages in other cases.
With increasing Mach number, a sufficiently small timestep becomes more important.
The scatter in velocity for the high $\mathcal{M}=100$ shock, as well as the small offset in the position of the shock front converge away with decreasing timesteps.

\begin{figure*}
    \centering
    \includegraphics[width=\textwidth]{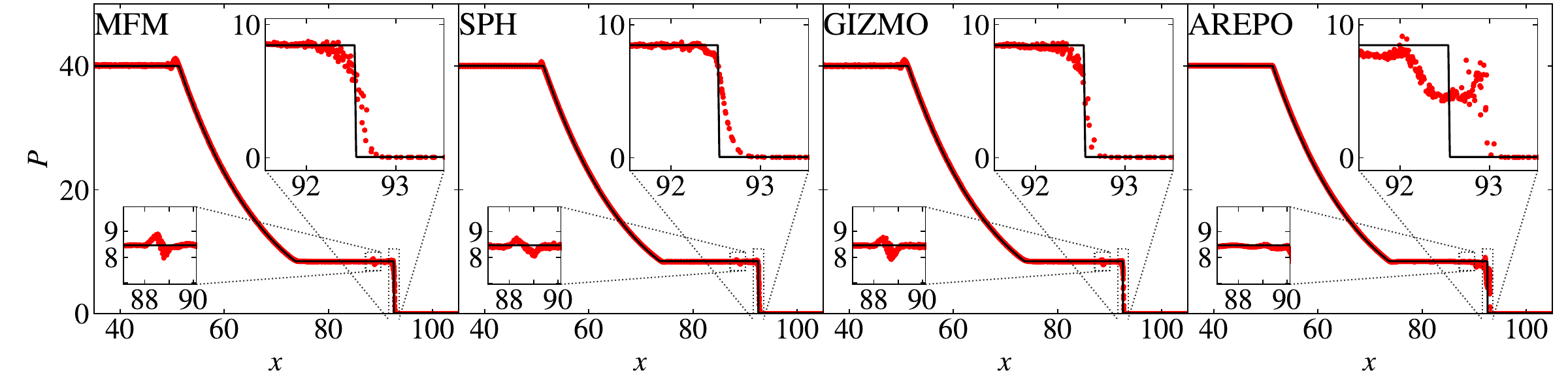}
    \caption{Pressure profile of the $\mathcal{M}=10$ shock tube at $t=2.5$, comparison between different hydro-methods. The different codes show different amount of surface tension and also slight differences in the position of the shock front due to different timestepping}
    \label{fig:shock_tube_methods}
\end{figure*}

The scatter in density present at all Mach numbers is a result of the choice of the ICs, which are setup in a glass-like configuration and designed for a higher neighbor number. It does not converge for low neighbor numbers, as chosen for MFM. The pressure profile shows the typical bump at the rarefaction fan, as well as the pressure blip at the contact discontinuity, shown in more detail in Fig.~\ref{fig:shock_tube_methods} for the intermediate $\mathcal{M}=10$ shock. This indicates the presence of surface tension-like error terms, introduced by the slope limiter.
As discussed in Sec.~\ref{sec:square} on the example of the hydrostatic square, these terms are present for SPH and both MFM implementations, but not for \textsc{Arepo}, manifesting also in the presence or absence of the pressure blip for the different methods.
The shock front is captured equally well for MFM and SPH, though less smoothed out for MFM due to the lower neighbor number. \textsc{Arepo} poorly captures the behavior at the shock front. Especially, it has troubles in the mesh reconstruction in this strongly anisotropic region, which leads to a shift in the position of the shockfront and to the oscillatory behavior in the shocked region. It could be improved using a static mesh, which would remove other advantages of this method, however.

\subsubsection{Sedov-Taylor Blastwave}
This very strong, radially symmetric shock has first been introduced by \citet{Sedov1946,Sedov1959}. Besides the capability to deal with jumps, \citet{Saitoh&Makino2009} describe how it can be used to analyze the timestep limiter and shows the need for the limiting to be non-local, as provided by the wakeup scheme. The test has become a popular benchmark for Supernova blast wave evolution in recent years \citep[e.g.][]{Kim&Ostriker2015, Steinwandel+2020}.

\begin{figure}
    \centering
    \includegraphics[width=\columnwidth]{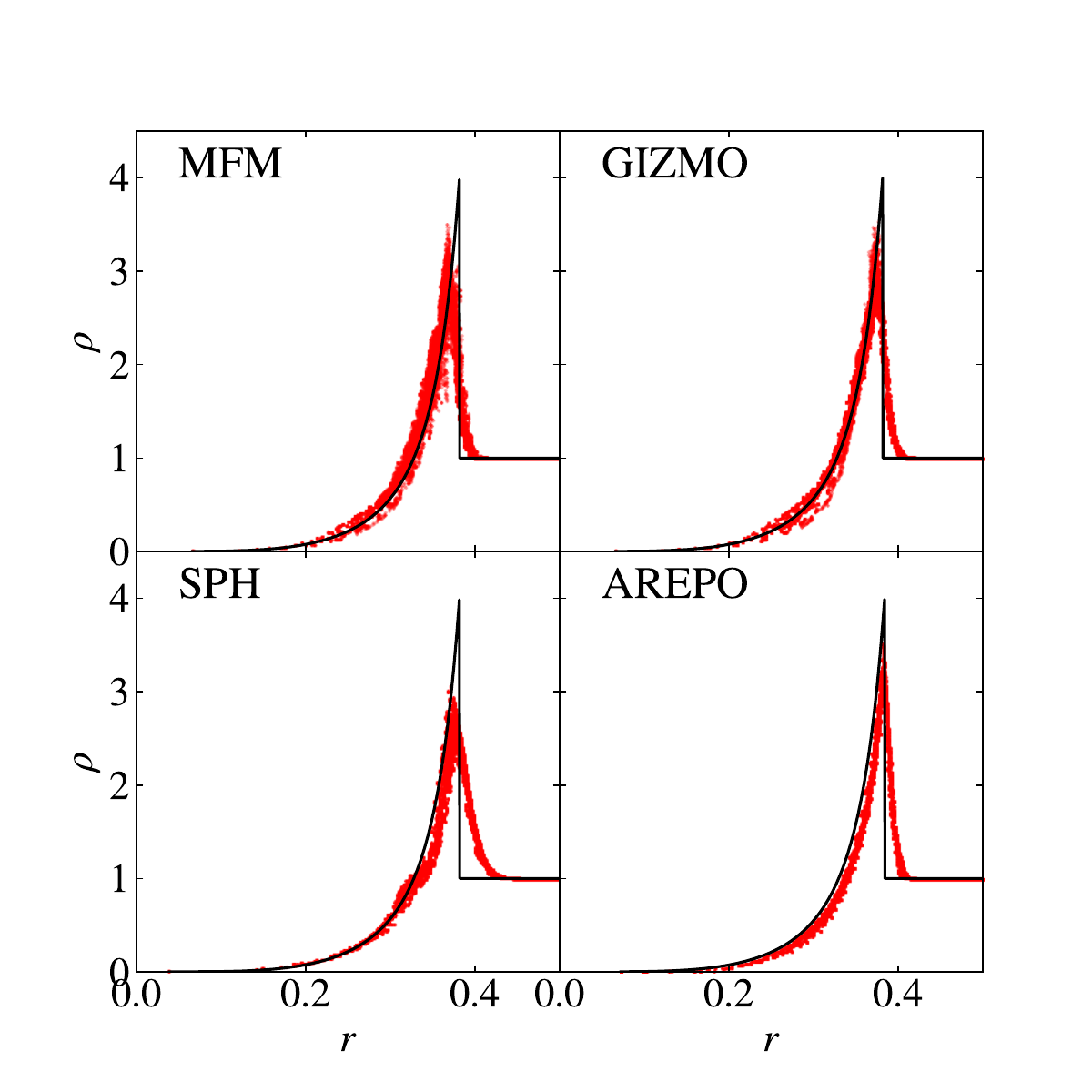}
    \caption{Sedov blast at $t=0.02$. Comparison between different methods. The analytical solution \citep{Sedov1946,Taylor1950,VonNeumann1961} is shown as reference. The main difference is the height of the peak, which is reduced due to smoothing of the jump.}
    \label{fig:sedov}
\end{figure}

As ICs, we setup a regular grid with $64^3$ particles and density $\rho=1$. While almost all particles exhibit a vanishing pressure $P_a=10^{-6}$, energy of $U=10$ is distributed equally into the eight central particles. A shock with very high $\mathcal{M}_i\gtrsim 2\e{4}$ arises, and quickly moves outwards.The radial density distribution is shown in Fig.~\ref{fig:sedov}.

All methods are able to capture the shock, though slightly smoothing it, thus underestimating the height of the density-peak.
SPH shows the strongest smoothing, followed by the two MFM implementations. \textsc{Arepo} is able to reproduce the height of the peak best.

The position of the peak is similar for all methods, with minor differences.
While \textsc{Arepo} and \textsc{gizmo}'s MFM implementation predict the peak position correctly, MFM and SPH in \textsc{OpenGadget3} lag slightly behind, which results in a more accurate position of the low-density side of the shock. This position strongly depends on the precise timestep settings, indicating differences in the timestepping between the codes.

\subsection{Including self-gravity}
In cosmological contexts, not only hydrodynamical forces, but also gravitational accelerations are of great importance. Gravity dominates the evolution on large scales due to its long-range character. It can lead to collapse of clouds, e.g. in the ISM for star formation, or balance thermal pressure and lead to hydrostatic equilibrium, such as in the global structure of galaxies or galaxy clusters. Thus, we analyze the interplay between hydrodynamical forces and gravity in the following.

\subsubsection{Gravitational Freefall}
\begin{figure}
    \centering
    \includegraphics[width=\columnwidth]{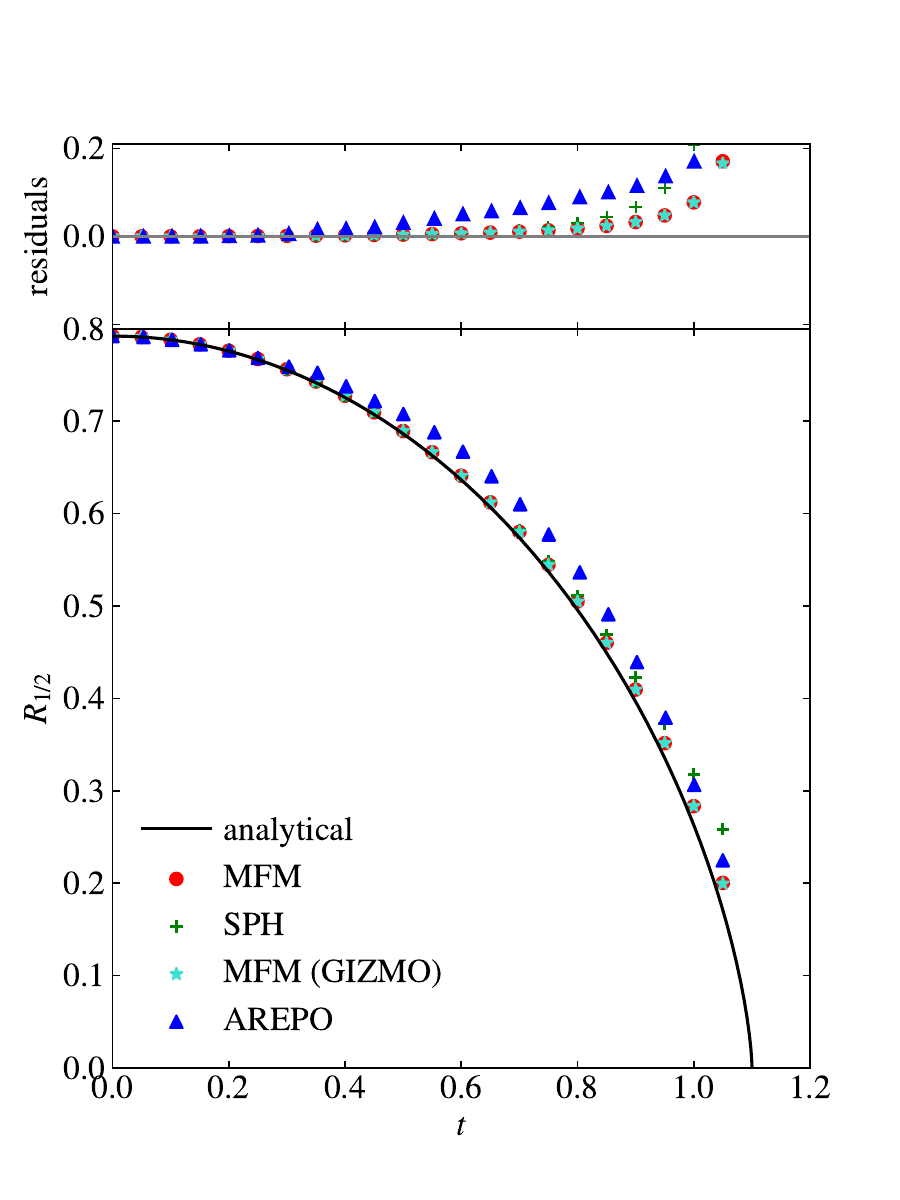}
    \caption{Evolution of the half-mass radius for the gravitational freefall test. All methods agree at early time, but deviate from the expected solution at later times when hydrodynamical contributions become more important.}
    \label{fig:grav_freefall}
\end{figure}

As a first test including self-gravity, we simulate a collapsing sphere. The ICs are set up on a regular grid of $20^3$ particles and cut out a sphere of radius $1$, which has a total mass of $M_\text{sphere}=1$ and a negligible pressure of $P=10^{-6}$. For the \textsc{Arepo} run, we fill the region not occupied by the sphere with low mass, low energy particles at resolution of only $8$ particles per unit length, arranged in a regular grid, in order to improve the mesh reconstruction at the boundary. We follow the evolution of the half-mass radius, to not be influenced by boundary effects as for the full radius, shown in  Fig.~\ref{fig:grav_freefall}.
Comparing to the analytic solution for a purely gravitational freefall
\begin{align}
    t(r) =~& \arccos\klammer{\sqrt{\frac{r}{r_0}} + \sqrt{\frac{r}{r_0}}\sqrt{1-\frac{r}{r_0}}}\cdot \frac{2}{\pi}\sqrt{\frac{3\pi}{32\rho_0}},
\end{align}
all methods agree at early times. At late times, pressure and thus effects of the hydro-scheme become more relevant, and deviations are visible. For all methods additional pressure contributions lead to an increase in radius as it would be expected. As the initial pressure is small, only a small deviation is expected. MFM lies closest to the ideal solution with both implementations being indistinguishable. The moving mesh code \textsc{Arepo} overestimates the radius already at early times, which can be explained by poor treatment of the non-periodic boundary conditions. While gravity is calculated in a non-periodic way, the mesh construction for the hydro-calculation requires the box to be treated periodically, which is not the case for all other methods. Including the low mass cells at the boundary already decreased the error by a factor of $2$ and it could be further decreased by enlarging the box. SPH lies in between the other methods except at very late times, when the deviation strongly increases due to over-smoothing.

\subsubsection{Hydrostatic Sphere} \label{sec:sphere}
\begin{figure*}
    \centering
    \includegraphics[width=\textwidth]{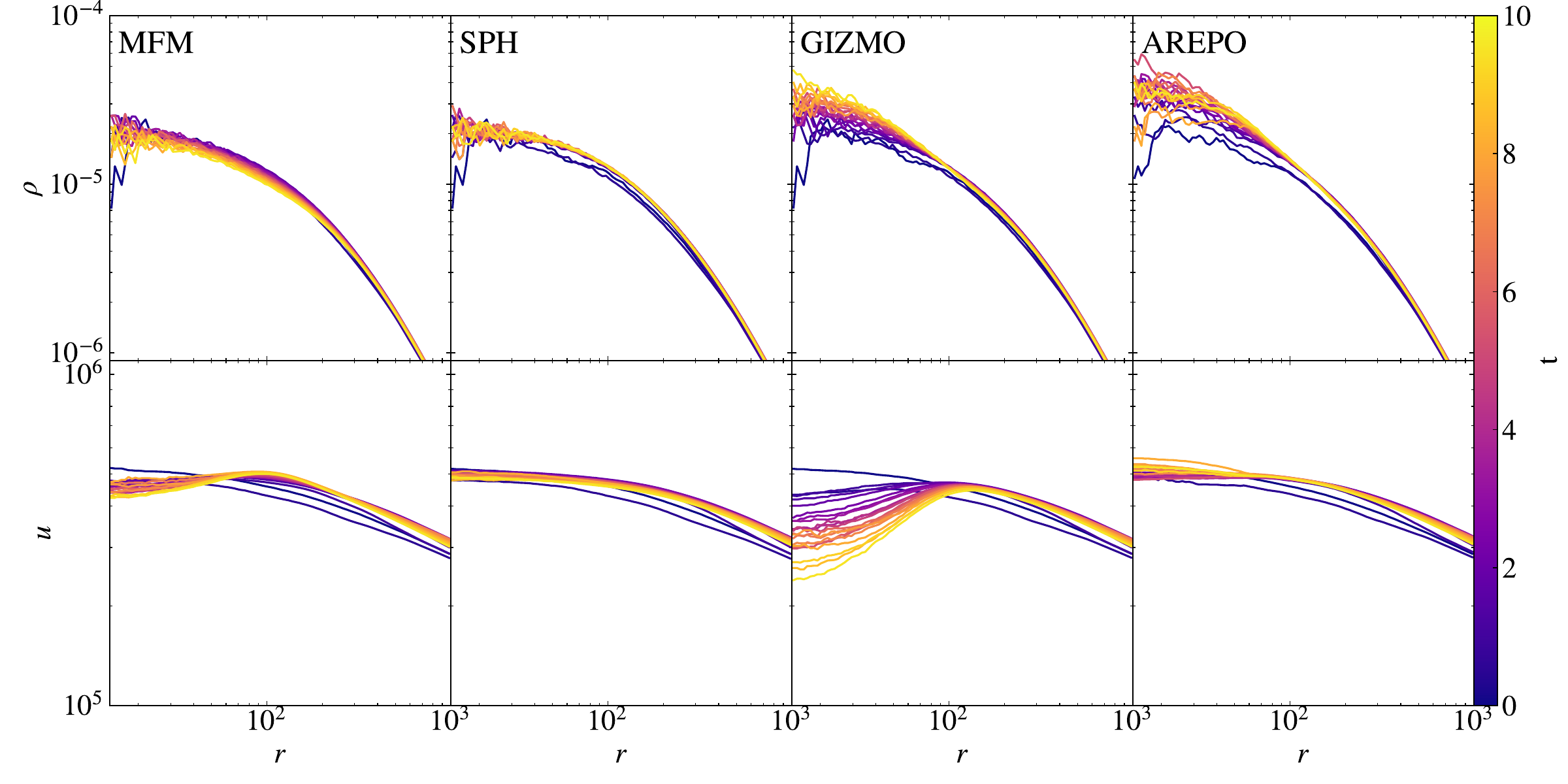}
    \caption{Evolution of gas density (top) and internal energy radial profiles (bottom) for the hydrostatic sphere for approximately $10$ dynamical times until $t=10$, colored by the time. Calculated using different hydro-methods. MFM shows a slightly larger numerical diffusivity, but overall still preserves the density profile.}
    \label{fig:hydro_sphere}
\end{figure*}

In cosmological contexts, e.g. for the ICM, the ability of the code to preserve hydrostatic equilibrium against gravity is of great importance. To test this, we calculate a hydrostatic sphere as a second test including self-gravity. It is also the first test including dark matter as second, only gravitationally interacting particle type.
The ICs have been created  
following \citet{Viola+2008}. 88088 DM particles are setup following an NFW profile \citep{Navarro+1997}, populated with 95156 gas particles in hydrostatic equilibrium. The corresponding density and internal energy profiles at different times are shown in Fig.~\ref{fig:hydro_sphere}.
After a short relaxation period, happening on a timescale approximately corresponding to the dynamical time $t_{\text{freefall}}\approx 1$ at $r=10^2$, we expect the gas to keep hydrostatic equilibrium.
SPH as well as our MFM implementation show the lowest deformation in density. MFM in \textsc{gizmo}, as well as \textsc{Arepo} show a slightly stronger increase in density, especially in the central region. 
While the density is only an indirect tracer of (numerical) diffusivity, the internal energy profile is more directly affected by it. Thus, it can give even more insight into the convergence over time.

For SPH, we observe a stable situation to be reached within one freefall time, and only minor changes to the initial profile. The same is true for MFM in \textsc{OpenGadget3}, where changes in internal energy are only marginally larger compared to SPH. For \textsc{Arepo}, changes compared to the initial profile are similar to the MFM result, as the initial conditions were designed assuming SPH. After a similar timescale for this relaxation, also for this method a stable situation is reached.
For MFM in \textsc{gizmo}, in contrast, an impact of the numerical diffusivity can be observed. Resulting mixing in the central region leads to a decrease in internal energy, leading to the observed increase in central density. Differences between the MFM implementations are results of the different Riemann solver in combination with fine differences of the implementation. Also in previous implementations of ours we observed a similar change as for \textsc{gizmo}.
Over very long timescales, the sphere would tend to become isothermal for MFM. Despite these findings, the effect on the density profile is quite small for all methods over the timescale considered.

\subsubsection{Zeldovich pancake}\label{sec:zeldovich}
The Zeldovich pancake is the first problem to test our implementation of co-moving integration. In addition, it is well suited to show effects of very high $\mathcal{M}$ flows, shocks, highly anisotropic particle arrangements, and also very low internal energies.
It has been introduced by \citet{Zeldovich1970}.
We start our calculation at $z_i=100$, setting up a single Fourier mode density perturbation. During the linear growth until the caustic formation at $z_c=1$, the evolution can be described by
\begin{align}
    x =~& x_i - \frac{1+z_c}{1+z}\frac{\sin(kx_i)}{k} \\
    \rho =~& \frac{\rho_0}{1-\frac{1+z_c}{1+z}\cos(kx_i)} \\
    \myvector{v}_\text{pec} =~& -H_0\frac{1+z_c}{\sqrt{1+z}\frac{\sin(kx_i)}{k}} \hat x \\
    T =~& T_i\klammer{\frac{1+z_c}{1+z}}^2\klammer{\frac{\rho(x,z)}{\rho_0}}^{2/3}
\end{align}
starting from the unperturbed position $x_i$. $\rho_c$ is the critical density, $H_0=h_0\cdot100$~km\,s${}^{-1}$\,Mpc${}^{-1}$ the Hubble parameter (today) with $h_0=1$, and $T_i=100$~K the initial temperature, such that pressure forces are negligible. The wavenumber $k=2\pi/(64~h^{-1}$\,Mpc$)$ corresponds to the first-order soundwave. We use the ICs provided by \citet{Hopkins2015}, with a resolution of $32^3$ particles.
After the linear growth, an accretion shock forms close to the center. 
As the scale factor increases, the background density decreases strongly and the background temperature decreases adiabatically. This causes a huge temperature contrast of $\approx 10$ orders of magnitude between the shocked region and the background. Due to the very low internal energy compared to other energy contributions $U\lesssim 10^{-3}E_{\text{kin}}$ and $U\lesssim10^{-2}E_{\text{pot}}$ in physical units, the evolution can be strongly dominated by numerical errors. Thus, the implementation of the energy-entropy switch described in Sec.~\ref{sec:switch} is important. The precise limits, when the switch is supposed to be active, are related to the numerical accuracy and details of the implementation such as the precision of the Riemann solver. The effect of the limits chosen on the evolution of the Zeldovich pancake is described further in Sec.~\ref{sec:switch_effect}.

\begin{figure*}
    \centering
    \includegraphics[width=\textwidth]{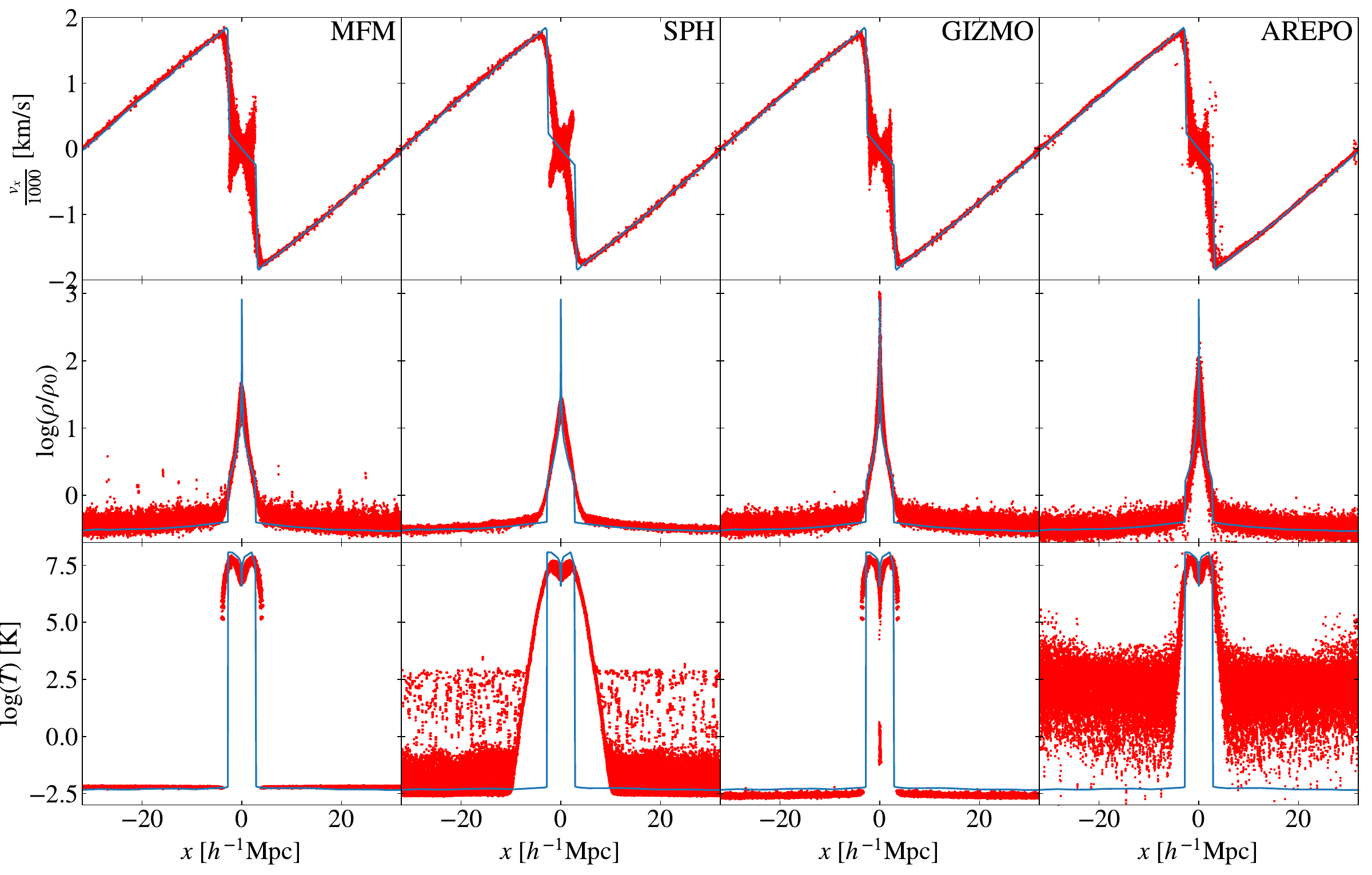}
    \caption{Zeldovich pancake at $z=0$ for different hydro-methods. As a comparison, a high resolution 1d simulation of \citet{Hopkins2015} is shown. While velocity and density profiles agree between the methods, strong deviations can be seen for the temperature profile. MFM performs best due to the energy-entropy switch employed.}
    \label{fig:zeldovich}
\end{figure*}
The resulting structure at $z=0$ is shown in Fig.~\ref{fig:zeldovich}. 
Again, we compare the performance of the different hydro-methods. The energy-entropy switch is included for MFM in \textsc{OpenGadget3} if $U<0.01E_\text{pot}$, corresponding also to the value implemented in \textsc{gizmo}. For \textsc{Arepo}, we had to use additional mesh regularization to avoid too irregular cell shapes in the highly unisotropically compressed shock region and allow the code to run until the end. All methods agree with the peculiar velocity profile with only slight differences. 
Compared to \citet{Hopkins2015} we find that all methods seem to have a too low viscosity and show particle over- or under-shooting compared to the  predicted velocity profile, as a result of a punch-through of some particles in the high $\mathcal{M}$ shock.
One difference between the methods is the height of the density peak. This is the lowest for \textsc{SPH}, which can be explained by the larger kernel for SPH, leading to stronger smoothing, compared to MFM. Thus, MFM in \textsc{OpenGadget3} is able to resolve the central region better and has a slightly higher peak. The \textsc{Arepo} run shows an even higher peak, contrarily to what \citet{Hopkins2015} found. Compared to the expected profile, all these methods over-smooth the central region. The \textsc{gizmo} code captures the density profile best, reaching the highest central peak.
Most difficult for all methods is to capture the temperature structure with its very strong contrasts. Both MFM implementations work very well, as the energy-entropy switch suppresses any numerical noise in the low-energy background and allows a clear jump between shocked and unshocked region. Slight differences in the implementation of the energy-entropy switch between \textsc{OpenGadget3} and \textsc{gizmo}, as well as a lower temperature in the central region in general result in more particles in the center being treated with the switch for \textsc{gizmo}, resulting in a larger number of cold particles. This difference can also explain the different height of the density peak. As we will show in Sec.~\ref{sec:switch_effect}, a less aggressive switch will result in a higher density peak. As no analytical solution exists for this test, is it unclear if this behavior is wanted.
The jump for SPH is more strongly smoothed in comparison to the other methods. In addition, amplified initial (numerical) noise causes a large scatter of several orders of magnitude in the very cold background.
For \textsc{Arepo}, we find that this behavior is much more drastic, and the background is dominated entirely by numerical noise. To properly resolve it, some energy-entropy switch would be required also in \textsc{Arepo}, which does not seem to be implemented in the public version. 

\subsubsection{Nifty Cluster}
\begin{figure*}
    \centering
    \includegraphics[width=\textwidth]{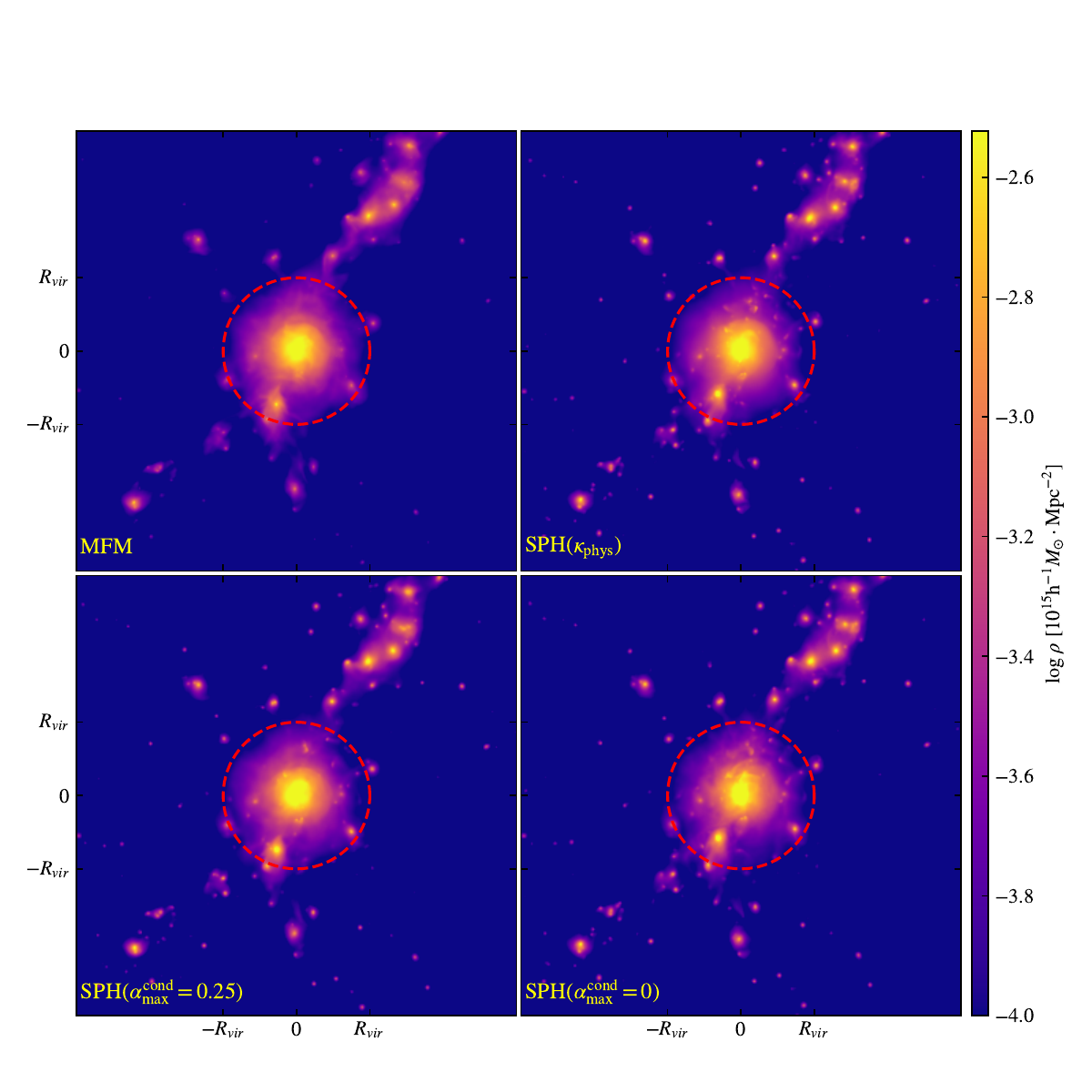}
    \caption{Projected surface density of the nifty cluster at $z=0$, comparison between MFM and SPH with usual amount ($\alpha_{\max}^{\text{cond}}=0.25$), physical ($\kappa_{\text{phys}}$), corresponding to an intermediate amount, and without artificial conductivity $\alpha_{\max}^{\text{cond}}=0$. The overall structure is very similar. Small sub-structures, however, appear less compact for MFM.}
    \label{fig:nifty_surface}
\end{figure*}

Finally, we apply our newly implemented method on more complex, cosmological cases. As an example, we re-simulate a cluster from the MUSIC-2 sample \citep{Prada+2012,Sembolini+2013,Sembolini+2014,Biffi+2014}, analyzed in detail with different codes by a collaboration formed during a nifty workshop \citep{Sembolini+2016}, thus called nifty cluster in the following. The cluster has a mass $M_{200\text{c}}=10^{15}~\text{M}_{\sun}$ with resolution $m_{\text{DM}}=9.01\e{8} h^{-1}\text{M}_{\sun}$ for dark matter and $m_{\text{gas}}=1.9\e{8} h^{-1}\text{M}_{\sun}$ for gas particles. The background cosmology has parameters $\Omega_{\text{M}}=0.27,\Omega_{\text{b}}=0.0469,\Omega_\Lambda=0.73,\sigma_8=0.82,n=0.95,h=0.7$ \citep[][]{Komatsu+2011}. The projected surface density at $z=0$ is shown in Fig.~\ref{fig:nifty_surface}, where the cluster center and virial radius are obtained using \textsc{Subfind} \citep{Springel+2001,Dolag+2009}.

We compare MFM to SPH with a different amount of artificial conductivity, ranging from the usually used amount $\alpha_{\max}=0.25$, $\alpha_{\min}=0$ \citep[notation following][]{Price2008}
over a run with physical conductivity at $1/20^{\text{th}}$ of the Spitzer value \citep[][]{Dolag+2004}, effectively corresponding to an intermediate amount, to more traditional SPH without artificial conductivity.
The usual amount is chosen to mimic the behavior of Godunov methods such as MFM, which have intrinsic numerical diffusivity due to the Riemann solver and thus allow for more mixing. The structure looks very similar between MFM and SPH with standard settings. This will change, however, for different values for artificial conductivity. For reduced artificial conductivity, structures are slightly less ``smeared out'', while the global structure does not change.

\begin{figure*}
    \centering
    \includegraphics[width=\textwidth]{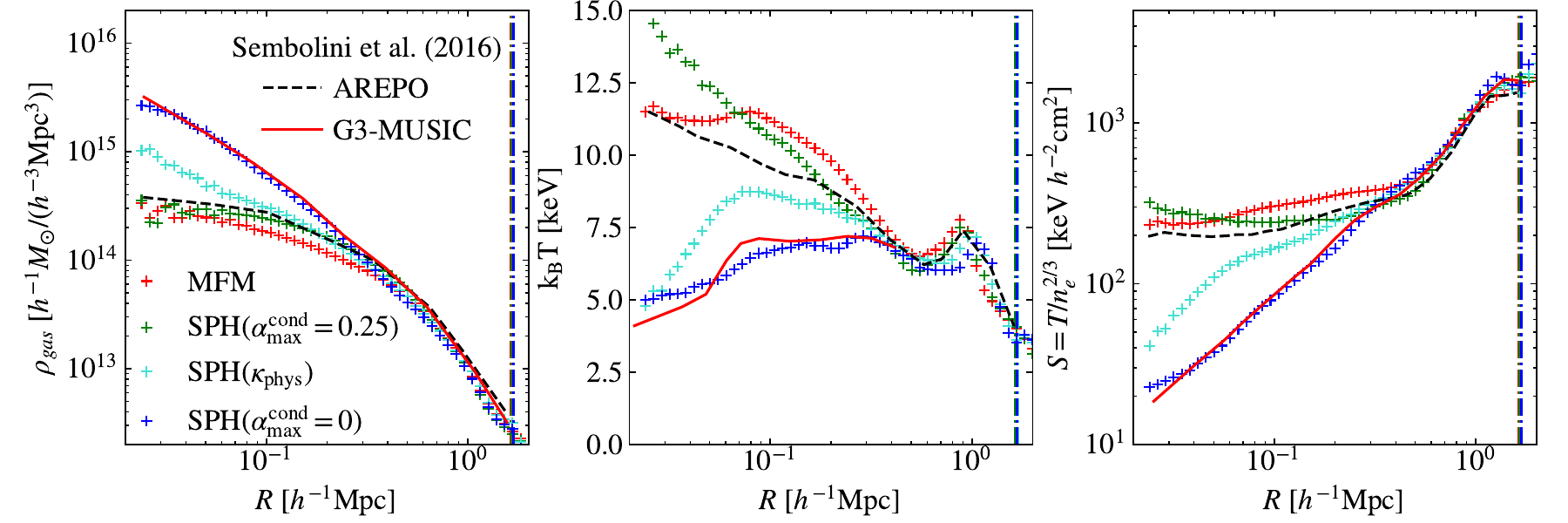}
    \caption{Gas density (left), temperature (middle) and entropy (right) radial profiles of the nifty cluster at $z=0$, comparison between different hydro methods, including our MFM implementation (red plus), SPH in OpenGadget with usual (green), physical, corresponding to an intermediate value, (turquoise) and without artificial conductivity (blue). As a comparison, the \textsc{Arepo} (black dashed) and G3-MUSIC (traditional) SPH line (red solid) from \citet{Sembolini+2016} are shown. The vertical line marks $R_{200}$.
    Our modern SPH run with sufficiently high artificial conductivity, as well as \textsc{Arepo} and MFM produce higher entropy cores with lower, less peaked density, while the central entropy is much lower for SPH with lower artificial conductivity.}
    \label{fig:nifty_profiles}
\end{figure*}

A more quantitative analysis can be done using gas radial density, temperature and entropy profiles shown in Fig.~\ref{fig:nifty_profiles}.
As a comparison, we provide lines from the nifty paper, obtained using \textsc{Arepo} and \textsc{Gadget3-MUSIC} as an example of a more traditional SPH code, which mark the range of solutions obtained.
SPH can span the whole range of possible solutions provided by \citet{Sembolini+2016}. By construction traditional SPH without artificial conductivity has no mixing and thus forms low entropy cores.
Subgrid mixing due to the Riemann solver for MFM and \textsc{Arepo} leads to mixing into the core, increasing the entropy compared traditional SPH. Thus, the central density is reduced.
By including artificial conductivity in SPH, it can reach the same profile as MFM, and also lie in between for effectively intermediate values by using physical conductivity.

\subsection{Decaying Subsonic Turbulence} \label{sec:turbulence}
In many astrophysical systems, ranging from the atmosphere over the ISM up to galaxy clusters, turbulence plays a crucial role. In the ICM, we expect subsonic turbulence with a turbulent energy fraction of $X\approx0.1$ to be excited, for instance after a merger \citep[compare, e.g.][]{Schuecker+2004,Subramanian+2006}.
The different hydro-schemes have problems to capture its full behavior. It has been shown that traditional SPH is not well suited to calculate sub-sonic turbulence \citep{Bauer&Springel2012}, but can be improved using modern SPH with more ideal settings for artificial diffusion terms \citep{Price2012}. While grid based methods produce better results, difficulties still remain for the evolution of turbulence within galaxy clusters.

\begin{figure*}
    \centering
    \includegraphics[width=\textwidth]{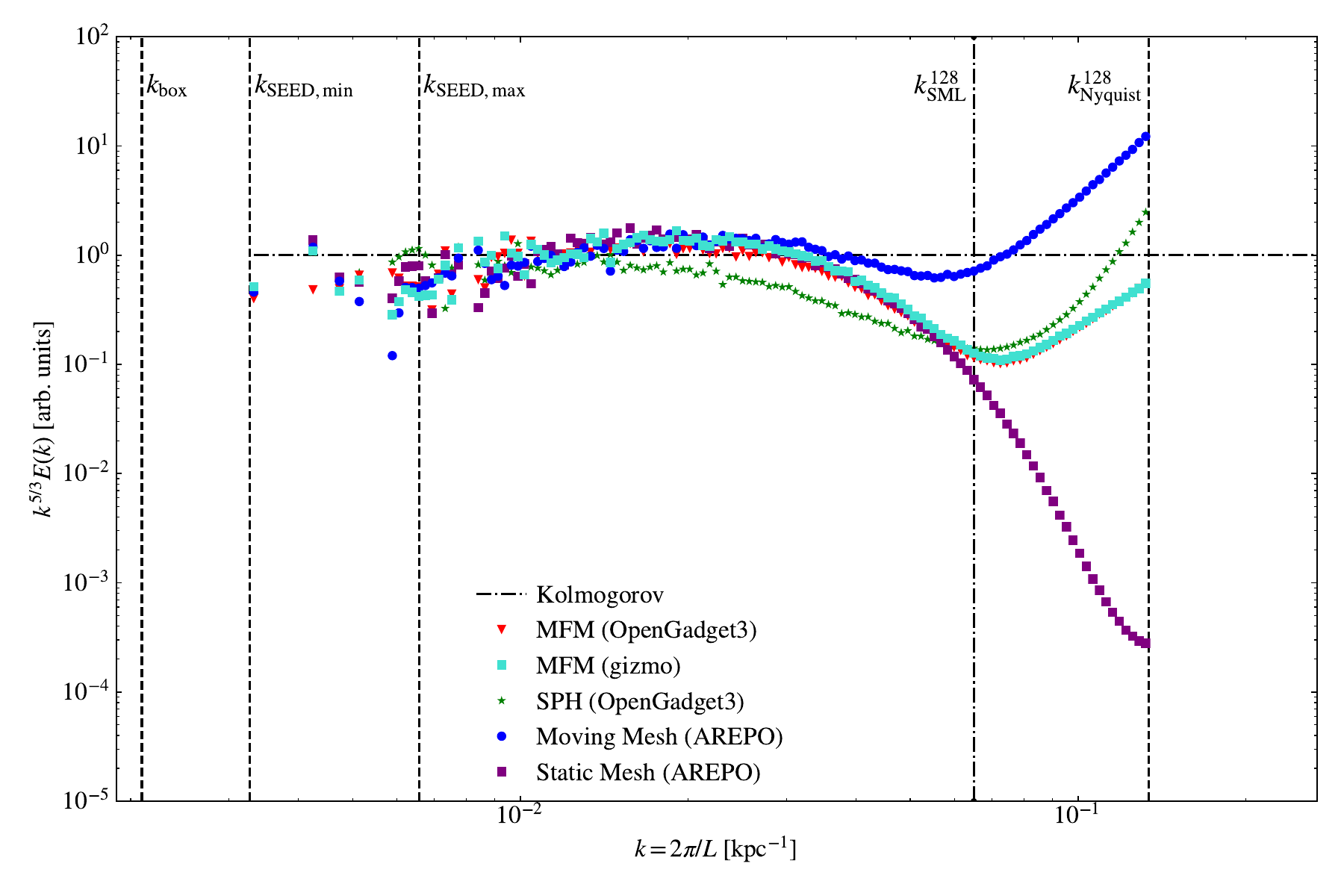}
    \caption{Normalized turbulent velocity power spectrum for different methods at $X_i=0.3$ and resolution $128^3$. All methods agree at large scales, but show a lack in energy at intermediate to small scales compared to the expected Kolmogorov-slope $P\sim k^{-5/3}$. Overall, all methods work very well} reproducing the expected spectrum.
    \label{fig:turbulence}
\end{figure*}

To test and compare the performance of our MFM implementation, we setup a $300$~kpc cubic box with varying number of particles, and seed the largest $\approx 70$ modes, similar to \citet{Bauer&Springel2012}. Due to the low initial density of $\rho\approx 1.5\e{-6}$, gravitational acceleration can be neglected. The initial turbulent energy fraction is varied between $X_i=E_{\text{turb},i}/E_{\text{therm},i}=0.3$, corresponding to a Mach number $\mathcal{M}_i\approx0.7$ and $X_i=0.00001$ corresponding to $\mathcal{M}_i\approx0.004$. In addition, the resolution is varied, ranging from $64^3$ up to $256^3$ particles.
We evolve the turbulence for $1.5$ sound-crossing-times. The turbulent kinetic energy cascades down to smaller scales, forming a turbulent power spectrum. In order to analyze the velocity power spectrum, the data are binned to a grid using the code Sph2Grid\footnote{Developed by J. Donnert, available at \url{https://github.com/jdonnert/Sph2Grid}}. From that, a power spectrum is calculated. We use a D20 sampling, to conserve energy \citep{Cui+2008}.
Theoretically, a Kolmogorov slope $E(k)\sim k^{-5/3}$ would be expected \citep{Kolmogorov1941}. In Fig.~\ref{fig:turbulence}, we compare the power spectra of the different methods, normalized by the expected slope. The wavenumber $k_{\text{box}}$ corresponds to a wavelength of the box size. Energy is seeded between $k_{\text{SEED,max}}$ and $k_{\text{SEED,min}}$. An estimate for the resolution limit is provided by $k_{\text{SML}}^{128}$, corresponding to the mean smoothing length for a Wendland C6 kernel at resolution $128^3$ in plots where an SPH run is included, otherwise to the mean smoothing length for a cubic spline kernel at resolution $128^3$, and $k_{\text{Nyquist}}^{128}$ denoting wavenumber of the initial grid-spacing and thus the smallest length to be possibly resolved.
\begin{figure*}
    \centering
    \includegraphics[width=\textwidth]{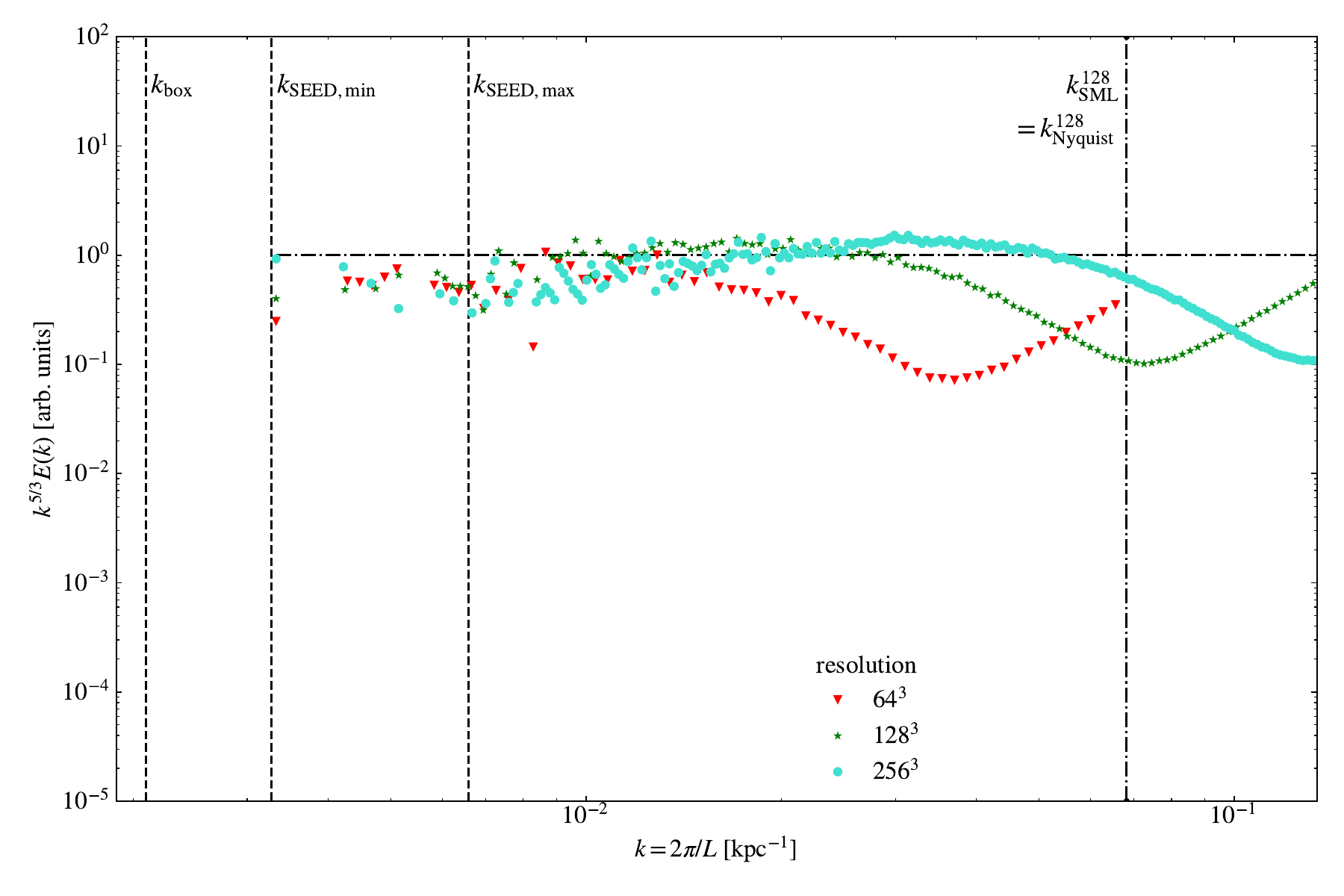}
    \caption{Normalized turbulent velocity power spectrum for MFM with different resolutions at $X_i=0.3$. MFM converges fast with resolution towards the expected Kolmogorov-slope $P\sim k^{-5/3}$.}
    \label{fig:turbulence_res}
\end{figure*}

In contrast to many previous findings \citep[compare, e.g.,][]{Padoan+2007,Bauer&Springel2012,Hopkins2015}, all methods are able to reproduce the expected Kolmogorov slope very well for such a mildly subsonic turbulence. SPH shows the strongest deviation, occurring already at intermediate scales, while for MFM in both implementations and also \textsc{Arepo} with moving and stationary mesh differences are present only at very small scales, approaching the resolution limit.

In addition, the MFM result converges quickly with resolution, shown in Fig.~\ref{fig:turbulence_res}. As the dip moves towards smaller scales, the overall spectrum becomes even closer to the Kolmogorov one over a wider range of scales.
At the highest resolution considered, it almost perfectly resembles the expected Kolmogorov slope over almost on order of magnitude of scales.

\begin{figure*}
    \centering
    \includegraphics[width=\textwidth]{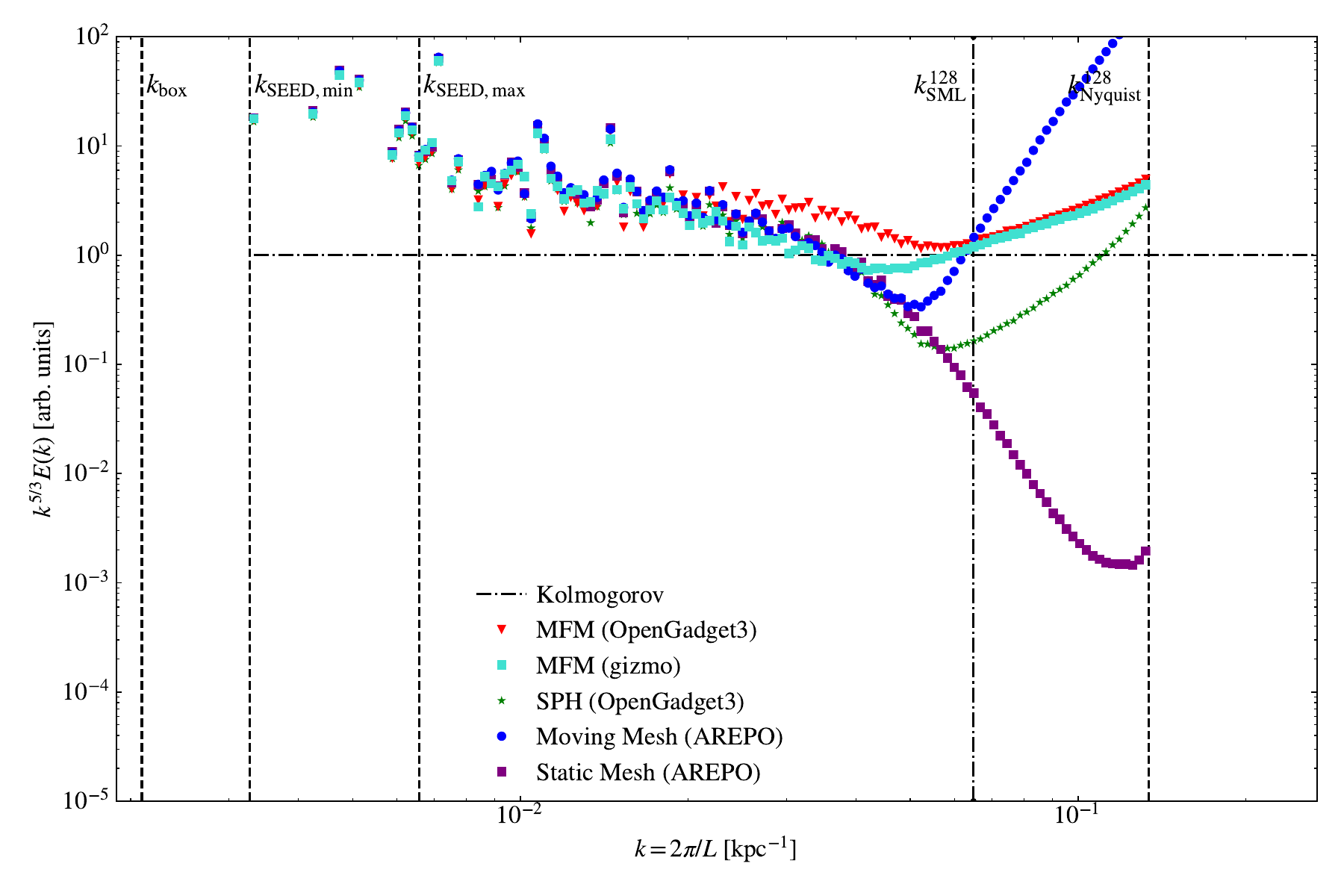}
    \caption{Normalized turbulent velocity power spectrum for different methods at $X_i=0.01$ and resolution $128^3$. At such low initial turbulent energy fraction, differences between the methods become more visible, where MFM works best overall reproducing the expected spectrum.}
    \label{fig:turbulence_lowx}
\end{figure*}
For even lower initial turbulent energy fractions, corresponding to even lower Mach numbers, more differences between the methods become visible. In Fig.~\ref{fig:turbulence_lowx}, we show the resulting spectrum for an initial turbulent energy fraction of $0.01$, corresponding to $\mathcal{M}_i\approx0.1$.
While all methods agree at large scales, where the energy was seeded, they show huge discrepancies at intermediate to small scales. \textsc{Arepo} shows deviations at the smallest scales compared to the other methods, underestimating the energy present at scales close to the resolution limit. SPH starts deviating at slightly larger scales, with a less deep dip in the power spectrum. For MFM the power spectrum shows a dip in energy at similar scales as the moving mesh code \textsc{Arepo}, but with a much shallower depth than in all other cases, thus being closer to the expected slope. Differences between the two MFM implementations can be attributed to different Riemann solvers used. Overall, advantages of MFM become clear for such very subsonic turbulence.

\begin{figure}
    \centering
    \includegraphics[width=\columnwidth]{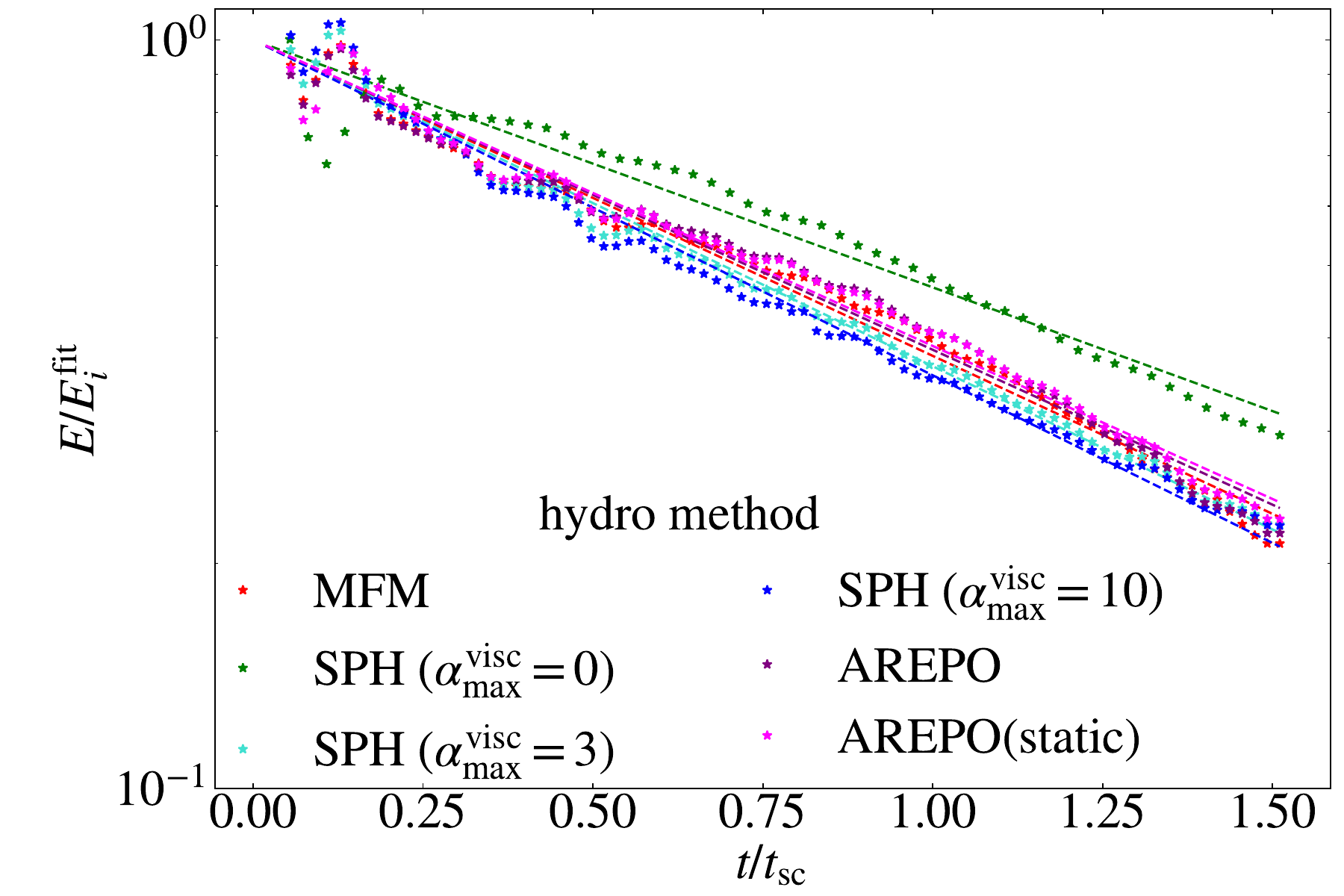}
    \caption{Decay time of turbulent energy for different methods at $X_i=0.3$. For SPH, the viscosity is varied between $\alpha_{\max}^{\text{visc}}=10$ and $\alpha_{\max}^{\text{visc}}=0$, where $\alpha_{\max}^{\text{visc}}=3$ is the value typically used \citep[notation following][]{Beck+2016}. \textsc{Arepo} has the highest decay time corresponding to the lowest numerical dissipation, while MFM and SPH at typical value of viscosity are on a similar order with a decay time of a few dynamical timescales.}
    \label{fig:decay_method}
\end{figure}

While the power spectrum builds up, energy is not only transported to smaller scales, but also partly converted into internal energy. We plot this decay of kinetic, turbulent energy, here labeled with $E$, in Fig.~\ref{fig:decay_method}, comparing the different hydro-methods. In order to better compare the slopes independent of initial turbulent energy fraction, we fit an exponential decay for each run and normalize by $E^{\text{fit}}_i=E^{\text{fit}}(t=0)$.
While in a physical situation the decay would depend on gas microphysics such as its viscosity, here we can use it to get an insight into the code behavior. The decay is mainly determined by numerical dissipation.
In all cases, the energy shows a periodic variation, caused by the ``ringing'' of the initially seeded modes.
The decay for SPH depends mildly on the artificial viscosity especially visible for the run excluding it. The power spectrum, in contrast, is only weakly influenced by the amount of artificial viscosity. In practical applications, it is tuned to a value of $\alpha_{\text{max}}^{\text{visc}}=3$, which leads to a similar decay rate as the other methods. The exponential decay time $t_{\text{dec}}$ roughly corresponds to the sound crossing time $t_{\text{sc}}=L_{\text{box}}/c_s$.

\begin{figure}
    \centering
    \includegraphics[width=\columnwidth]{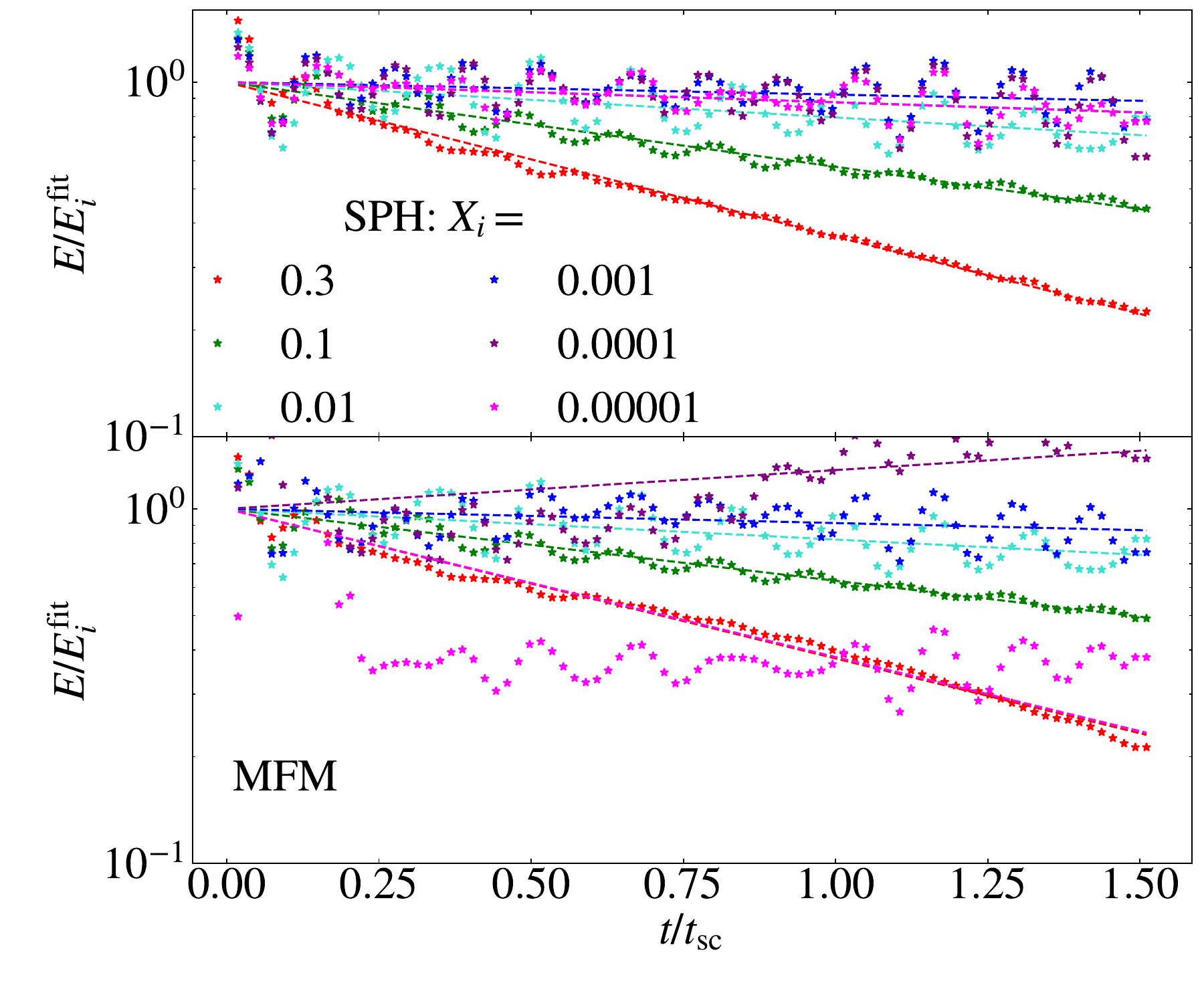}
    \caption{As Fig.~\ref{fig:decay_method}, but for varying initial turbulent energy fractions $X_i$, corresponding to variations in the turbulent Mach number. The decay is consistent for all $X_i$ for SPH, and down to $X_i=0.003$ for MFM, when numerical artifacts lead to an unphysical increase in energy.}
    \label{fig:decay_x}
\end{figure}

A comparison for the decay at different initial turbulent energy fractions, corresponding to variations in the Mach number, is shown in Fig.~\ref{fig:decay_x} for MFM and SPH.
The variation between $0.3$ and $0.00001$ for the initial turbulent energy fraction corresponds to a range of Mach numbers from $0.7$ down to below $0.004$.
For SPH the decay is decreasing with initial turbulent energy fraction down to $X_i=0.01$ ($\mathcal{M}\approx0.1$), and stays independent of the Mach number afterwards, as one would expect, so it is for \textsc{Arepo}.
For MFM, the same initial trend can be observed. For very low initial turbulent energy fraction $X_i<0.0001$ ($\mathcal{M}\lesssim0.01$), however, an unphysical increase of turbulent energy occurs.
At the same point also the density pdf deviates from the Gaussian shape together with the velocity power spectrum becoming strongly non-Kolmogorov, indicating the evolution is dominated by numerical artifacts for such low Mach numbers. Specifically, while we find that for this test total energy is conserved the internal energy only accounts for a small fraction, such that even evolving internal energy instead of total energy will be dominated by numerical errors. Applying an energy entropy switch could in principle alleviate this problem. On the other hand, it would be unclear if this would remove other advantages for MFM in this test problem.

\subsection{Runtime Analysis}

The precise difference in wallclock runtime between MFM and SPH depends on the problem that is considered, as this can include different additional physics and might trigger different timestep-limiters. Therefore, we provide an overview of all runtimes for the tests that we have run, including the MPI and OpenMP configuration in Tab.~\ref{tab:runtimes}.
\begin{table*}
    \centering
    \begin{tabular}{l|cc|cc|cc|cc}
        test & MPI & OpenMP & \multicolumn{2}{|c|}{runtime [s]} & \multicolumn{2}{|c|}{time per timestep [s]} & \multicolumn{2}{|c|}{\#steps} \\
         & & & MFM & SPH & MFM & SPH & MFM & SPH \\
        \hline
        soundwave (res 128) & 4 & 14 & $4.3\e{4}$ & $3.2\e{4}$ & $5.3\e{0}$ & $7.9\e{0}$ & $8.1\e{3}$ & $4.1\e{3}$ \\
        kepler disk & 4 & 14 & $4.1\e{5}$ & $5.4\e{5}$ & $8.1\e{0}$ & $8.0\e{0}$ & $5.1\e{4}$ & $6.8\e{4}$ \\
        Rayleigh-Taylor instability & 4 & 14 & $8.4\e{3}$ & $3.4\e{3}$ & $2.6\e{-1}$ & $1.1\e{-1}$ & $3.2\e{4}$ & $3.0\e{4}$ \\
        Kelvin-Helmholtz instability & 8 & 12 & $6.7\e{5}$ & $2.9\e{5}$ & $1.0\e{1}$ & $8.8\e{0}$ & $6.5\e{4}$ & $3.2\e{4}$ \\
        Hydrostatic Square & 4 & 14 & $1.9\e{2}$ & $7.1\e{1}$ & $2.3\e{-2}$ & $1.6\e{-2}$ & $8.2\e{3}$ & $4.2\e{3}$ \\
        blob test & 8 & 12 & $2.5\e{6}$ & $1.5\e{6}$ & $1.3\e{2}$ & $1.4\e{2}$ & $1.9\e{4}$ & $1.1\e{4}$ \\
        shock tube ($\mathcal{M}=10$) & 4 & 14 & $8.9\e{3}$ & $1.7\e{3}$ & $1.1\e{0}$ & $7.7\e{-1}$ & $8.2\e{3}$ & $2.1\e{3}$ \\
        Sedov blast & 4 & 14 & $1.1\e{2}$ & $1.9\e{2}$ & $2.1\e{-1}$ & $2.5\e{-1}$ & $5.3\e{2}$ & $7.8\e{2}$ \\
        gravitational freefall & 4 & 14 & $1.3\e{1}$ & $1.7\e{1}$ & $5.1\e{-2}$ & $6.6\e{-2}$ & $2.6\e{2}$ & $2.7\e{2}$ \\
        hydrostatic sphere & 4 & 14 & $1.3\e{4}$ & $1.9\e{4}$ & $8.2\e{-1}$ & $1.2\e{0}$ & $1.6\e{4}$ & $1.6\e{4}$ \\
        zeldovich pancake & 4 & 14 & $1.6\e{3}$ & $2.5\e{3}$ & $1.1\e{0}$ & $1.6\e{0}$ & $1.5\e{3}$ & $1.5\e{3}$ \\
        nifty cluster & 16 & 28 & $2.6\e{4}$ & $2.3\e{4}$ & $1.7\e{0}$ & $1.4\e{0}$ & $1.5\e{4}$ & $1.6\e{4}$ \\
        turbulence (resolution 128) & 8 & 12 &$1.2\e{5}$ & $4.9\e{4}$ & $2.4\e{1}$ & $1.6\e{1}$ & $5.1\e{3}$ & $3.1\e{3}$
    \end{tabular}
    \caption{Comparison of the runtime between MFM and SPH in \textsc{OpenGadget3}. Different test are run with different MPI and OpenMPI configurations and also on different machines. In general, we observe an increase in runtime between MFM and SPH, varying between a factor of $\approx 2$ for pure hydrodynamical tests to only a factor of $1.1$ for the nifty cluster.}
    \label{tab:runtimes}
\end{table*}
Overall, the computational costs for running MFM are comparable to those of SPH.
For pure hydrodynamical problems the number of timesteps required increases together with the effective spatial resolution by a factor of $\lesssim2$. It can be even larger for strong shocks due to the more strict timestep limiting at higher spatial resolution. The CPU-time required per timestep is very similar between the methods. While the Riemann solver is more expensive than the calculation of hydrodynamical accelerations for SPH, and also additional neighbor loops are required for MFM, the lower neighbor number leads to a decrease in computational costs. Depending on the problem, the time required per timestep can be smaller or larger than the time required for the SPH comparison run. In combination, these effects on average lead to slightly larger total runtimes by a factor of $\approx 2$ for MFM.

If gravity is included, the simulation timestep is dominated by the gravitational interactions in many cases, such that a similar number of timesteps is required, independent of the hydro-method. Also the time spent per timestep becomes even more similar, as the computation of gravitational interaction, which are calculated in the same way for MFM and SPH, are contributing as well.

If even more modules are activated, such as \textsc{Subfind}, as it is done for the nifty cluster, the runtime is mainly determined by the precise evolution. For the nifty cluster we found a slight increase in runtime for MFM, but it can be different for other objects simulated. If even more physics was turned on, we expect differences to become even smaller.

Memory requirements are mainly defined by the size of the particle structures. As MFM holds more variables in the gas particle structure, it is larger by a factor of $\approx 2$ compared to SPH. It could be improved by making more efficient use of existing SPH data and avoiding duplication which are currently still present. The total memory requirement is thus larger by a factor of $\approx 1.3-2$ for pure hydro problems and $\approx 1.5$ for the nifty cluster. Including more physics, the difference would become negligible.

\subsection{Effects of Numerical Parameters} \label{sec:parameters}
The performance of the numerical methods strongly depends on the precise parameters used. Effects of neighbor number and kernel have already been analyzed in detail by various authors \citep[compare, e.g.,][]{Dehnen&Aly2012,Tricco&Price2013,Hu+2014} for SPH.
To this end, we focus on two other parameters that play a major role for MFM, namely the slope-limiting scheme and the energy-entropy-switch.

\subsubsection{Slope-Limiter} \label{sec:slope_limiter}
The different slope-limiting procedures, which are implemented in our code, differ not only in how aggressively they limit the slope, but also in how much numerical diffusivity they introduce.
In general, different limiters are shown to produce different results for specific test-cases \citep[compare e.g.][]{Barth&Jespersen1989,Balsara2004,May&Berger2013,Hubber+2018,AlonsoAsensio+2023}.

\begin{figure}
    \centering
    \includegraphics[width=\columnwidth]{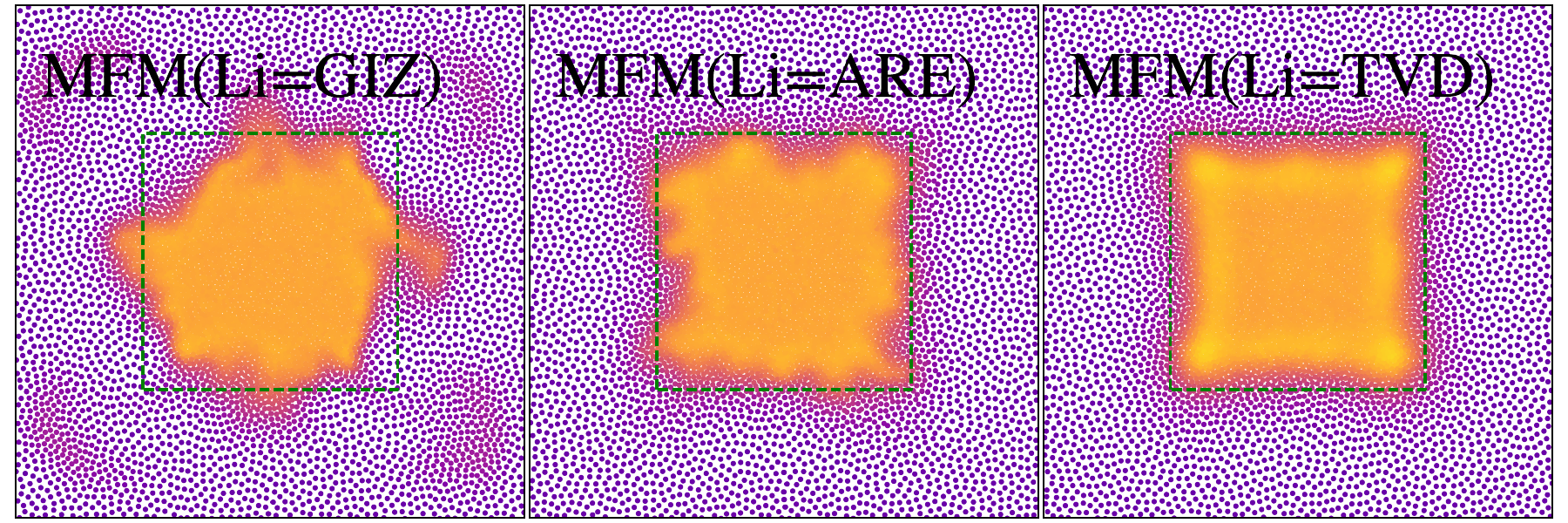}
    \includegraphics[width=\columnwidth]{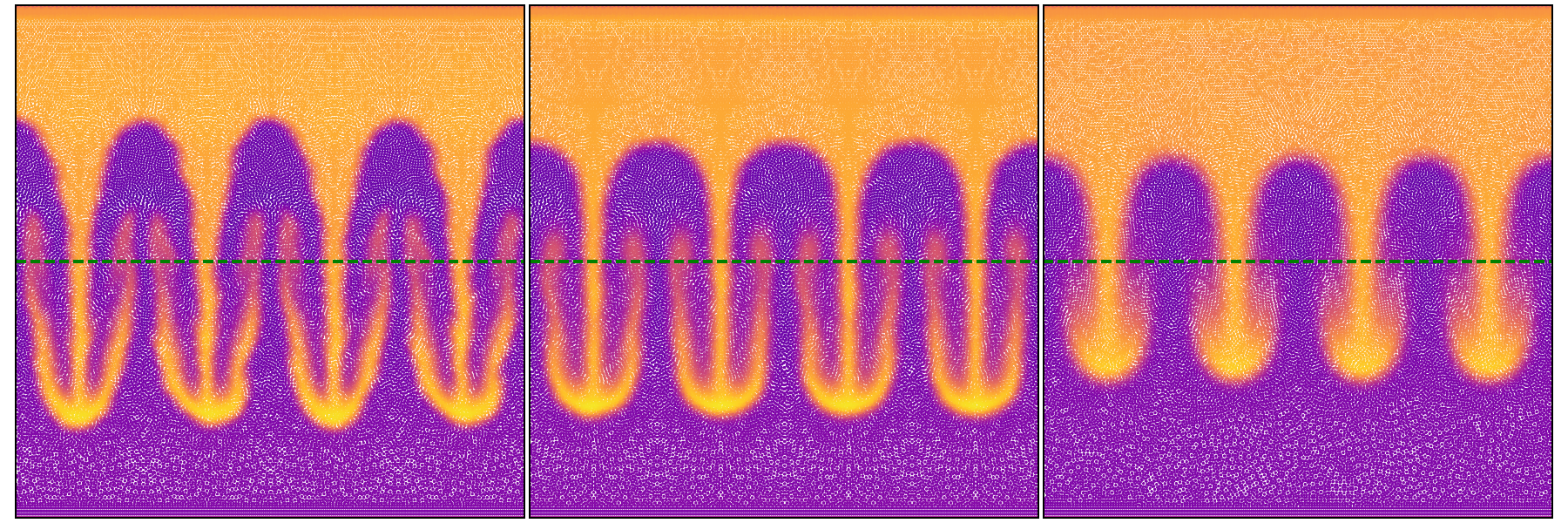}
    \caption{Hydrostatic square (top) and Rayleigh-Taylor instability (bottom), developed using different slope-limiters, the \textsc{gizmo} limiter we usually use (left), compared the the same test, but evolved using the \textsc{Arepo} limiter (middle) and TVD limiter (right). Depending on the test, different slope-limiters could be preferred.}
    \label{fig:limiter}
\end{figure}

In the following, we compare the three cases of the limiter from \textsc{gizmo} as described by Eqn.~(\ref{eq:gizmo_limiter}) in combination with their pariwise limiter (Eqn.~(\ref{eq:gizmp_pairwise_limiter1}/\ref{eq:gizmp_pairwise_limiter2})), that we usually use, the \textsc{Arepo} (Eqn.~(\ref{eq:arepo_limiter})) and, the TVD limiter (Eqn.~(\ref{eq:tvd_limiter})). The \textsc{gizmo} and TVD limiters are the most extreme cases of the limiters implemented, with lowest and highest numerical diffusivity, respectively. The \textsc{Arepo} limiter lies in between.
We analyze the effect on the hydrostatic square (compare also Sec.~\ref{sec:square}) and the Rayleigh-Taylor instability (Sec.~\ref{sec:rt}).
The results are shown in Fig.~\ref{fig:limiter}.

While for the Rayleigh-Taylor instability the much less diffusive \textsc{gizmo} limiter performs best, evolving a much finer structure, this causes the strongest deformation of the hydrostatic square.
The \textsc{Arepo} limiter is slightly more diffusive, leading to less strong secondary instabilities for the Rayleigh-Taylor instability and slightly less deformation of the square, especially at the edges.
The TVD limiter has an even higher numerical diffusivity, thus strongly smooths the Rayleigh-Taylor instability, not only preventing secondary instabilities to form, but also noticeably reducing the overall growth of the instability. The hydrostatic square, however, is preserved best, due to lower surface-tension like errors, which also manifest in the presence of absence of the pressure-blib for shocks.

Combining the results, we show that it is not always clear which slope-limiting procedure would be the overall preferred choice. As in most cases the \textsc{gizmo} limiter performs best, we chose this as our reference method.

\subsubsection{Energy-Entropy-Switch} \label{sec:switch_effect}
To avoid numerical errors to dominate the evolution of the internal energy, an energy-entropy switch as described in Sec.~\ref{sec:switch} has to be used in specific problems such as the Zeldovich pancake. Especially, the numerical noise should be suppressed in the very cold, unshocked region, while the shock should not be influenced at all.

\begin{figure}
    \centering
    \includegraphics[width=\columnwidth]{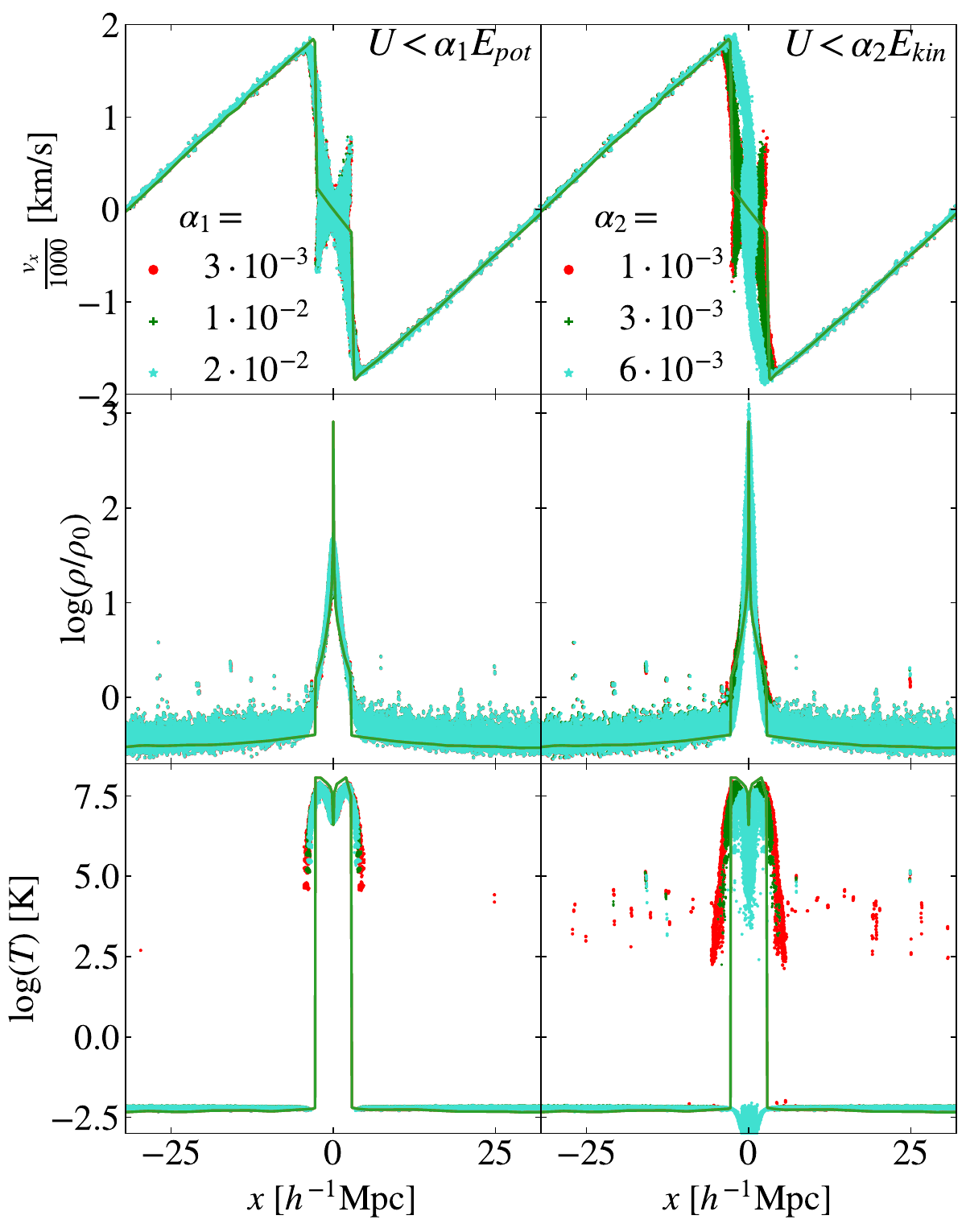}
    \caption{Effect of the choice of the energy-entropy switch on the Zeldovich pancake. Comparison between the switch based on kinetic and potential energy, each with three different $\alpha$-values. The switch based on potential energy is much more stable.}
    \label{fig:switch}
\end{figure}

The resulting structure at $z=0$, comparing different possibilities for the switch based on potential and kinetic energy estimates (compare also Eqn.~(\ref{eq:switch})), is shown in Fig.~\ref{fig:switch}.
We increase the tuned values ($\alpha_1=10^{-2}$ for the potential energy and $\alpha_2=3\e{-3}$ for kinetic energy) by a factor $2$ and decrease them by a factor $\approx 3$.

A more strict switch (larger $\alpha$) causes less particles to be treated with the adiabatic approximation. For the kinetic energy switch, this difference causes strong variations in the temperature profiles. While for $\alpha_2=1\e{-3}$ more extended wings form and some scatter in the low-temperature background close to the peak appears, the increased value of $6\e{-3}$ treats even particles inside the peaked region with the adiabatic approximation and causes too low temperatures. A very fine-tuned choice of $\alpha_2$ is necessary to accurately capture all particles, both the shocked ones and the low-temperature ones.

Compared to that, a variation of $\alpha_{1}$ within the switch based on potential energy influences the temperature profile only weakly. Thus, it seems to be much more stable and should be the preferred option. We thus used $\alpha_1=10^{-2}$ and $\alpha_2=0$ for the calculations in Sec.~\ref{sec:zeldovich}.

\section{Discussion and Conclusions} \label{sec:discussion}

We presented a new MFM implementation into \textsc{OpenGadget3} as an alternative hydro-solver to the currently used modern SPH. We verified its capabilities, both in idealized and more complex, cosmological test cases. Tests range from smooth, simple situations, mixing instabilities, shocks, tests including self-gravity, to the nifty cluster as cosmological example and decaying, subsonic turbulence.
A comparison has been preformed between MFM and SPH in \textsc{OpenGadget3}, the MFM implementation in \textsc{gizmo} and the moving mesh code \textsc{Arepo}.
In addition, two parameters have been analyzed in more detail.

Overall, we find very good agreement between the MFM implementation in \textsc{OpenGadget3} and that in \textsc{gizmo}. Minor differences are found in the precise appearance, while global properties are indistinguishable in most test cases. Even without further tuning, MFM reproduces the expected behavior in all test cases considered.
The soundwave test is well suited for a convergence analysis, as an analytical solution exists. MFM shows a very good convergence behavior between first and second order for dispersion errors. Diffusion errors as well as the scatter converge second order. While the convergence is better than for SPH and a moving mesh, these methods show lower errors at low resolution, especially for the dispersion error.

An important advantage of MFM over SPH is the capability to accurately evolve mixing instabilities without additional artificial viscosity or conductivity as for SPH. In addition, a lower neighbor number compared to SPH is sufficient. MFM as well as a moving mesh even show secondary instabilities to occur.
The blob test as combination between mixing and shocks emphasizes the ability of MFM to allow for more mixing. The decay rate of the cloud is similar to that of a moving mesh simulation and larger than for SPH. Compared to the more traditional SPH implementation shown by \citet{Hopkins2015}, the modern SPH implementation \textsc{OpenGadget3} allows for more mixing and leads to a faster decay of the cloud.
As this test is designed to mimic ram-pressure stripping, we expect this effect to be modeled more accurately using MFM compared to SPH. This should also lead to an overall more accurate evolution of galaxies in the environment of galaxy clusters.
To fully understand and follow the evolution of such gas blob in cosmological contexts more physics such as cooling, and, depending on the context, star formation, is necessary. \citet{Gronke&Oh2018,Gronke&Oh2020,Gronke&Oh2022} have analyzed this test in detail with such additional physics and found a great importance of the cooling.

In addition, MFM can model shocks for a wide range of Mach numbers. For the shock tube tests MFM performs especially well for lower Mach numbers, while effects of surface tension due to the choice of the slope-limiter are visible at higher Mach numbers. Nevertheless, it is still able to capture the main features of the shock including the position of the shock front, the contact discontinuity, and the rarefaction fan. Different methods lead to differences in the smoothing of the shock front. The lower neighbor number in MFM compared to SPH increases the effective spatial resolution by a factor of $\approx 2$. 
For \textsc{Arepo}, the shock front is dominated by numerical artifacts due to difficulties in the mesh reconstruction in such highly anisotropic region.

The Sedov blast works well for all methods, verifying the capability of the wakeup scheme as non-local timestep criterion. Main differences are the smoothing and resulting lower amplitude of the density peak, revealing an even smaller smoothing for the moving mesh compared to MFM.
The narrower shock front will help e.g. for shock detection in cosmological simulations \citep[compare, e.g., ][]{Pfrommer+2006,Beck+2016a}.

In general, MFM is able to preserve hydrostatic equilibrium accurately, as well as preserving stable orbits. The better stability of the Kepler disk compared to SPH will improve results for simulations of e.g. isolated galaxies. For this case, a moving mesh leads to even better results, but requires additional boundary particles.

The hydrostatic sphere test showed that our MFM implementation coupled to gravity leads to stable hydrodynamical equilibrium, as for SPH and a moving mesh. Depending on the details of the implementation, however, numerical diffusivity can be introduced. Thus, one could expect isolated galaxies or also the core of galaxy clusters to be more compact and cooler in the center. The timescales, on which these changes would happen, are, however, very long.

Also for the nifty galaxy cluster we saw that there is no difference between MFM, \textsc{Arepo} and modern SPH in the global structure.
Numerical diffusivity introduced by the Riemann solver allows mixing of entropy into the core, thus decreasing the central density compared to traditional SPH, which suppresses any mixing. Modern SPH mimics the same effect by applying artificial conductivity, while a variation of the precise amount introduced can lead to significant changes in the structure.
As observed galaxy clusters show a wide range of central entropy profiles \citep{Cavagnolo+2009}, both results are consistent with observations. Especially, we expect a more complex interplay with cooling, as well as stellar and AGN feedback to influence the entropy-evolution of the core \citep[compare, e.g.][]{Pearce+2000,Borgani+2005,Rasia+2015}. These effects lead to the whole range of possible central profiles, dominating over effects of the hydro-solver. Thus, further studies including such processes would be necessary.

In the intra cluster medium, we expect turbulence at low Mach number to be seeded e.g. by mergers at large scales. It will then decay and build up a turbulent power spectrum.
Such decaying, subsonic turbulence is a very challenging problem for many hydro-methods.
MFM is able to recover the turbulent power spectrum best compared to SPH and a moving and stationary mesh, best visible at very low initial turbulent energy fractions. Only a small lack of energy at intermediate to small scales close to the resolution limit -- similar to where this occurs also for \textsc{Arepo} -- is present. This ``dip'' in energy moves to smaller scales for higher resolution, overall leading to fast convergence towards the expected Kolmogorov spectrum.

The decay rate of turbulent energy due to numerical dissipation is on the same order as for modern SPH, and decreases towards higher resolution. The results are consistent down to very small initial turbulent energy fractions $X_i=0.0001$, corresponding to small Mach numbers $\mathcal{M}=0.01$. For smaller $X_i<0.0001$ numerical effects dominate and lead to unphysical increase in turbulent energy.
Overall, the results are very promising for the accurate evolution of turbulence also within galaxy clusters.

An energy-entropy switch is of great importance to accurately evolve the temperature profile for the Zeldovich pancake. When it is included, MFM yields the best results, having a clear jump in the temperature. Comparing different possible values for such a switch, we found that careful tuning is required. In general, the switch based on potential energy produces more stable results.

\textsc{Arepo} misses the implementation of such a switch in the public version, such that the low-temperature region is entirely dominated by numerical noise. SPH also shows noise in the low-temperature region, originating from the amplification of noise present in the ICs, and also much broader wings around the peak.
All methods show some punch-through in the temperature profile, indicating a too low viscosity.

In addition to comparing different methods, we used two tests to analyze the impact of the slope-limiter. Depending on the problem, different slope-limiters can be preferred. While the \textsc{gizmo} limiter performs best in most test-cases, having a much lower numerical diffusivity, specific cases such as the hydrostatic square and also strong shocks work better using a more diffusive TVD-limiter. The \textsc{Arepo} limiter has an intermediate diffusivity and lies in between the two other results.

Overall, our implementation of MFM produces accurate results for the cases considered. It avoids some of the disadvantages of SPH, while requiring 
a similar computational cost per timestep. 
The total number of timesteps and thus the total runtime increases as a result of the smaller smoothing length and effectively higher spatial resolution.
A faster, approximate Riemann solver can further decrease the computational costs in some cases, but has the drawback of introducing more numerical diffusivity, as discussed in App.~\ref{app:Riemann}.
Compared to MFM, a moving mesh requires a very expensive tessellation to be performed, such that the required computational costs for many tests are drastically increased.

Overall, MFM is a promising alternative for cosmological simulations.

\subsection{Outlook -- possible extensions in the future}
To make use of the full advantages of \textsc{OpenGadget3}, it will be useful to couple MFM not only to gravity, but also to include more physical processes, such as cooling, star formation and stellar feedback, AGN feedback, physical conductivity and viscosity. For these, we can make use of already existing implementations in \textsc{OpenGadget3}.

Finally, MFM can be expanded to an MHD method, including magnetic fields. This will also allow to include the existing implementation of cosmic rays.

For many of these extensions, coupling can be done in a similar way as for SPH, while others such as magnetic fields will require more significant changes including another Riemann solver.

In principle, also a general-relativistic (GR) extension would be possible, which has been implemented both for SPH \citep{Liptai&Price2019,Rosswog&Diener2021} and a moving mesh \citep{Chang&Etienne2020,Lioutas+2022} and also exists for MFM within the \textsc{gizmo} code \citep{Lupi2022a}. As GR is mainly important in extreme situations such as accretion discs around black holes, this would also make use of the fact that our MFM implementation is originally based on \textsc{GANDALF}, which itself was designed to deal with star and planet formation, and thus we would expect also our implementation to be well suited for calculations of disks.

\section*{Acknowledgements}

We thank the anonymous referee for the detailed and constructive feedback which helped to improve the quality of this paper.
FG and KD acknowledge support by the COMPLEX project from the European Research Council (ERC) under the European Union’s Horizon 2020 research and innovation program grant agreement ERC-2019-AdG 882679. UPS is supported by a Flatiron Research Fellowship at the Center for Computational Astrophysics (CCA) of the Flatiron Institute. The Flatiron Institute is supported by the Simons Foundation. FG, MV and KD acknowledges support by the Deutsche Forschungsgemeinschaft (DFG, German Research Foundation) under Germany’s Excellence Strategy - EXC-2094 - 390783311. MV is supported by the Alexander von Humboldt Stiftung and the Carl Friedrich von Siemens Stiftung. We are especially grateful for the support by M. Petkova through the Computational Center for Particle and Astrophysics (C2PAP) under the project pn68va. Some calculations for the hydrodynamical simulations were carried out at the Leibniz Supercomputer Center (LRZ) under the project pr86re (SuperCast). We thank C.~Alig for a turbulence \textsc{Arepo} setup and L.~B\"oss for providing ICs for the shock-tubes. The analysis was performed mainly in julia \citep[][]{Bezanson+2014}, including the package GadgetIO by \citet{GadgetIO}.
The surface density of the nifty cluster was calculated using Smac \citep{Dolag+2005}. We thank the developers of \textsc{gizmo} and \textsc{Arepo} for making the codes publicly available. We thank R. Pakmor for the help in analyzing and improving the convergence of the soundwave with \textsc{Arepo}.

\section*{Data Availability}

The setup for the different tests are publicly available at \url{https://github.com/fgroth/hydro_tests}. This includes parameter and config files for the different codes used, as well as our analysis routines. If applicable, routines to create ICs are also included. Other data will be shared upon reasonable request to the corresponding author. \textsc{OpenGadget3} is a non-public developer version of \textsc{Gadget-2}. It is available upon reasonable request from K.~Dolag.



\bibliographystyle{mnras}
\bibliography{quellen}

\begin{thebibliography}{}
\makeatletter
\relax
\def\mn@urlcharsother{\let\do\@makeother \do\$\do\&\do\#\do\^\do\_\do\%\do\~}
\def\mn@doi{\begingroup\mn@urlcharsother \@ifnextchar [ {\mn@doi@}
  {\mn@doi@[]}}
\def\mn@doi@[#1]#2{\def\@tempa{#1}\ifx\@tempa\@empty \href
  {http://dx.doi.org/#2} {doi:#2}\else \href {http://dx.doi.org/#2} {#1}\fi
  \endgroup}
\def\mn@eprint#1#2{\mn@eprint@#1:#2::\@nil}
\def\mn@eprint@arXiv#1{\href {http://arxiv.org/abs/#1} {{\tt arXiv:#1}}}
\def\mn@eprint@dblp#1{\href {http://dblp.uni-trier.de/rec/bibtex/#1.xml}
  {dblp:#1}}
\def\mn@eprint@#1:#2:#3:#4\@nil{\def\@tempa {#1}\def\@tempb {#2}\def\@tempc
  {#3}\ifx \@tempc \@empty \let \@tempc \@tempb \let \@tempb \@tempa \fi \ifx
  \@tempb \@empty \def\@tempb {arXiv}\fi \@ifundefined
  {mn@eprint@\@tempb}{\@tempb:\@tempc}{\expandafter \expandafter \csname
  mn@eprint@\@tempb\endcsname \expandafter{\@tempc}}}

\bibitem[\protect\citeauthoryear{Agertz et~al.,}{Agertz
  et~al.}{2007}]{Agertz+2007}
Agertz O.,  et~al., 2007, \mn@doi [Monthly Notices of the Royal Astronomical
  Society] {10.1111/j.1365-2966.2007.12183.x}, 380, 963

\bibitem[\protect\citeauthoryear{Alonso~Asensio, Dalla~Vecchia, Potter  \&
  Stadel}{Alonso~Asensio et~al.}{2023}]{AlonsoAsensio+2023}
Alonso~Asensio I.,  Dalla~Vecchia C.,  Potter D.,   Stadel J.,  2023, \mn@doi
  [Monthly Notices of the Royal Astronomical Society] {10.1093/mnras/stac3447},
  519, 300

\bibitem[\protect\citeauthoryear{Appel}{Appel}{1985}]{Appel1985}
Appel A.~W.,  1985, SIAM Journal on Scientific and Statistical Computing, 6, 85

\bibitem[\protect\citeauthoryear{Arth, Dolag, Beck, Petkova  \& Lesch}{Arth
  et~al.}{2014}]{Arth+2014}
Arth A.,  Dolag K.,  Beck A.~M.,  Petkova M.,   Lesch H.,  2014, Anisotropic
  Thermal Conduction in Galaxy Clusters with {{MHD}} in {{Gadget}}

\bibitem[\protect\citeauthoryear{Balsara}{Balsara}{1998}]{Balsara1998}
Balsara D.~S.,  1998, \mn@doi [The Astrophysical Journal Supplement Series]
  {10.1086/313093}, 116, 133

\bibitem[\protect\citeauthoryear{Balsara}{Balsara}{2004}]{Balsara2004}
Balsara D.~S.,  2004, \mn@doi [The Astrophysical Journal Supplement Series]
  {10.1086/381377}, 151, 149

\bibitem[\protect\citeauthoryear{Barnes \& Hut}{Barnes \&
  Hut}{1986}]{Barnes&Hut1986}
Barnes J.,  Hut P.,  1986, \mn@doi [Nature] {10.1038/324446a0}, 324, 446

\bibitem[\protect\citeauthoryear{Barth \& Jespersen}{Barth \&
  Jespersen}{1989}]{Barth&Jespersen1989}
Barth T.,  Jespersen D.,  1989, in 27th {{Aerospace Sciences Meeting}}.
  {American Institute of Aeronautics and Astronautics},
  \mn@doi{10.2514/6.1989-366}

\bibitem[\protect\citeauthoryear{Bauer \& Springel}{Bauer \&
  Springel}{2012}]{Bauer&Springel2012}
Bauer A.,  Springel V.,  2012, \mn@doi [Monthly Notices of the Royal
  Astronomical Society] {10.1111/j.1365-2966.2012.21058.x}, 423, 2558

\bibitem[\protect\citeauthoryear{Beck et~al.,}{Beck et~al.}{2016a}]{Beck+2016}
Beck A.~M.,  et~al., 2016a, \mn@doi [Monthly Notices of the Royal Astronomical
  Society] {10.1093/mnras/stv2443}, 455, 2110

\bibitem[\protect\citeauthoryear{Beck, Dolag  \& Donnert}{Beck
  et~al.}{2016b}]{Beck+2016a}
Beck A.~M.,  Dolag K.,   Donnert J. M.~F.,  2016b, \mn@doi [Monthly Notices of
  the Royal Astronomical Society] {10.1093/mnras/stw487}, 458, 2080

\bibitem[\protect\citeauthoryear{Berlok}{Berlok}{2022}]{Berlok2022a}
Berlok T.,  2022, \mn@doi [Monthly Notices of the Royal Astronomical Society]
  {10.1093/mnras/stac1882}, 515, 3492

\bibitem[\protect\citeauthoryear{Bezanson, Edelman, Karpinski  \&
  Shah}{Bezanson et~al.}{2014}]{Bezanson+2014}
Bezanson J.,  Edelman A.,  Karpinski S.,   Shah V.~B.,  2014, Julia: {{A Fresh
  Approach}} to {{Numerical Computing}}

\bibitem[\protect\citeauthoryear{Biffi, Sembolini, De~Petris, Valdarnini, Yepes
   \& Gottl{\"o}ber}{Biffi et~al.}{2014}]{Biffi+2014}
Biffi V.,  Sembolini F.,  De~Petris M.,  Valdarnini R.,  Yepes G.,
  Gottl{\"o}ber S.,  2014, \mn@doi [Monthly Notices of the Royal Astronomical
  Society] {10.1093/mnras/stu018}, 439, 588

\bibitem[\protect\citeauthoryear{Bode, Ostriker  \& Xu}{Bode
  et~al.}{2000}]{Bode+2000}
Bode P.,  Ostriker J.~P.,   Xu G.,  2000, \mn@doi [The Astrophysical Journal
  Supplement Series] {10.1086/313398}, 128, 561

\bibitem[\protect\citeauthoryear{Borgani, Finoguenov, Kay, Ponman, Springel,
  Tozzi  \& Voit}{Borgani et~al.}{2005}]{Borgani+2005}
Borgani S.,  Finoguenov A.,  Kay S.~T.,  Ponman T.~J.,  Springel V.,  Tozzi P.,
    Voit G.~M.,  2005, \mn@doi [Monthly Notices of the Royal Astronomical
  Society] {10.1111/j.1365-2966.2005.09158.x}, 361, 233

\bibitem[\protect\citeauthoryear{B{\"o}ss \& Valenzuela}{B{\"o}ss \&
  Valenzuela}{2022}]{GadgetIO}
B{\"o}ss L.~M.,  Valenzuela L.~M.,  2022, {{LudwigBoess}}/{{GadgetIO}}.Jl:
  V0.6.2, Zenodo, \mn@doi{10.5281/zenodo.7055005}

\bibitem[\protect\citeauthoryear{B{\"o}ss, Steinwandel, Dolag  \&
  Lesch}{B{\"o}ss et~al.}{2023}]{Boss+2023}
B{\"o}ss L.~M.,  Steinwandel U.~P.,  Dolag K.,   Lesch H.,  2023, \mn@doi
  [Monthly Notices of the Royal Astronomical Society] {10.1093/mnras/stac3584},
  519, 548

\bibitem[\protect\citeauthoryear{Bryan, Norman, Stone, Cen  \& Ostriker}{Bryan
  et~al.}{1995}]{Bryan+1995}
Bryan G.~L.,  Norman M.~L.,  Stone J.~M.,  Cen R.,   Ostriker J.~P.,  1995,
  \mn@doi [Computer Physics Communications] {10.1016/0010-4655(94)00191-4}, 89,
  149

\bibitem[\protect\citeauthoryear{Bryan et~al.,}{Bryan
  et~al.}{2014}]{Bryan+2014}
Bryan G.~L.,  et~al., 2014, \mn@doi [The Astrophysical Journal Supplement
  Series] {10.1088/0067-0049/211/2/19}, 211, 19

\bibitem[\protect\citeauthoryear{Cavagnolo, Donahue, Voit  \& Sun}{Cavagnolo
  et~al.}{2009}]{Cavagnolo+2009}
Cavagnolo K.~W.,  Donahue M.,  Voit G.~M.,   Sun M.,  2009, \mn@doi [The
  Astrophysical Journal Supplement Series] {10.1088/0067-0049/182/1/12}, 182,
  12

\bibitem[\protect\citeauthoryear{Cha \& Whitworth}{Cha \&
  Whitworth}{2003}]{Cha&Whitworth2003}
Cha S.~H.,  Whitworth A.~P.,  2003, \mn@doi [Monthly Notices of the Royal
  Astronomical Society] {10.1046/j.1365-8711.2003.06266.x}, 340, 73

\bibitem[\protect\citeauthoryear{Chang \& Etienne}{Chang \&
  Etienne}{2020}]{Chang&Etienne2020}
Chang P.,  Etienne Z.~B.,  2020, \mn@doi [Monthly Notices of the Royal
  Astronomical Society] {10.1093/mnras/staa1532}, 496, 206

\bibitem[\protect\citeauthoryear{Cui, Liu, Yang, Wang, Feng  \& Springel}{Cui
  et~al.}{2008}]{Cui+2008}
Cui W.,  Liu L.,  Yang X.,  Wang Y.,  Feng L.,   Springel V.,  2008, \mn@doi
  [The Astrophysical Journal] {10.1086/592079}, 687, 738

\bibitem[\protect\citeauthoryear{Dehnen \& Aly}{Dehnen \&
  Aly}{2012}]{Dehnen&Aly2012}
Dehnen W.,  Aly H.,  2012, \mn@doi [Monthly Notices of the Royal Astronomical
  Society] {10.1111/j.1365-2966.2012.21439.x}, 425, 1068

\bibitem[\protect\citeauthoryear{Dolag}{Dolag}{2015}]{Dolag2015}
Dolag K.,  2015, in {{IAU General Assembly}}. p. 2250156

\bibitem[\protect\citeauthoryear{Dolag \& Stasyszyn}{Dolag \&
  Stasyszyn}{2009}]{Dolag&Stasyszyn2009}
Dolag K.,  Stasyszyn F.,  2009, \mn@doi [Monthly Notices of the Royal
  Astronomical Society] {10.1111/j.1365-2966.2009.15181.x}, 398, 1678

\bibitem[\protect\citeauthoryear{Dolag, Jubelgas, Springel, Borgani  \&
  Rasia}{Dolag et~al.}{2004}]{Dolag+2004}
Dolag K.,  Jubelgas M.,  Springel V.,  Borgani S.,   Rasia E.,  2004, \mn@doi
  [The Astrophysical Journal] {10.1086/420966}, 606, L97

\bibitem[\protect\citeauthoryear{Dolag, Hansen, Roncarelli  \&
  Moscardini}{Dolag et~al.}{2005a}]{Dolag+2005}
Dolag K.,  Hansen F.~K.,  Roncarelli M.,   Moscardini L.,  2005a, \mn@doi
  [Monthly Notices of the Royal Astronomical Society]
  {10.1111/j.1365-2966.2005.09452.x}, 363, 29

\bibitem[\protect\citeauthoryear{Dolag, Vazza, Brunetti  \& Tormen}{Dolag
  et~al.}{2005b}]{Dolag+2005b}
Dolag K.,  Vazza F.,  Brunetti G.,   Tormen G.,  2005b, \mn@doi [Monthly
  Notices of the Royal Astronomical Society]
  {10.1111/j.1365-2966.2005.09630.x}, 364, 753

\bibitem[\protect\citeauthoryear{Dolag, Borgani, Murante  \& Springel}{Dolag
  et~al.}{2009}]{Dolag+2009}
Dolag K.,  Borgani S.,  Murante G.,   Springel V.,  2009, \mn@doi [Monthly
  Notices of the Royal Astronomical Society]
  {10.1111/j.1365-2966.2009.15034.x}, 399, 497

\bibitem[\protect\citeauthoryear{Duffell \& MacFadyen}{Duffell \&
  MacFadyen}{2011}]{Duffell&MacFadyen2011}
Duffell P.~C.,  MacFadyen A.~I.,  2011, \mn@doi [The Astrophysical Journal
  Supplement Series] {10.1088/0067-0049/197/2/15}, 197, 15

\bibitem[\protect\citeauthoryear{Eastwood \& Hockney}{Eastwood \&
  Hockney}{1974}]{Eastwood&Hockney1974}
Eastwood J.~W.,  Hockney R.~W.,  1974, \mn@doi [Journal of Computational
  Physics] {10.1016/0021-9991(74)90044-8}, 16, 342

\bibitem[\protect\citeauthoryear{Federrath, Klessen  \& Schmidt}{Federrath
  et~al.}{2008}]{Federrath+2008}
Federrath C.,  Klessen R.~S.,   Schmidt W.,  2008, \mn@doi [The Astrophysical
  Journal] {10.1086/595280}, 688, L79

\bibitem[\protect\citeauthoryear{Federrath, Klessen  \& Schmidt}{Federrath
  et~al.}{2009}]{Federrath+2009}
Federrath C.,  Klessen R.~S.,   Schmidt W.,  2009, \mn@doi [The Astrophysical
  Journal] {10.1088/0004-637X/692/1/364}, 692, 364

\bibitem[\protect\citeauthoryear{Federrath, {Roman-Duval}, Klessen, Schmidt  \&
  Mac~Low}{Federrath et~al.}{2010}]{Federrath+2010}
Federrath C.,  {Roman-Duval} J.,  Klessen R.~S.,  Schmidt W.,   Mac~Low M.-M.,
  2010, \mn@doi [Astronomy and Astrophysics, Volume 512, id.A81,
  {$<$}NUMPAGES{$>$}28{$<$}/NUMPAGES{$>$} pp.] {10.1051/0004-6361/200912437},
  512, A81

\bibitem[\protect\citeauthoryear{Federrath, Klessen, Iapichino  \&
  Beattie}{Federrath et~al.}{2021}]{Federrath+2021}
Federrath C.,  Klessen R.~S.,  Iapichino L.,   Beattie J.~R.,  2021, \mn@doi
  [Nature Astronomy] {10.1038/s41550-020-01282-z}, 5, 365

\bibitem[\protect\citeauthoryear{Fischer, Br{\"u}ggen, {Schmidt-Hoberg}, Dolag,
  Kahlhoefer, Ragagnin  \& Robertson}{Fischer et~al.}{2022}]{Fischer+2022}
Fischer M.~S.,  Br{\"u}ggen M.,  {Schmidt-Hoberg} K.,  Dolag K.,  Kahlhoefer
  F.,  Ragagnin A.,   Robertson A.,  2022, arXiv:2205.02243 [astro-ph,
  physics:hep-ph]

\bibitem[\protect\citeauthoryear{Frigo \& Johnson}{Frigo \&
  Johnson}{2005}]{FFTW3}
Frigo M.,  Johnson S.~G.,  2005, Proceedings of the IEEE, 93, 216

\bibitem[\protect\citeauthoryear{Fryxell et~al.,}{Fryxell
  et~al.}{2000}]{Fryxell+2000}
Fryxell B.,  et~al., 2000, \mn@doi [The Astrophysical Journal Supplement
  Series] {10.1086/317361}, 131, 273

\bibitem[\protect\citeauthoryear{Gaburov \& Nitadori}{Gaburov \&
  Nitadori}{2011}]{Gaburov&Nitadori2011}
Gaburov E.,  Nitadori K.,  2011, \mn@doi [Monthly Notices of the Royal
  Astronomical Society] {10.1111/j.1365-2966.2011.18313.x}, 414, 129

\bibitem[\protect\citeauthoryear{Godunov}{Godunov}{1959}]{Godunov1959}
Godunov S.,  1959, Matematicheski\textbackslash u \i{} Sbornik. Novaya Seriya,
  47(89), 271

\bibitem[\protect\citeauthoryear{Gronke \& Oh}{Gronke \&
  Oh}{2018}]{Gronke&Oh2018}
Gronke M.,  Oh S.~P.,  2018, \mn@doi [Monthly Notices of the Royal Astronomical
  Society] {10.1093/mnrasl/sly131}, 480, L111

\bibitem[\protect\citeauthoryear{Gronke \& Oh}{Gronke \&
  Oh}{2020}]{Gronke&Oh2020}
Gronke M.,  Oh S.~P.,  2020, \mn@doi [Monthly Notices of the Royal Astronomical
  Society] {10.1093/mnrasl/slaa033}, 494, L27

\bibitem[\protect\citeauthoryear{Gronke \& Oh}{Gronke \&
  Oh}{2022}]{Gronke&Oh2022}
Gronke M.,  Oh S.~P.,  2022, Cooling Driven Coagulation

\bibitem[\protect\citeauthoryear{Hernquist \& Katz}{Hernquist \&
  Katz}{1989}]{Hernquist&Katz1989}
Hernquist L.,  Katz N.,  1989, \mn@doi [The Astrophysical Journal Supplement
  Series] {10.1086/191344}, 70, 419

\bibitem[\protect\citeauthoryear{Hess \& Springel}{Hess \&
  Springel}{2010}]{Hess&Springel2010}
Hess S.,  Springel V.,  2010, \mn@doi [Monthly Notices of the Royal
  Astronomical Society] {10.1111/j.1365-2966.2010.16892.x}, 406, 2289

\bibitem[\protect\citeauthoryear{Hirschmann, Dolag, Saro, Bachmann, Borgani  \&
  Burkert}{Hirschmann et~al.}{2014}]{Hirschmann+2014}
Hirschmann M.,  Dolag K.,  Saro A.,  Bachmann L.,  Borgani S.,   Burkert A.,
  2014, \mn@doi [Monthly Notices of the Royal Astronomical Society]
  {10.1093/mnras/stu1023}, 442, 2304

\bibitem[\protect\citeauthoryear{Hopkins}{Hopkins}{2013}]{Hopkins2013}
Hopkins P.~F.,  2013, \mn@doi [Monthly Notices of the Royal Astronomical
  Society] {10.1093/mnras/sts210}, 428, 2840

\bibitem[\protect\citeauthoryear{Hopkins}{Hopkins}{2015}]{Hopkins2015}
Hopkins P.~F.,  2015, \mn@doi [Monthly Notices of the Royal Astronomical
  Society] {10.1093/mnras/stv195}, 450, 53

\bibitem[\protect\citeauthoryear{Hu, Naab, Walch, Moster  \& Oser}{Hu
  et~al.}{2014}]{Hu+2014}
Hu C.-Y.,  Naab T.,  Walch S.,  Moster B.~P.,   Oser L.,  2014, \mn@doi
  [Monthly Notices of the Royal Astronomical Society] {10.1093/mnras/stu1187},
  443, 1173

\bibitem[\protect\citeauthoryear{Hubber, Rosotti  \& Booth}{Hubber
  et~al.}{2018}]{Hubber+2018}
Hubber D.~A.,  Rosotti G.~P.,   Booth R.~A.,  2018, \mn@doi [Monthly Notices of
  the Royal Astronomical Society] {10.1093/mnras/stx2405}, 473, 1603

\bibitem[\protect\citeauthoryear{Iapichino \& Niemeyer}{Iapichino \&
  Niemeyer}{2008}]{Iapichino&Niemeyer2008a}
Iapichino L.,  Niemeyer J.~C.,  2008, \mn@doi [Monthly Notices of the Royal
  Astronomical Society] {10.1111/j.1365-2966.2008.13518.x}, 388, 1089

\bibitem[\protect\citeauthoryear{Iapichino, Federrath  \& Klessen}{Iapichino
  et~al.}{2017}]{Iapichino+2017}
Iapichino L.,  Federrath C.,   Klessen R.~S.,  2017, \mn@doi [Monthly Notices
  of the Royal Astronomical Society] {10.1093/mnras/stx882}, 469, 3641

\bibitem[\protect\citeauthoryear{Idelsohn, O{\~n}ate, Calvo  \&
  Del~Pin}{Idelsohn et~al.}{2003}]{Idelsohn+2003}
Idelsohn S.~R.,  O{\~n}ate E.,  Calvo N.,   Del~Pin F.,  2003, \mn@doi
  [International Journal for Numerical Methods in Engineering]
  {10.1002/nme.798}, 58, 893

\bibitem[\protect\citeauthoryear{Inutsuka}{Inutsuka}{2002}]{Inutsuka2002}
Inutsuka S.-I.,  2002, \mn@doi [Journal of Computational Physics]
  {10.1006/jcph.2002.7053}, 179, 238

\bibitem[\protect\citeauthoryear{Junk, Walch, Heitsch, Burkert, Wetzstein,
  Schartmann  \& Price}{Junk et~al.}{2010}]{Junk+2010}
Junk V.,  Walch S.,  Heitsch F.,  Burkert A.,  Wetzstein M.,  Schartmann M.,
  Price D.,  2010, \mn@doi [Monthly Notices of the Royal Astronomical Society]
  {10.1111/j.1365-2966.2010.17039.x}, 407, 1933

\bibitem[\protect\citeauthoryear{Kim \& Ostriker}{Kim \&
  Ostriker}{2015}]{Kim&Ostriker2015}
Kim C.-G.,  Ostriker E.~C.,  2015, \mn@doi [The Astrophysical Journal]
  {10.1088/0004-637X/802/2/99}, 802, 99

\bibitem[\protect\citeauthoryear{Kitsionas et~al.,}{Kitsionas
  et~al.}{2009}]{Kitsionas+2009}
Kitsionas S.,  et~al., 2009, \mn@doi [Astronomy and Astrophysics, Volume 508,
  Issue 1, 2009, pp.541-560] {10.1051/0004-6361/200811170}, 508, 541

\bibitem[\protect\citeauthoryear{Kolmogorov}{Kolmogorov}{1941}]{Kolmogorov1941}
Kolmogorov A.~N.,  1941, Akademiia Nauk SSSR Doklady, 32, 16

\bibitem[\protect\citeauthoryear{Komatsu et~al.,}{Komatsu
  et~al.}{2011}]{Komatsu+2011}
Komatsu E.,  et~al., 2011, \mn@doi [The Astrophysical Journal Supplement
  Series] {10.1088/0067-0049/192/2/18}, 192, 18

\bibitem[\protect\citeauthoryear{Kritsuk, Yee, Sj{\"o}green  \& Kotov}{Kritsuk
  et~al.}{2020}]{Kritsuk+2020}
Kritsuk A.~G.,  Yee H.~C.,  Sj{\"o}green B.,   Kotov D.,  2020, \mn@doi
  [Journal of Physics: Conference Series] {10.1088/1742-6596/1623/1/012010},
  1623, 012010

\bibitem[\protect\citeauthoryear{Lanson \& Vila}{Lanson \&
  Vila}{2008a}]{Lanson&Vila2008}
Lanson N.,  Vila J.-P.,  2008a, \mn@doi [SIAM Journal on Numerical Analysis]
  {10.1137/S0036142903427718}, 46, 1912

\bibitem[\protect\citeauthoryear{Lanson \& Vila}{Lanson \&
  Vila}{2008b}]{Lanson&Vila2008a}
Lanson N.,  Vila J.-P.,  2008b, \mn@doi [SIAM Journal on Numerical Analysis]
  {10.1137/S003614290444739X}, 46, 1935

\bibitem[\protect\citeauthoryear{Lioutas, Bauswein, Soultanis, Pakmor, Springel
   \& R{\"o}pke}{Lioutas et~al.}{2022}]{Lioutas+2022}
Lioutas G.,  Bauswein A.,  Soultanis T.,  Pakmor R.,  Springel V.,   R{\"o}pke
  F.~K.,  2022, General Relativistic Moving-Mesh Hydrodynamics Simulations with
  {{AREPO}} and Applications to Neutron Star Mergers (\mn@eprint {arxiv}
  {arXiv:2208.04267}), \mn@doi{10.48550/arXiv.2208.04267}

\bibitem[\protect\citeauthoryear{Liptai \& Price}{Liptai \&
  Price}{2019}]{Liptai&Price2019}
Liptai D.,  Price D.~J.,  2019, \mn@doi [Monthly Notices of the Royal
  Astronomical Society] {10.1093/mnras/stz111}, 485, 819

\bibitem[\protect\citeauthoryear{Lodato \& Price}{Lodato \&
  Price}{2010}]{Lodato&Price2010}
Lodato G.,  Price D.~J.,  2010, \mn@doi [Monthly Notices of the Royal
  Astronomical Society] {10.1111/j.1365-2966.2010.16526.x}, 405, 1212

\bibitem[\protect\citeauthoryear{Lupi}{Lupi}{2022}]{Lupi2022a}
Lupi A.,  2022, \mn@doi [Monthly Notices of the Royal Astronomical Society]
  {10.1093/mnras/stac3574}, 519, 1115

\bibitem[\protect\citeauthoryear{Maier, Iapichino, Schmidt  \& Niemeyer}{Maier
  et~al.}{2009}]{Maier+2009a}
Maier A.,  Iapichino L.,  Schmidt W.,   Niemeyer J.~C.,  2009, \mn@doi [The
  Astrophysical Journal] {10.1088/0004-637X/707/1/40}, 707, 40

\bibitem[\protect\citeauthoryear{{Marin-Gilabert}, Valentini, Steinwandel  \&
  Dolag}{{Marin-Gilabert} et~al.}{2022}]{Marin-Gilabert+2022}
{Marin-Gilabert} T.,  Valentini M.,  Steinwandel U.~P.,   Dolag K.,  2022, The
  Role of Physical and Numerical Viscosity in Hydrodynamical Instabilities

\bibitem[\protect\citeauthoryear{Martel \& Shapiro}{Martel \&
  Shapiro}{1998}]{Martel&Shapiro1998}
Martel H.,  Shapiro P.~R.,  1998, \mn@doi [Monthly Notices of the Royal
  Astronomical Society] {10.1046/j.1365-8711.1998.01497.x}, 297, 467

\bibitem[\protect\citeauthoryear{May \& Berger}{May \&
  Berger}{2013}]{May&Berger2013}
May S.,  Berger M.,  2013, \mn@doi [SIAM Journal on Scientific Computing]
  {10.1137/120875624}, 35, A2163

\bibitem[\protect\citeauthoryear{McNally, Lyra  \& Passy}{McNally
  et~al.}{2012}]{McNally+2012a}
McNally C.~P.,  Lyra W.,   Passy J.-C.,  2012, \mn@doi [The Astrophysical
  Journal Supplement Series] {10.1088/0067-0049/201/2/18}, 201, 18

\bibitem[\protect\citeauthoryear{Miniati}{Miniati}{2014}]{Miniati2014}
Miniati F.,  2014, \mn@doi [The Astrophysical Journal]
  {10.1088/0004-637X/782/1/21}, 782, 21

\bibitem[\protect\citeauthoryear{Miniati}{Miniati}{2015}]{Miniati2015}
Miniati F.,  2015, \mn@doi [The Astrophysical Journal]
  {10.1088/0004-637X/800/1/60}, 800, 60

\bibitem[\protect\citeauthoryear{Mohapatra, Federrath  \& Sharma}{Mohapatra
  et~al.}{2021}]{Mohapatra+2021}
Mohapatra R.,  Federrath C.,   Sharma P.,  2021, \mn@doi [Monthly Notices of
  the Royal Astronomical Society] {10.1093/mnras/staa3564}, 500, 5072

\bibitem[\protect\citeauthoryear{Mohapatra, Jetti, Sharma  \&
  Federrath}{Mohapatra et~al.}{2022}]{Mohapatra+2022}
Mohapatra R.,  Jetti M.,  Sharma P.,   Federrath C.,  2022, \mn@doi [Monthly
  Notices of the Royal Astronomical Society] {10.1093/mnras/stab3429}, 510,
  2327

\bibitem[\protect\citeauthoryear{Monaghan \& Lattanzio}{Monaghan \&
  Lattanzio}{1985}]{Monaghan&Lattanzio1985}
Monaghan J.~J.,  Lattanzio J.~C.,  1985, Astronomy and Astrophysics, 149, 135

\bibitem[\protect\citeauthoryear{Morris}{Morris}{1996}]{Morris1996}
Morris J.~P.,  1996, \mn@doi [Publications of the Astronomical Society of
  Australia] {10.1017/S1323358000020610}, 13, 97

\bibitem[\protect\citeauthoryear{Murante, Monaco, Giovalli, Borgani  \&
  Diaferio}{Murante et~al.}{2010}]{Murante+2010}
Murante G.,  Monaco P.,  Giovalli M.,  Borgani S.,   Diaferio A.,  2010,
  \mn@doi [Monthly Notices of the Royal Astronomical Society]
  {10.1111/j.1365-2966.2010.16567.x}, 405, 1491

\bibitem[\protect\citeauthoryear{Murante, Monaco, Borgani, Tornatore, Dolag  \&
  Goz}{Murante et~al.}{2015}]{Murante+2015}
Murante G.,  Monaco P.,  Borgani S.,  Tornatore L.,  Dolag K.,   Goz D.,  2015,
  \mn@doi [Monthly Notices of the Royal Astronomical Society]
  {10.1093/mnras/stu2400}, 447, 178

\bibitem[\protect\citeauthoryear{Navarro, Frenk  \& White}{Navarro
  et~al.}{1997}]{Navarro+1997}
Navarro J.~F.,  Frenk C.~S.,   White S. D.~M.,  1997, \mn@doi [The
  Astrophysical Journal] {10.1086/304888}, 490, 493

\bibitem[\protect\citeauthoryear{Padoan, Nordlund, Kritsuk, Norman  \&
  Li}{Padoan et~al.}{2007}]{Padoan+2007}
Padoan P.,  Nordlund {\AA}.,  Kritsuk A.~G.,  Norman M.~L.,   Li P.~S.,  2007,
  \mn@doi [The Astrophysical Journal] {10.1086/516623}, 661, 972

\bibitem[\protect\citeauthoryear{Pakmor}{Pakmor}{2010}]{Pakmor2010}
Pakmor R.~M.,  2010, PhD thesis, Technical University of Munich

\bibitem[\protect\citeauthoryear{Pakmor, Edelmann, R{\"o}pke  \&
  Hillebrandt}{Pakmor et~al.}{2012}]{Pakmor+2012}
Pakmor R.,  Edelmann P.,  R{\"o}pke F.~K.,   Hillebrandt W.,  2012, \mn@doi
  [Monthly Notices of the Royal Astronomical Society]
  {10.1111/j.1365-2966.2012.21383.x}, 424, 2222

\bibitem[\protect\citeauthoryear{Pearce, Thomas, Couchman  \& Edge}{Pearce
  et~al.}{2000}]{Pearce+2000}
Pearce F.~R.,  Thomas P.~A.,  Couchman H. M.~P.,   Edge A.~C.,  2000, \mn@doi
  [Monthly Notices of the Royal Astronomical Society]
  {10.1046/j.1365-8711.2000.03773.x}, 317, 1029

\bibitem[\protect\citeauthoryear{Pfrommer, Springel, En{\ss}lin  \&
  Jubelgas}{Pfrommer et~al.}{2006}]{Pfrommer+2006}
Pfrommer C.,  Springel V.,  En{\ss}lin T.~A.,   Jubelgas M.,  2006, \mn@doi
  [Monthly Notices of the Royal Astronomical Society]
  {10.1111/j.1365-2966.2005.09953.x}, 367, 113

\bibitem[\protect\citeauthoryear{Prada, Klypin, Cuesta, {Betancort-Rijo}  \&
  Primack}{Prada et~al.}{2012}]{Prada+2012}
Prada F.,  Klypin A.~A.,  Cuesta A.~J.,  {Betancort-Rijo} J.~E.,   Primack J.,
  2012, \mn@doi [Monthly Notices of the Royal Astronomical Society]
  {10.1111/j.1365-2966.2012.21007.x}, 423, 3018

\bibitem[\protect\citeauthoryear{Price}{Price}{2008}]{Price2008}
Price D.~J.,  2008, \mn@doi [Journal of Computational Physics]
  {10.1016/j.jcp.2008.08.011}, 227, 10040

\bibitem[\protect\citeauthoryear{Price}{Price}{2012}]{Price2012}
Price D.~J.,  2012, \mn@doi [Monthly Notices of the Royal Astronomical Society]
  {10.1111/j.1745-3933.2011.01187.x}, 420, L33

\bibitem[\protect\citeauthoryear{Price \& Federrath}{Price \&
  Federrath}{2010}]{Price&Federrath2010}
Price D.~J.,  Federrath C.,  2010, \mn@doi [Monthly Notices of the Royal
  Astronomical Society] {10.1111/j.1365-2966.2010.16810.x}, 406, 1659

\bibitem[\protect\citeauthoryear{Price et~al.,}{Price
  et~al.}{2018}]{Price+2018}
Price D.~J.,  et~al., 2018, \mn@doi [Publications of the Astronomical Society
  of Australia] {10.1017/pasa.2018.25}, 35, e031

\bibitem[\protect\citeauthoryear{Ragagnin, Dolag, Wagner, Gheller, Roffler,
  Goz, Hubber  \& Arth}{Ragagnin et~al.}{2020}]{Ragagnin+2020}
Ragagnin A.,  Dolag K.,  Wagner M.,  Gheller C.,  Roffler C.,  Goz D.,  Hubber
  D.,   Arth A.,  2020, Gadget3 on {{GPUs}} with {{OpenACC}}

\bibitem[\protect\citeauthoryear{Rasia et~al.,}{Rasia
  et~al.}{2015}]{Rasia+2015}
Rasia E.,  et~al., 2015, \mn@doi [The Astrophysical Journal]
  {10.1088/2041-8205/813/1/L17}, 813, L17

\bibitem[\protect\citeauthoryear{Roe}{Roe}{1981}]{Roe1981}
Roe P.~L.,  1981, \mn@doi [Journal of Computational Physics]
  {10.1016/0021-9991(81)90128-5}, 43, 357

\bibitem[\protect\citeauthoryear{Roettiger \& Burns}{Roettiger \&
  Burns}{1999}]{Roettiger&Burns1999}
Roettiger K.,  Burns J.~O.,  1999, in American {{Astronomical Society Meeting
  Abstracts}}. p. 13.04

\bibitem[\protect\citeauthoryear{Rosswog}{Rosswog}{2020}]{Rosswog2020}
Rosswog S.,  2020, \mn@doi [Monthly Notices of the Royal Astronomical Society]
  {10.1093/mnras/staa2591}, 498, 4230

\bibitem[\protect\citeauthoryear{Rosswog \& Diener}{Rosswog \&
  Diener}{2021}]{Rosswog&Diener2021}
Rosswog S.,  Diener P.,  2021, \mn@doi [Classical and Quantum Gravity]
  {10.1088/1361-6382/abee65}, 38, 115002

\bibitem[\protect\citeauthoryear{Ryu, Ostriker, Kang  \& Cen}{Ryu
  et~al.}{1993}]{Ryu+1993}
Ryu D.,  Ostriker J.~P.,  Kang H.,   Cen R.,  1993, \mn@doi [The Astrophysical
  Journal] {10.1086/173051}, 414, 1

\bibitem[\protect\citeauthoryear{Ryu, Miniati, Jones  \& Frank}{Ryu
  et~al.}{1998}]{Ryu+1998}
Ryu D.,  Miniati F.,  Jones T.~W.,   Frank A.,  1998, \mn@doi [The
  Astrophysical Journal] {10.1086/306481}, 509, 244

\bibitem[\protect\citeauthoryear{Saitoh \& Makino}{Saitoh \&
  Makino}{2009}]{Saitoh&Makino2009}
Saitoh T.~R.,  Makino J.,  2009, \mn@doi [The Astrophysical Journal]
  {10.1088/0004-637X/697/2/L99}, 697, L99

\bibitem[\protect\citeauthoryear{Sayers, Sereno, Ettori, Rasia, Cui, Golwala,
  Umetsu  \& Yepes}{Sayers et~al.}{2021}]{Sayers+2021}
Sayers J.,  Sereno M.,  Ettori S.,  Rasia E.,  Cui W.,  Golwala S.,  Umetsu K.,
    Yepes G.,  2021, \mn@doi [Monthly Notices of the Royal Astronomical
  Society] {10.1093/mnras/stab1542}, 505, 4338

\bibitem[\protect\citeauthoryear{Schekochihin, Cowley, Maron  \&
  Malyshkin}{Schekochihin et~al.}{2001}]{Schekochihin+2001}
Schekochihin A.,  Cowley S.,  Maron J.,   Malyshkin L.,  2001, \mn@doi
  [Physical Review E] {10.1103/PhysRevE.65.016305}, 65, 016305

\bibitem[\protect\citeauthoryear{Schekochihin, Cowley, Taylor, Maron  \&
  McWilliams}{Schekochihin et~al.}{2004}]{Schekochihin+2004}
Schekochihin A.~A.,  Cowley S.~C.,  Taylor S.~F.,  Maron J.~L.,   McWilliams
  J.~C.,  2004, \mn@doi [The Astrophysical Journal] {10.1086/422547}, 612, 276

\bibitem[\protect\citeauthoryear{Schuecker, Finoguenov, Miniati, B{\"o}hringer
  \& Briel}{Schuecker et~al.}{2004}]{Schuecker+2004}
Schuecker P.,  Finoguenov A.,  Miniati F.,  B{\"o}hringer H.,   Briel U.~G.,
  2004, \mn@doi [Astronomy and Astrophysics, v.426, p.387-397 (2004)]
  {10.1051/0004-6361:20041039}, 426, 387

\bibitem[\protect\citeauthoryear{Sedov}{Sedov}{1946}]{Sedov1946}
Sedov L.~I.,  1946, Journal of Applied Mathematics and Mechanics, 10, 241

\bibitem[\protect\citeauthoryear{Sedov}{Sedov}{1959}]{Sedov1959}
Sedov L.~I.,  1959, Similarity and {{Dimensional Methods}} in {{Mechanics}}.
{New York: Academic Press}

\bibitem[\protect\citeauthoryear{Sembolini, Yepes, De~Petris, Gottl{\"o}ber,
  Lamagna  \& Comis}{Sembolini et~al.}{2013}]{Sembolini+2013}
Sembolini F.,  Yepes G.,  De~Petris M.,  Gottl{\"o}ber S.,  Lamagna L.,   Comis
  B.,  2013, \mn@doi [Monthly Notices of the Royal Astronomical Society]
  {10.1093/mnras/sts339}, 429, 323

\bibitem[\protect\citeauthoryear{Sembolini, De~Petris, Yepes, Foschi, Lamagna
  \& Gottl{\"o}ber}{Sembolini et~al.}{2014}]{Sembolini+2014}
Sembolini F.,  De~Petris M.,  Yepes G.,  Foschi E.,  Lamagna L.,
  Gottl{\"o}ber S.,  2014, \mn@doi [Monthly Notices of the Royal Astronomical
  Society] {10.1093/mnras/stu554}, 440, 3520

\bibitem[\protect\citeauthoryear{Sembolini et~al.,}{Sembolini
  et~al.}{2016}]{Sembolini+2016}
Sembolini F.,  et~al., 2016, \mn@doi [Monthly Notices of the Royal Astronomical
  Society] {10.1093/mnras/stw250}, 457, 4063

\bibitem[\protect\citeauthoryear{Sod}{Sod}{1978}]{Sod1978}
Sod G.~A.,  1978, \mn@doi [Journal of Computational Physics]
  {10.1016/0021-9991(78)90023-2}, 27, 1

\bibitem[\protect\citeauthoryear{Springel}{Springel}{2005}]{Springel2005}
Springel V.,  2005, \mn@doi [Monthly Notices of the Royal Astronomical Society]
  {10.1111/j.1365-2966.2005.09655.x}, 364, 1105

\bibitem[\protect\citeauthoryear{Springel}{Springel}{2010}]{Springel2010}
Springel V.,  2010, \mn@doi [Monthly Notices of the Royal Astronomical Society]
  {10.1111/j.1365-2966.2009.15715.x}, 401, 791

\bibitem[\protect\citeauthoryear{Springel \& Hernquist}{Springel \&
  Hernquist}{2002}]{Springel&Hernquist2002}
Springel V.,  Hernquist L.,  2002, \mn@doi [Monthly Notices of the Royal
  Astronomical Society] {10.1046/j.1365-8711.2002.05445.x}, 333, 649

\bibitem[\protect\citeauthoryear{Springel \& Hernquist}{Springel \&
  Hernquist}{2003}]{Springel&Hernquist2003a}
Springel V.,  Hernquist L.,  2003, \mn@doi [Monthly Notices of the Royal
  Astronomical Society] {10.1046/j.1365-8711.2003.06206.x}, 339, 289

\bibitem[\protect\citeauthoryear{Springel, Yoshida  \& White}{Springel
  et~al.}{2001}]{Springel+2001}
Springel V.,  Yoshida N.,   White S. D.~M.,  2001, \mn@doi [New Astronomy]
  {10.1016/S1384-1076(01)00042-2}, 6, 79

\bibitem[\protect\citeauthoryear{Springel, Pakmor, Zier  \& Reinecke}{Springel
  et~al.}{2021}]{Springel+2021}
Springel V.,  Pakmor R.,  Zier O.,   Reinecke M.,  2021, \mn@doi [Monthly
  Notices of the Royal Astronomical Society] {10.1093/mnras/stab1855}, 506,
  2871

\bibitem[\protect\citeauthoryear{Stasyszyn, Dolag  \& Beck}{Stasyszyn
  et~al.}{2013}]{Stasyszyn+2013}
Stasyszyn F.~A.,  Dolag K.,   Beck A.~M.,  2013, \mn@doi [Monthly Notices of
  the Royal Astronomical Society] {10.1093/mnras/sts018}, 428, 13

\bibitem[\protect\citeauthoryear{Steinborn, Dolag, Hirschmann, Prieto  \&
  Remus}{Steinborn et~al.}{2015}]{Steinborn+2015}
Steinborn L.~K.,  Dolag K.,  Hirschmann M.,  Prieto M.~A.,   Remus R.-S.,
  2015, \mn@doi [Monthly Notices of the Royal Astronomical Society]
  {10.1093/mnras/stv072}, 448, 1504

\bibitem[\protect\citeauthoryear{Steinwandel, Moster, Naab, Hu  \&
  Walch}{Steinwandel et~al.}{2020}]{Steinwandel+2020}
Steinwandel U.~P.,  Moster B.~P.,  Naab T.,  Hu C.-Y.,   Walch S.,  2020,
  \mn@doi [Monthly Notices of the Royal Astronomical Society]
  {10.1093/mnras/staa821}, 495, 1035

\bibitem[\protect\citeauthoryear{Steinwandel, Boess, Dolag  \&
  Lesch}{Steinwandel et~al.}{2021}]{Steinwandel+2021}
Steinwandel U.~P.,  Boess L.~M.,  Dolag K.,   Lesch H.,  2021, arXiv:2108.07822
  [astro-ph]

\bibitem[\protect\citeauthoryear{Stone \& Norman}{Stone \&
  Norman}{1992}]{Stone&Norman1992}
Stone J.~M.,  Norman M.~L.,  1992, \mn@doi [The Astrophysical Journal
  Supplement Series] {10.1086/191680}, 80, 753

\bibitem[\protect\citeauthoryear{Stone, Gardiner, Teuben, Hawley  \&
  Simon}{Stone et~al.}{2008}]{Stone+2008}
Stone J.~M.,  Gardiner T.~A.,  Teuben P.,  Hawley J.~F.,   Simon J.~B.,  2008,
  \mn@doi [The Astrophysical Journal Supplement Series] {10.1086/588755}, 178,
  137

\bibitem[\protect\citeauthoryear{Stone, Tomida, White  \& Felker}{Stone
  et~al.}{2020}]{Stone+2020}
Stone J.~M.,  Tomida K.,  White C.~J.,   Felker K.~G.,  2020, \mn@doi [The
  Astrophysical Journal Supplement Series] {10.3847/1538-4365/ab929b}, 249, 4

\bibitem[\protect\citeauthoryear{Subramanian, Shukurov  \& Haugen}{Subramanian
  et~al.}{2006}]{Subramanian+2006}
Subramanian K.,  Shukurov A.,   Haugen N. E.~L.,  2006, \mn@doi [Monthly
  Notices of the Royal Astronomical Society]
  {10.1111/j.1365-2966.2006.09918.x}, 366, 1437

\bibitem[\protect\citeauthoryear{Taylor}{Taylor}{1950}]{Taylor1950}
Taylor G.,  1950, \mn@doi [Proceedings of the Royal Society of London Series A]
  {10.1098/rspa.1950.0049}, 201, 159

\bibitem[\protect\citeauthoryear{Teyssier}{Teyssier}{2002}]{Teyssier2002}
Teyssier R.,  2002, \mn@doi [Astronomy and Astrophysics]
  {10.1051/0004-6361:20011817}, 385, 337

\bibitem[\protect\citeauthoryear{Tornatore, Borgani, Springel, Matteucci, Menci
   \& Murante}{Tornatore et~al.}{2003}]{Tornatore+2003}
Tornatore L.,  Borgani S.,  Springel V.,  Matteucci F.,  Menci N.,   Murante
  G.,  2003, \mn@doi [Monthly Notices of the Royal Astronomical Society]
  {10.1046/j.1365-8711.2003.06631.x}, 342, 1025

\bibitem[\protect\citeauthoryear{Tornatore, Borgani, Matteucci, Recchi  \&
  Tozzi}{Tornatore et~al.}{2004}]{Tornatore+2004}
Tornatore L.,  Borgani S.,  Matteucci F.,  Recchi S.,   Tozzi P.,  2004,
  \mn@doi [Monthly Notices of the Royal Astronomical Society]
  {10.1111/j.1365-2966.2004.07689.x}, 349, L19

\bibitem[\protect\citeauthoryear{Tornatore, Borgani, Dolag  \&
  Matteucci}{Tornatore et~al.}{2007}]{Tornatore+2007}
Tornatore L.,  Borgani S.,  Dolag K.,   Matteucci F.,  2007, \mn@doi [Monthly
  Notices of the Royal Astronomical Society]
  {10.1111/j.1365-2966.2007.12070.x}, 382, 1050

\bibitem[\protect\citeauthoryear{Toro}{Toro}{2009}]{Toro2009}
Toro E.~F.,  2009, Riemann Solvers and Numerical Methods for Fluid Dynamics: A
  Practical Introduction, 3rd ed edn.
{Springer}, {Dordrecht ; New York}

\bibitem[\protect\citeauthoryear{Tricco \& Price}{Tricco \&
  Price}{2013}]{Tricco&Price2013}
Tricco T.,  Price D.,  2013, A {{Switch}} for {{Artificial Resistivity}} and
  {{Other Dissipation Terms}}

\bibitem[\protect\citeauthoryear{Valentini, Murante, Borgani, Monaco, Bressan
  \& Beck}{Valentini et~al.}{2017}]{Valentini+2017}
Valentini M.,  Murante G.,  Borgani S.,  Monaco P.,  Bressan A.,   Beck A.~M.,
  2017, \mn@doi [Monthly Notices of the Royal Astronomical Society]
  {10.1093/mnras/stx1352}, 470, 3167

\bibitem[\protect\citeauthoryear{Valentini et~al.,}{Valentini
  et~al.}{2020}]{Valentini+2020}
Valentini M.,  et~al., 2020, \mn@doi [Monthly Notices of the Royal Astronomical
  Society] {10.1093/mnras/stz3131}, 491, 2779

\bibitem[\protect\citeauthoryear{Vandenbroucke \& De~Rijcke}{Vandenbroucke \&
  De~Rijcke}{2016}]{Vandenbroucke&DeRijcke2016}
Vandenbroucke B.,  De~Rijcke S.,  2016, \mn@doi [Astronomy and Computing]
  {10.1016/j.ascom.2016.05.001}, 16, 109

\bibitem[\protect\citeauthoryear{Vazza, Brunetti, Kritsuk, Wagner, Gheller  \&
  Norman}{Vazza et~al.}{2009}]{Vazza+2009}
Vazza F.,  Brunetti G.,  Kritsuk A.,  Wagner R.,  Gheller C.,   Norman M.,
  2009, \mn@doi [Astronomy \& Astrophysics] {10.1051/0004-6361/200912535}, 504,
  33

\bibitem[\protect\citeauthoryear{Vazza, Angelinelli, Jones, Eckert,
  Br{\"u}ggen, Brunetti  \& Gheller}{Vazza et~al.}{2018}]{Vazza+2018}
Vazza F.,  Angelinelli M.,  Jones T.~W.,  Eckert D.,  Br{\"u}ggen M.,  Brunetti
  G.,   Gheller C.,  2018, \mn@doi [Monthly Notices of the Royal Astronomical
  Society] {10.1093/mnrasl/sly172}, 481, L120

\bibitem[\protect\citeauthoryear{Verlet}{Verlet}{1967}]{Verlet1967}
Verlet L.,  1967, \mn@doi [Physical Review] {10.1103/PhysRev.159.98}, 159, 98

\bibitem[\protect\citeauthoryear{Vila}{Vila}{1999}]{Vila1999}
Vila J.~P.,  1999, \mn@doi [Mathematical Models \& Methods in Applied Sciences
  - M3AS] {10.1142/S0218202599000117}, 09, 161

\bibitem[\protect\citeauthoryear{Viola, Monaco, Borgani, Murante  \&
  Tornatore}{Viola et~al.}{2008}]{Viola+2008}
Viola M.,  Monaco P.,  Borgani S.,  Murante G.,   Tornatore L.,  2008, \mn@doi
  [Monthly Notices of the Royal Astronomical Society]
  {10.1111/j.1365-2966.2007.12598.x}, 383, 777

\bibitem[\protect\citeauthoryear{Wadsley, Stadel  \& Quinn}{Wadsley
  et~al.}{2004}]{Wadsley+2004}
Wadsley J.~W.,  Stadel J.,   Quinn T.,  2004, \mn@doi [New Astronomy]
  {10.1016/j.newast.2003.08.004}, 9, 137

\bibitem[\protect\citeauthoryear{Wadsley, Keller  \& Quinn}{Wadsley
  et~al.}{2017}]{Wadsley+2017}
Wadsley J.~W.,  Keller B.~W.,   Quinn T.~R.,  2017, \mn@doi [Monthly Notices of
  the Royal Astronomical Society] {10.1093/mnras/stx1643}, 471, 2357

\bibitem[\protect\citeauthoryear{Weinberger, Springel  \& Pakmor}{Weinberger
  et~al.}{2020}]{Weinberger+2020}
Weinberger R.,  Springel V.,   Pakmor R.,  2020, \mn@doi [The Astrophysical
  Journal Supplement Series] {10.3847/1538-4365/ab908c}, 248, 32

\bibitem[\protect\citeauthoryear{Wendland}{Wendland}{1995}]{Wendland1995}
Wendland H.,  1995, \mn@doi [Advances in Computational Mathematics]
  {10.1007/BF02123482}, 4, 389

\bibitem[\protect\citeauthoryear{Xu}{Xu}{1995}]{Xu1995}
Xu G.,  1995, \mn@doi [The Astrophysical Journal Supplement Series]
  {10.1086/192166}, 98, 355

\bibitem[\protect\citeauthoryear{Zel'dovich}{Zel'dovich}{1970}]{Zeldovich1970}
Zel'dovich Y.~B.,  1970, Astronomy and Astrophysics, 5, 84

\bibitem[\protect\citeauthoryear{{von Neumann}}{{von
  Neumann}}{1961}]{VonNeumann1961}
{von Neumann} J.,  1961, Collected Works

\makeatother
\end{thebibliography}



\appendix

\section{Slope-Limiters in \textsc{OpenGadget3}} \label{app:limiters}

We implemented seven different slope-limiters and therein variants of their specific parameters in \textsc{OpenGadget3}. The main concept is described in Sec.~\ref{sec:slope_limiter}. In general, we substitute $\nabla W_{i,k}\to \alpha_{i,k}\nabla W_{i,k}$ for each particle $i$ and component $k$, for the face interpolation, with $\alpha_{i,k}\in [0,1]$. In the following, we briefly describe the implemented limiters.

The simplest option are to use a zeroth order interpolation setting
\begin{align}
    \alpha_{i,k}^{\text{ZERO SLOPES}} =~& 0
\end{align}
or to include no slope-limiter
\begin{align}
    \alpha_{i,k}^{\text{NULL}} =~& 1.
\end{align}
Alternatively, we implemented several more complex limiters. A commonly used one is a TVD scalar limiter \citep{Duffell&MacFadyen2011}, which is designed to produce good results especially for strong shocks. Compared to the other limiters implemented, it is the most diffusive one. It sets
\begin{align}
    \alpha_{i,k}^{\text{TVD SCALAR}} =~& \min_{j\in \text{Ngb}}\max\mathcases{0 \\ \min\mathcases{1\\ \Delta W_{ij,k}/\dif W_{ij,k}}} \label{eq:tvd_limiter}
\end{align}
where $\Delta \myvector{W}_{ij}=\myvector{W}_j - \myvector{W}_i$, $\dif \myvector{W}_{ij}=\dif \myvector{r}_{ij}\cdot \nabla\outerprod \myvector{W}$.

An alternative is the scalar limiter which is a modified version of the \citet{Balsara2004,Gaburov&Nitadori2011} limiter, with relaxed constraints, as presented in the \textsc{GANDALF} code \citep{Hubber+2018}. It looses the TVD behavior but is less diffusive. Only the extreme values are used over the neighbors in the numerator, and the maximum possible values to be reconstructed in the denominator, thus avoiding an additional neighbor loop, leading to the limiter
\begin{align}
    \alpha_{i,k}^{\text{SCALAR}} =~& \max\mathcases{0 \\ \min\mathcases{1 \\ \min\mathcases{\frac{\Delta W_{i\max,k}}{\abs{\dif r}_{\max}\abs{\nabla W_{k}}}\\ \frac{\Delta W_{i\min,k}}{\abs{\dif r}_{\max}\abs{\nabla W_{k}}} }}} \label{eq:scalar_limiter}
\end{align}
where 
$\Delta W_{i\min/\max,k}= \abs{W_{i,k}-\min/\max_{j\in \text{Ngb}}W_{j,k}}$, and $\abs{\dif r}_{\max}= \max\left( \max_{j\in\text{Ngb}}\abs{r_{ij}}, h_i \right)$. In contrast to the TVD limiter, only the global neighbor distribution is considered. Thus, values calculated from all neighbors individually for the TVD limiter are calculated in an approximate way. Finally, we implemented the limiters used both in the \textsc{Arepo} and \textsc{gizmo} code.
In the \textsc{Arepo} code \citep{Springel2010}, the slope is limited using the \citet{Barth&Jespersen1989} limiter
\begin{align}
    \alpha_{i,k}^{\text{\textsc{Arepo}}} =~& \min_{j\in \text{Ngb}}\mathcases{\Delta W_{i\max,k}/\dif W_{ij,k}~&\text{if }\dif W_{ij,k}>0\\\Delta W_{i\min,k}/\dif W_{ij,k}~&\text{if }\dif W_{ij,k}<0 \\ 1& \text{if }\dif W_{ij,k}=0.} \label{eq:arepo_limiter}
\end{align}
It lies in between the TVD and scalar limiter, as only the dividend is approximated from the global neighbor distribution, while the divisor is still calculated for all neighbors individually.

In \textsc{gizmo} \citep{Hopkins2015} a general limiter is introduced described by 
\begin{align}
    \alpha_{i,k}^{\text{\textsc{gizmo}}} =~& \min\mathcases{1\\ \beta_i\min\mathcases{\frac{\dif W_{i\max,k}}{0.5h_i\abs{\nabla W_{k}}}\\ \frac{\dif W_{i\min,k}}{0.5h_i\abs{\nabla W_{k}}}.}} \label{eq:gizmo_limiter}
\end{align}
Also this limiter has the advantage of avoiding an additional neighbor loop.
The parameter $\beta$ has to be $\beta_i>0.5$ to ensure second order stability. A higher number corresponds to a more aggressive, less diffusive and less stable limiter. We use the suggested value $\beta=2$ of \citet{Hopkins2015}, which is a compromise to reduce numerical diffusivity while still working for very strong interacting shocks. While they suggest this value to be used only for particle distributions being isotropic enough based on the condition number, we use this value always as we found hardly any differences.
For $\beta=2$, this limiter is also similar to the scalar limiter with the difference that the theoretically possible distance between neighbors is defined only by the smoothing length. In addition, \citet{Hopkins2015} provide a pairwise limiter, acting on only one specific interaction, instead of all neighbors. For this, it uses already limited slopes for the interpolation.
The pairwise limiter described by \citet{Hopkins2015} limits the already interpolated face values. The aim is to directly calculate the face value $W_{ij,k}^{\text{new}}$, starting from the extrapolated value $W_{ij,k}^{\text{frame}}$ according to Eqn.~(\ref{eq:mfm_face_value}), possible already with limited gradients. If $W_{i,k}=W_{j,k}$, the face value is just chosen the same as the particle values $W_{ij,k}^{\text{new}}=W_{i,k}$.
Otherwise, the values
\begin{align}
    \delta_1 =~& \psi_1\abs{W_{i,k}-W_{j,k}} \label{eq:gizmo_pairwise_gamma1}\\
    \delta_2 =~& \psi_2\abs{W_{i,k}-W_{j,k}}
\end{align}
are calculated. The free parameters $\psi_{1/2}$ are tuned to $\psi_1=0.5$, $\psi_2=0.25$.
A simple intermediate value used later is given by
\begin{align}
    \bar W_{ij,k} =~& W_{i,k}+\frac{\dif r_{ij}}{\dif r_{i}^{\text{frame}}}(W_{j,k}-W_{i,k}).
\end{align}
The maximum/minimum value is $W_{\min/\max,k}=\min/\max(W_{i,k},W_{j,k})$.
Depending on how the two face values compare, the new face value is calculated:
If $W_{i,k}<W_{j,k}$, then
\begin{align}
    W_{ij,k}^{\text{new}} =~& \max\mathcases{\mathcases{W_{\min,k}-\delta_1~&\text{if }\text{SIGN}(W_{\min,k}-\delta_1)=\text{SIGN}(W_{\min,k}) \\ \frac{W_{\min,k}}{1+\frac{\delta_1}{\abs{W_{\min,k}}}}~&\text{else}} \\ \min\mathcases{W_{ij,k}^{\text{frame}} \\ \bar W_{ij,k}+\delta_2.}} \label{eq:gizmp_pairwise_limiter1}
\end{align}
If $W_{i,k}\ge W_{j,k}$, then
\begin{align}
    W_{ij,k}^{\text{new}} =~& \min\mathcases{\mathcases{W_{\max,k}+\delta_1~&\text{if }\text{SIGN}(W_{\max,k}+\delta_1)=\text{SIGN}(W_{\max,k}) \\ \frac{W_{\max,k}}{1+\frac{\delta_1}{\abs{W_{\max,k}}}}~&\text{else}}\\
    \max\mathcases{W_{ij,k}^\text{frame} \\ \bar W_{ij,k}-\delta_2.}}  \label{eq:gizmp_pairwise_limiter2}
\end{align}
The same limiter is applied for particle $j$. Finally, the \textsc{gizmo} code uses a slightly different pairwise limiter. Depending on the tolerance $t$ chosen as input parameter for the run with a typical value of 1, the parameters
\begin{align}
    \psi_1 =~& \mathcases{0 &t=0 \\ 0.5 &t=1 \\ 0.75~~&t=2}\\
    \psi_2 =~& \mathcases{0 &t=0 \\ 0.4 &t=1 \\ 0.375~&t=2}
\end{align}
are defined. To calculate $\bar W_{ij,k}$, the factor $\dif r_{ij}/\dif r_i^{\text{frame}}$ is approximated by the first order value $0.5$. Except these differences, the limiter is identical to the already described one.
In our implementation, we apply the limiter in the reference frame of the interface, such that the velocity is a relative velocity. This makes the limiter Lagrangian and increases the symmetry between different directions within symmetric flows such as in the Zeldovich pancake.

\section{Effect of the Riemann solver} \label{app:Riemann}

In \textsc{OpenGadget3} we use an exact, iterative Riemann solver \citep{Toro2009} by default. This is however, computationally expensive as up to eight iterations are used to get close to the exact solution. An alternative is using approximate Riemann solvers, where we implemented a Roe solver \citep{Roe1981}, the HLL solver \citep{Toro2009}, and the HLLC \citep{Toro2009} solver.

The strongest effect in runtime is present for strong shocks, where we find an speedup of up to $20$~per cent in total. For more smooth problems, where less iterations of the exact solver are necessary, the speedup for the calculation of the fluxes itself is $20$~per cent, resulting on an overall speedup of only up to $9$~per cent for such problems dominated by hydrodynamical calculations. The effect becomes less important when using gravity and possibly even more extensions in cosmological applications. Already for the hydrostatic sphere, there is no significant difference in runtime, or even a slight increase.

\begin{figure*}
    \centering
    \includegraphics[width=\textwidth]{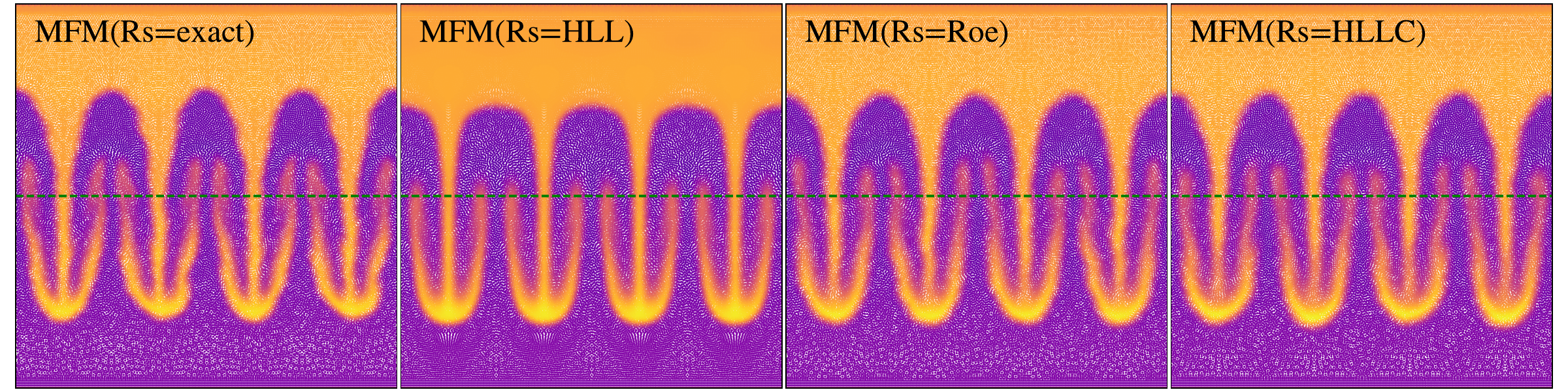}
    \caption{Rayleight-Taylor instability at time $t=3.6$, evolved using different Riemann solvers. Secondary instabilities are more or less pronounced, depending on the solver used.}
    \label{fig:riemann_rt}
\end{figure*}

As the Riemann solver introduces numerical diffusivity, the evolution will be different. This can be seen in various test problems, such as the Rayleigh-Taylor instability, shown in Fig.~\ref{fig:riemann_rt}.
While the instability evolved using the exact Riemann solver shows the most prominent secondary instabilities, closely followed by the Roe solver and the HLLC solver, the HLL Riemann solver leads to a suppression of any asymmetries in the final shape of the instability. This is a result of the additional numerical diffusivity introduced by the Riemann solver, as discussed in Sec.~\ref{sec:sphere}.

While for specific problems these alternative solvers could lead to faster results, we in general use the most accurate exact Riemann solver. The increase in runtime is compensated by the gain in accuracy.

\section{Soundwave convergence with \textsc{Arepo}} \label{app:soundwave_arepo}
As described in Sec.~\ref{sec:soundwave}, we would expect \textsc{Arepo} to have better convergence than observed. A first reason is the mesh regularization. If triggered, the position of the cells are shifted, introducing a small numerical noise. In addition, small interfaces which contribute by less than $10^{-5}$ to the total interface are neglected. While this makes the code more stable in extreme environments, it introduces small errors (R. Pakmor, 2023, priv. comm.), which will be relevant for the very small deviations analyzed here.

\begin{figure*}
    \centering
    \includegraphics[width=0.8\textwidth]{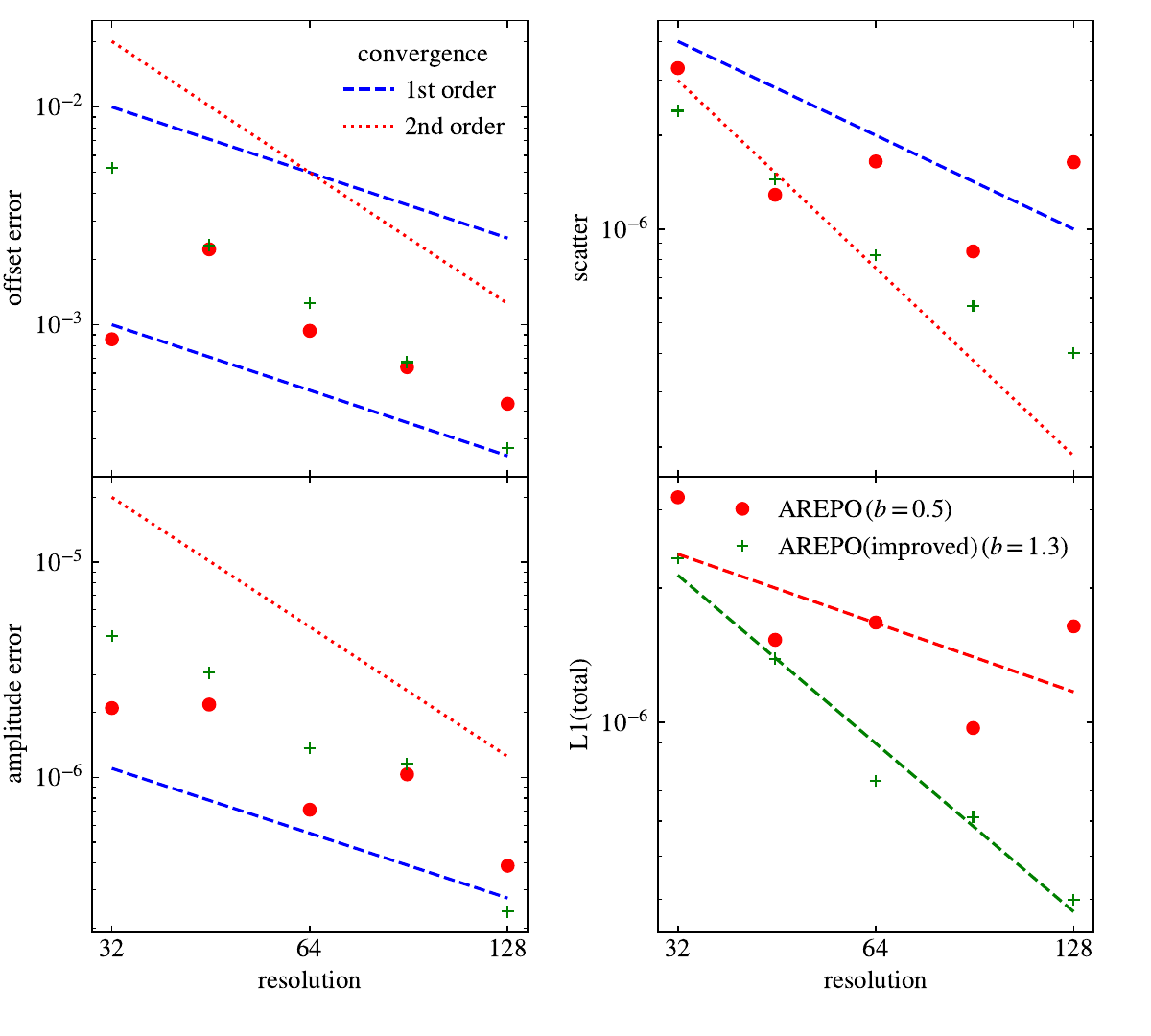}
    \caption{Offset-, amplitude-, scatter and total L1-errors of the density of a soundwave at $t=\frac{2}{c_s}$. Applying the changes described in the text reduces especially the scatter error, thus increasing the order of convergence and making also the determination of the other components more reliable. Thus, also the overall order of convergence increases.}
    \label{fig:soundwave_arepo}
\end{figure*}

Turning off the mesh regularization and only skipping interfaces which contribute by less than $10^{-8}$, \textsc{Arepo} shows much better convergence behavior, close to what is expected from the analysis by \citet{Weinberger+2020}, as shown in Fig.~\ref{fig:soundwave_arepo}. Especially, the scatter error drastically decreases, also making the determination of the other error components more reliable.

While these changes can lead to a better convergence behavior, they will cause other problems in cosmological simulations, such that in physical applications rather the non-optimized behavior would be observed.

\bsp	
\label{lastpage}
\end{document}